\documentclass[11pt,a4paper]{article}
\pdfoutput=1
\usepackage{jheppub}
\usepackage[T1]{fontenc}
\usepackage[english]{babel}
\usepackage{lmodern}
\usepackage[colorlinks = true]{hyperref}
\usepackage{amsfonts}
\usepackage{amsmath}
\usepackage{bbm}
\usepackage{bm}
\usepackage{color}
\usepackage{slashed}
\usepackage{graphicx}
\usepackage{dsfont}
\usepackage{bbold}
\usepackage{braket}
\usepackage{caption}

\allowdisplaybreaks

\makeatletter \@addtoreset{equation}{section} \makeatother
\DeclareMathOperator{\arccot}{arccot}

\title{Effective field theories for dark matter pairs in the early universe: cross sections and widths}

\author[a]{S.~Biondini,}
\author[b,c,d]{N.~Brambilla,}
\author[b]{G.~Qerimi}
\author[b]{and A. Vairo}

\affiliation[a]{Department of Physics, University of Basel,\\
Klingelbergstr. 82, CH-4056 Basel, Switzerland}
\affiliation[b]{Physik-Department, Technical University Munich,\\
James-Franck-Str.  1, 85748 Garching, Germany}
\affiliation[c]{Institute for Advanced Study, Technical University Munich, \\
Lichtenbergstrasse 2 a, 85748 Garching, Germany}
\affiliation[d]{Munich Data Science Institute, Technical University Munich, \\
Walther-von-Dyck-Strasse 10, 85748 Garching, Germany}

\emailAdd{simone.biondini@unibas.ch}
\emailAdd{nora.brambilla@ph.tum.de}
\emailAdd{gramos.qerimi@tum.de}
\emailAdd{antonio.vairo@ph.tum.de}

\abstract{
  In order to predict the cosmological abundance of dark matter, an estimation of particle rates in an expanding thermal environment is needed.
  For thermal dark matter, the non-relativistic regime sets the stage for the freeze-out of the dark matter energy density. 
  We compute transition widths and annihilation, bound-state formation, and dissociation cross sections of dark matter fermion pairs 
  in the unifying framework of non-relativistic effective field theories at finite temperature, 
  with the thermal bath modeling the thermodynamical behaviour of the early universe.
  We reproduce and extend some known results  for the paradigmatic case of a dark fermion species coupled to dark gauge bosons.
  The effective field theory framework allows to highlight their range of validity and consistency, and to identify some possible improvements.
}

\begin{document}
\maketitle

\section{Introduction}
One of the major challenges in cosmology and particle physics is to understand the matter content of the universe.
Notably, visible ordinary matter appears to be only a small fraction of the matter in the cosmos,
whereas the bulk seems to come in the form of non-luminous and non-baryonic particles, dubbed dark matter (DM).
Complementary measurements of large scale structures, galaxy formation, gravitational lensing and the cosmic
microwave background (CMB) suggest that more than 80\% of the matter in the cosmos consists of DM.
The most accurate determination for the DM energy density is provided by anisotropies
in the CMB and amounts to $\Omega_{\hbox{\tiny DM}} h^2 = 0.1200 \pm 0.0012$~\cite{Planck:2018nkj}, where $h$ is the reduced Hubble constant.

Although this is not the only viable option, the interpretation of DM as due to new, yet undiscovered, particles has been put forward in a plethora of models,  
see e.g.~\cite{Bertone:2004pz,Feng:2010gw} for extensive reviews.
One can categorize DM particles according to their production mechanism in the early universe.

The thermal freeze-out has been widely adopted to infer the present-day abundance of a DM candidate.
It allows to effectively link the particle model parameters, such as couplings and DM mass, with the observed relic density.
The freeze-out mechanism has been extensively used for Weakly Interacting Massive Particles (WIMPs), however it equally applies
to cases where interactions are stronger. One assumes an initial thermal abundance for the dark species,
whose evolution is governed by the interplay between the thermally averaged annihilation cross section $\langle \sigma_{\hbox{\scriptsize ann}} v_{\hbox{\scriptsize rel} } \rangle$
and the expansion rate of the universe, $H$, namely the Hubble rate.
The standard rate equation is a Boltzmann equation of the form~\cite{Lee:1977ua,Gondolo:1990dk,Griest:1990kh}\footnote{
In the case of annihilation of identical particles, e.g. Majorana fermions, the factor $1/2$ on the right-hand side of eq.~\eqref{Boltzmann_1} should be replaced by $1$.
}
\begin{equation}
(\partial_t + 3H) n = - \frac{1}{2}\langle \sigma_{\hbox{\scriptsize ann}} v_{\hbox{\scriptsize rel} } \rangle (n^2-n^2_{{\rm{eq}}}) \, ,
\label{Boltzmann_1}
\end{equation} 
where $n$ is the total number density of DM particles ($n_{{\rm{eq}}}$ is that in equilibrium\footnote{
  We understand thermal equilibrium as chemical $\emph{and}$ kinetic equilibrium.})
and $v_{\hbox{\scriptsize rel}} = |\bm{v}_1-\bm{v}_2|$ is the relative velocity of the annihilating pair.
For a DM particle of mass $M$, the chemical freeze-out occurs at a temperature $T \approx M/25$.
Therefore, at freeze-out the dark matter particles are \textit{non-relativistic}.\footnote{
  The freeze-out temperature is estimated by equating the expansion with the  annihilation cross section
  $H \sim n_{{\rm{eq}}} \langle \sigma_{\hbox{\scriptsize ann}} v_{\hbox{\scriptsize rel} } \rangle$,
  namely $T^2/M_{\hbox{\tiny Pl}} \sim \left( \frac{MT}{2\pi}\right)^{3/2} e^{-M/T} \alpha^2/M^2$ where $\alpha$
  is some fine structure constant and $M_{\hbox{\tiny Pl}} \simeq 1.2 \times 10^{19}$ GeV.
  After the chemical equilibrium is lost, which corresponds to the chemical freeze-out,
  kinetic equilibrium is usually kept for longer times and provides a thermal distribution for the dark matter momenta (and velocities).} 

It is crucial to calculate $\langle \sigma_{\hbox{\scriptsize ann}} v_{\hbox{\scriptsize rel} } \rangle$ accurately because the present-day DM energy density,
as predicted by a given model, depends on it through the solution of eq.~\eqref{Boltzmann_1}.
The DM mass is in turn fixed as a function of the other model parameters to reproduce $\Omega_{\hbox{\scriptsize DM}} h^2$.
A solid prediction for DM mass benchmarks compatible with the relic density is needed to establish the viability of models, 
guide the experimental searches and put DM phenomenology on a sound theoretical ground.
However, determining $\langle \sigma_{\hbox{\scriptsize ann}} v_{\hbox{\scriptsize rel} } \rangle$ by including the full features of each model,
and the thermal environment, is not straightforward. 

In a variety of theories, DM interacts with gauge bosons or scalars that induce long-range interactions because of repeated soft exchanges.
While a successful WIMP-like DM candidate should be weakly interacting with the Standard Model (SM), we cannot say much on the interactions
between the DM particles themselves and/or among additional degrees of freedom in the dark sector.
In general, there may be non-negligible forces between DM particles mediated by light particles leading to the formation of 
bound states of genuine WIMPs, sometimes referred to as \emph{wimponium}~\cite{MarchRussell:2008tu,Shepherd:2009sa}.
The main motivation that makes self-interacting DM welcome are some compelling inconsistencies between the predictions of collisionless cold DM and the observed large-scale structure
of the universe~\cite{Spergel:1999mh}, in the numbers of the galactic satellite halos,
and in the DM density profiles in the galaxies~\cite{Markevitch:2003at,Randall:2007ph,Feng:2008mu,Rocha:2012jg,Foot:2013nea,Foot:2013lxa,10.1111/j.1365-2966.2012.21182.x,Tulin:2013teo,Kahlhoefer:2013dca,Harvey:2015hha,Kaplinghat:2015aga}. 

Bound-state effects cannot be avoided in the context of next-to-minimal DM models,
where a mediator between the visible and dark sector is charged under some of the SM gauge groups.
In coannihilation scenarios, the presence of additional massive states can drastically affect the thermal freeze-out
when the coannihilating partner is close in mass with the DM particle~\cite{Griest:1990kh,Edsjo:1997bg}.
Consequently, one has to track also its (co)annihilations. For example, when the partner particle is charged under QCD,
long-range interactions mediated by gluons affect severely the annihilation process
and the formation of (many) bound states has to be included~\cite{Ellis:2014ipa,Liew:2016hqo,Kim:2016kxt,Mitridate:2017izz,Garny:2021qsr}.

Accordingly, in recent years there has been some effort in revisiting dark-matter pair annihilation by encompassing near-threshold effects induced by repeated soft exchange.
First, the so-called Sommerfeld enhancement has been included in DM freeze-out calculations for several models,
resulting in mass benchmarks that give DM energy densities compatible with observation 
rather different from the case with no threshold 
effects~\cite{Hisano:2006nn,Cirelli:2007xd,Cirelli:2008id,Feng:2009mn,Cirelli:2009uv,Feng:2010zp,deSimone:2014pda,Beneke:2014gja,Beneke:2014hja,Ibarra:2015nca}.
Typically, the Sommerfeld enhancement increases the annihilation cross section,
for a pair in an attractive channel, and leads to a larger dark matter mass compatible with the observed relic density.
More recently, bound-state effects have shown a relevant impact on DM annihilations as well, as pointed out originally in refs.~\cite{Feng:2009mn,vonHarling:2014kha}.
Indeed, DM bound states contribute as an efficient annihilation channel and dilute further the overall DM population.
The presence of meta-stable bound states is simply another manifestation of repeated soft exchanges:
in the spectrum of a two-particle system there is an above threshold continuum of states along with bound states below threshold.

Treating interacting non-relativistic particle pairs in a thermal environment is complicated by the presence of several energy scales. 
The energy scales dynamically generated by the relative motion are 
the momentum transfer, which is also proportional to the inverse of the typical size of the pair, and the kinetic/binding energy of the pair.
Such scales are hierarchically ordered as $M \gg M v_{\hbox{\scriptsize rel} } \gg Mv_{\hbox{\scriptsize rel} }^2$ for 
near threshold particles moving with relative velocity $v_{\hbox{\scriptsize rel} }$.
For Coulombic bound states, the relative velocity of the pair is fixed by the virial theorem to be $v_{\hbox{\scriptsize rel} } \sim \alpha$,
hence the corresponding hierarchy is $M \gg M \alpha \gg M \alpha^2$, 
$\alpha = g^2 / (4 \pi)$ being the fine structure constant in terms of the coupling $g$ between the DM particle and the force mediator. 
The thermodynamical scales include the plasma temperature $T$ and the Debye mass $m_D$, which is the inverse of the chromoelectric screening length;
in a weakly-coupled plasma $m_D \sim gT$. 
For kinetically equilibrated particles and pairs the typical momentum is of order $\sqrt{MT}$ and the typical relative velocity is then $v_{\hbox{\scriptsize rel} } \sim \sqrt{T/M}$;
  clearly, after freeze-out $v_{\hbox{\scriptsize rel} } \ll 1$.
Depending on the plasma temperature, the scale $\sqrt{MT}$ may be larger or smaller than the scales $M\alpha$ and $M\alpha^2$.

\begin{figure}[ht]
    \centering
    \includegraphics[scale=0.58]{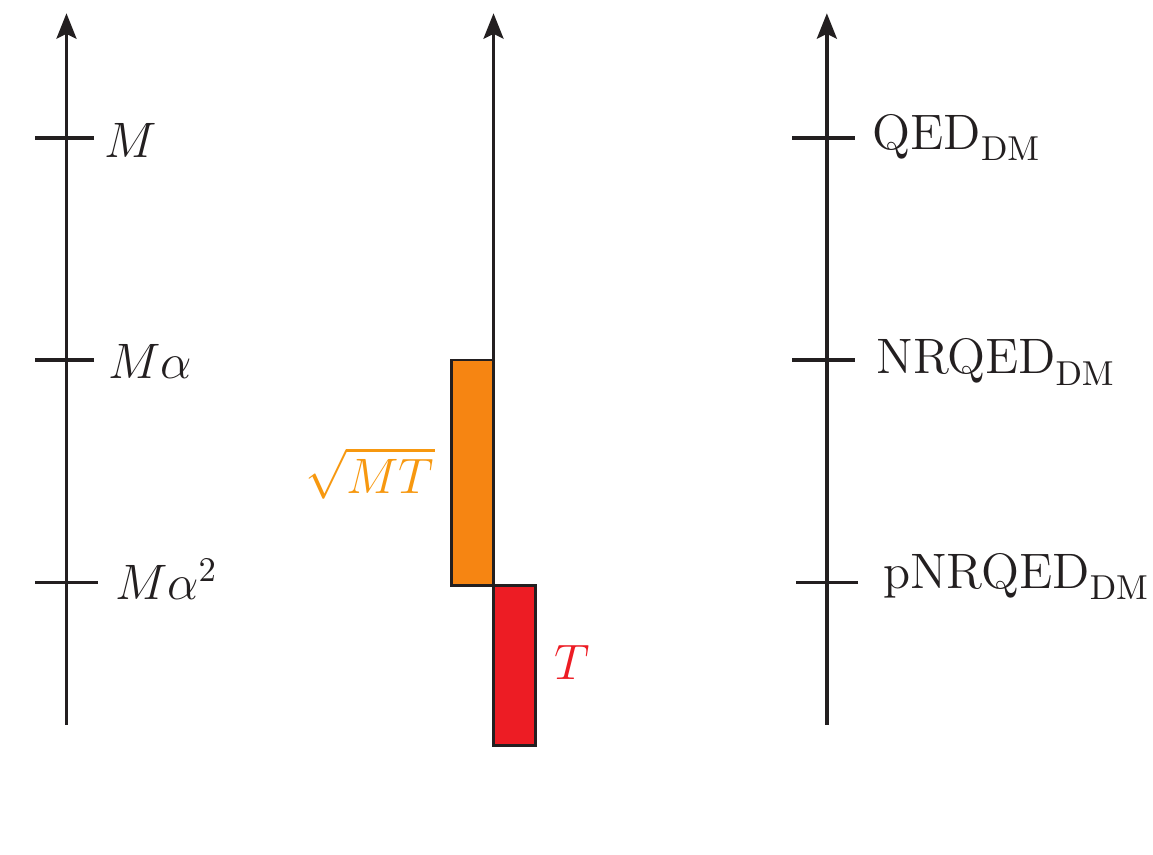}
    \caption{
      Hierarchy of energy scales and effective field theories considered in this work for the DM Lagrangian density defined in eq.~\eqref{lag_mod_0}. A similar tower of EFTs applies for the non-abelian model given in eq.~\eqref{non_ab_model} and the corresponding hierarchy of scales considered in section~\ref{sec:non_abelian_model}. 
       }
    \label{fig:intro_1}
\end{figure}

In this work, we compute production cross sections and decay widths of near threshold weakly-coupled dark matter particle-antiparticle states
in a thermal bath of SM particles describing the early universe. 
We exploit the hierarchy of energy scales typical of near threshold states and a weakly-coupled plasma 
by replacing the fundamental DM theory with a sequence of non-relativistic effective field theories (EFTs). 
In most of the paper, from section~\ref{sec:NREFTs} to section~\ref{sec:numerics}, we consider a minimal scenario that consists in having a DM sector made of a new species of fermions, \textit{dark fermions},
and photons, \textit{dark photons}, that interact like in QED, see section~\ref{sec:NREFTs}.
To remark the new particle content of the theory, we call it QED$_\textrm{DM}$.
The EFT that follows from  QED$_\textrm{DM}$ by integrating out modes associated with the energy scale $M$
is non-relativistic QED ($\textrm{NRQED}_{\textrm{DM}}$)~\cite{Caswell:1985ui}.
The EFT that follows from $\textrm{NRQED}_{\textrm{DM}}$ by integrating out dark photons 
of energy or momentum of order $M v_{\hbox{\scriptsize rel}}$ is potential non-relativistic QED 
($\textrm{pNRQED}_{\textrm{DM}}$)~\cite{Pineda:1997bj,Pineda:1998kn,Brambilla:2004jw}.
We sketch the tower of energy scales and EFTs considered in this work in figure~\ref{fig:intro_1}.
Potential $\textrm{NRQED}_{\textrm{DM}}$ is made of low-energy particles only and well suited to compute 
near threshold observables. Many of them have been computed also elsewhere, hence the main motivation of this part of the work 
is to present a systematic and coherent framework for these computations highlighting at the same time 
their regime of validity and possible improvements.
In the final part of the paper, i.e. section~\ref{sec:non_abelian_model}, we consider the case of a dark sector made of dark fermions coupled to SU($N$) dark gauge bosons. Here, the prototype low-energy EFT that we adopt is a pNRQCD-like EFT~\cite{Pineda:1997bj,Brambilla:1999xf}, which describes non-relativistic fermion-antifermion pairs interacting through non-abelian gauge bosons.

\begin{figure}[ht]
    \centering
    \includegraphics[scale=0.53]{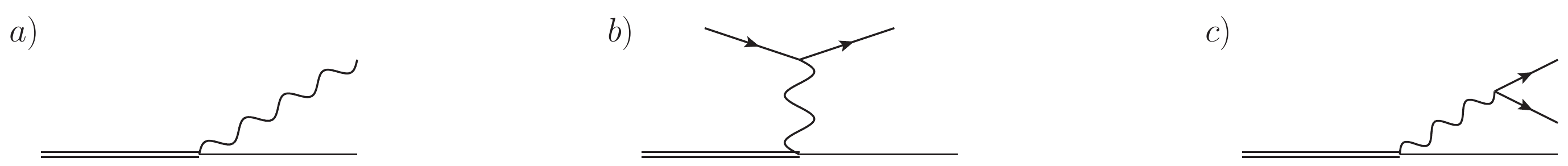}
    \caption{Bound-state formation processes.
      Solid double lines stand for dark matter pairs in a scattering state, 
      solid lines for bound states, wiggly lines for gauge bosons (e.g.~dark photons), arrowed thin lines for light plasma constituents
      if the gauge boson couples to some of them.}
    \label{fig:intro_2}
\end{figure}

Near threshold effects not only affect directly the annihilation cross 
section in eq.~\eqref{Boltzmann_1}, but also add to it new production and 
decay mechanisms that may eventually modify the abundance of DM particles significantly.
We summarize them in figure \ref{fig:intro_2}. 
The diagram labeled \emph{a)} shows a transition from a scattering state to a bound state via a gauge boson
emission~\cite{vonHarling:2014kha,Petraki:2015hla,Asadi:2016ybp,Beneke:2016ync,Ellis:2015vna,Liew:2016hqo,Mitridate:2017izz,Cirelli:2016rnw,Beneke:2016jpw,Harz:2018csl}. 
The diagram labeled \emph{b)} shows an inelastic collision of a scattering
state with a constituent of the thermal bath that turns it into a bound
state~\cite{Kim:2016kxt,Biondini:2017ufr,Biondini:2018pwp,Biondini:2018xor,Binder:2018znk,Biondini:2018ovz,Biondini:2019zdo,Biondini:2019int,Binder:2019erp,Binder:2020efn,Binder:2021otw}.
Finally, diagram \emph{c)} shows the decay of the emitted gauge boson mediator, 
if this couples to some light degrees of freedom in the thermal bath of the early universe~\cite{Binder:2020efn,Binder:2021otw}.
The latter two processes entail further model-dependent details on the interactions among the mediator and the additional light degrees of freedom. 
The relative importance of the various mechanisms depends on the temperature.
In this work, we focus on the bound-state formation/dissociation as induced by the radiative emission of a gauge vector boson, namely process \emph{a)}.
In the hierarchy of scales that we assume in our work (cf.~eq.~\eqref{scale_arrang}), this is expected to be the dominant process~\cite{Brambilla:2008cx}.
Moreover, we do not include in our model couplings of the dark gauge boson with light degrees of freedom, this excludes explicitly the possibility of having the processes \emph{b)} and \emph{c)}.
It also excludes the generation of a Debye mass.
Recent investigations of near-threshold effects induced by scalar mediators and their applications to dark matter models can be found in
refs.~\cite{Harz:2017dlj,Harz:2019rro,Oncala:2019yvj,Oncala:2018bvl,Biondini:2021ccr,Biondini:2021ycj}. 

The outline of the paper is as follows.
In section~\ref{sec:NREFTs}, we introduce an abelian DM model, and 
in sections~\ref{sec:NREFT_part} and~\ref{sec:pNREFT_part}
its low-energy effective field theories upon integrating out modes carrying energies and momenta of order $M$ and $M v_{\textrm{rel}}$.
In section~\ref{sec:sommerfeld}, we consider annihilations, 
and derive the corresponding observables in section~\ref{sec:nrqed_annihilation}
and~\ref{sec:pnrqed_annihilation}.
Electric transitions are covered in section~\ref{sec:Electric_transitions}.
Using the optical theorem, we derive the bound-state formation cross section in section~\ref{sec:Electric_transitions_form} and the bound-state dissociation 
thermal width in section~\ref{sec:Electric_transitions_diss}. 
Details of the derivations and the treatment of excitations, de-excitations and bremsstrahlung effects are collected in the appendices~\ref{sec:app_A}, \ref{sec:app_B} and \ref{sec:app_C}.
In section~\ref{sec:numerics}, we scrutinize the interplay between the coupling and velocity expansion, and the inclusion of excited states on the dark matter energy density. 
Moreover, we present contour plots for the dark matter mass and coupling. 
In section~\ref{sec:non_abelian_model}, we discuss a non-abelian SU($N$) model with fermionic dark matter, 
whose electric dipole matrix elements are collected in appendix~\ref{sec:app_D}.
Finally, conclusions and outlook are in section~\ref{sec:concl}.

\section{EFTs for non-relativistic dark matter pairs}
\label{sec:NREFTs}
In this and in the next three sections,
we consider a simple model where the dark sector consists of a dark Dirac fermion $X$
that is charged under an abelian gauge group~\cite{Feldman:2006wd,Fayet:2007ua,Goodsell:2009xc,Morrissey:2009ur,Andreas:2011in}.
We denote the corresponding dark photon with $\gamma$. The Lagrangian density reads
\begin{equation}
\mathcal{L}=\bar{X} (i \slashed {D} -M) X -\frac{1}{4} F_{\mu \nu} F^{\mu \nu} + \mathcal{L}_{\textrm{portal}} \, ,
\label{lag_mod_0}
\end{equation}
where the covariant derivative is $D_\mu=\partial_\mu + i g A_\mu$, with $A_\mu$ the dark photon field and $F_{\mu \nu} = \partial_\mu A_\nu - \partial_\nu A_\mu$;
we define the fine structure constant as $\alpha \equiv g^2/(4 \pi)$.
The term $\mathcal{L}_{\textrm{portal}}$ comprises additional interactions coupling the dark photon with the SM degrees of freedom.
A popular interaction is a mixing with the neutral components of the SM gauge fields \cite{Holdom:1985ag,Foot:1991kb} (see also \cite{Koren:2019iuv}).
Such interactions are responsible for the eventual decay of the dark photons,
in this way avoiding that their number density dominates the universe at later stages.
Moreover, additional fermionic or scalar degrees of freedom may be coupled to the dark photon.
In an abelian theory, and at temperatures $T \ll M$, they are responsible for quantum corrections to the dark photon propagator, 
whose pole develops a real and an imaginary part~\cite{Braaten:1991gm,Escobedo:2008sy,Escobedo:2010tu}.\footnote{
  The situation is somewhat different in the non-abelian case, where quantum corrections to the dark gluon propagator
  may be induced also by dark gluon self interactions \cite{Brambilla:2008cx,Brambilla:2010vq,Brambilla:2011sg,Brambilla:2013dpa}.}
The real part introduces a screening (Debye) mass $m_D$ of order $gT$ for the temporal dark photon,
whereas the imaginary part of the pole originates from $2 \to 2$ scatterings with plasma constituents, also referred to as Landau damping.
It is beyond the scope of this work to elaborate further either on $\mathcal{L}_{\textrm{portal}}$
or on the physical effects of additional fermionic or scalar degrees of freedom.
From now on we set $\mathcal{L}_{\textrm{portal}}=0$.

The Lagrangian \eqref{lag_mod_0} describes also processes involving two dark fer\-mi\-ons close to threshold,
i.e. processes where the fermions are non-relativistic and move with relative velocity $v_{\hbox{\scriptsize rel} } \ll 1$.
For $v_{\text{rel}} \sim \alpha$, (ladder) photons exchanged between the pair contribute with a relative factor of order $\alpha/v_{\text{rel}} \sim 1$ and need to be resummed.
The resummation generates bound-state poles of order $M\alpha^2$ at negative energies and a continuous scattering spectrum at positive energies.
The typical momentum exchanged between the pair when $v_{\text{rel}} \sim \alpha$ is $M\alpha$, which is of the order of the inverse Bohr radius of the bound state.
The dynamically generated scales $M\alpha$ and $M\alpha^2$ are the more separated the smaller $\alpha$ is: $M \gg M\alpha \gg M\alpha^2$.
We call them \emph{soft} and \emph{ultrasoft} scales, respectively, to distinguish them from the \emph{hard} scale associated with the mass $M$.
These energy scales affect significantly various processes in the near threshold momentum region,
like dark fermion pair annihilation, formation and transition via emission or absorption of photons. 
The use of the full Lagrangian \eqref{lag_mod_0} to compute near threshold observables is in general unpractical 
as all energy scales get entangled in the amplitudes.
It is more convenient, instead, to take advantage of the fact that the energy scales are hierarchically ordered 
and replace systematically \eqref{lag_mod_0} with a hierarchy of (non-relativistic) effective field theories along what has been done
for near threshold fermion pairs in QED and QCD~\cite{Caswell:1985ui,Pineda:1998kn,Brambilla:2004jw}.

Another relevant energy scale is the inverse correlation length characterizing the me\-dium made of dark fermions, dark photons and SM particles.
We assume the medium to be thermalized and identify the inverse of its correlation length with the temperature $T$. 
If the dark fermions are also thermalized then $v_{\text{rel}} \sim \sqrt{T/M}$.

In the following, we compute near threshold observables affecting the evolution of the dark matter density in the early universe.
In particular, we compute annihilation, dissociation and formation cross sections of dark matter fermion-antifermion pairs.
We compute these quantities by means of the tower of non-relativistic effective field theories depicted in figure~\ref{fig:intro_1}.
We include thermal effects due to the medium assuming that the temperature $T$ is about the ultrasoft scale $M\alpha^2$ or smaller.
  This guarantees that thermal effects do not enter the potential, which may be taken as Coulombic.
  The typical momentum of the thermalized dark fermions is then $M v_{\text{rel}} \sim \sqrt{MT} \lesssim M \alpha$, which implies $v_{\text{rel}} \lesssim \alpha$.
  Moreover, in the temperature ranges considered in this work it also holds that $\sqrt{MT}\gtrsim M\alpha^2$
  (more precisely $\sqrt{MT}\gtrsim M\alpha^2/4$).
  These conditions qualify $M v_{\text{rel}}$ as a soft scale and $M v_{\text{rel}}^2 \sim T$ as an ultrasoft scale.
  Our ensemble of thermalized dark fermions and antifermions realizes, therefore, the hierarchy of energy scales shown in figure~\ref{fig:intro_1}:
\begin{equation}
M \gg M\alpha \gtrsim \sqrt{MT} \gg M\alpha^2 \gtrsim T \,.
\label{scale_arrang}
\end{equation}

The hierarchy \eqref{scale_arrang} is of phenomenological interest for the study of near threshold effects in the minimal dark matter model
under consideration~\cite{vonHarling:2014kha,Cirelli:2016rnw}.
Indeed, since the decoupling from chemical equilibrium happens at around $T/M \approx 1/25$,
the condition \eqref{scale_arrang} may be fulfilled for most of the time after the decoupling.\footnote{
  If, more conservatively, we identify the absolute value of the ground state energy, $M\alpha^2/4$,
  with the ultrasoft scale, the condition $M\alpha^2/4 \gtrsim T$ is fulfilled for all times after decoupling if $\alpha \gtrsim 0.4$.}
In the model \eqref{lag_mod_0} with $\mathcal{L}_{\textrm{portal}}=0$, bound-state formation happens via radiative emission.
For complementary temperature regimes and bound-state formation processes, the latter triggered by additional degrees of freedom added to the model \eqref{lag_mod_0},
see also refs.~\cite{Biondini:2017ufr,Biondini:2018pwp,Binder:2018znk,Binder:2019erp,Binder:2020efn}.

\subsection{NRQED$_{\textrm{DM}}$}
\label{sec:NREFT_part}
At energies much smaller than $M$, the effective degrees of freedom are non-relativistic dark fermions and antifermions, low energy dark photons and SM particles.
The effective field theory that follows from \eqref{lag_mod_0} by integrating out dark photons and fermions of energy or momentum of order $M$ has the form of NRQED~\cite{Caswell:1985ui}.
It is organized as an expansion in $1/M$ and $\alpha$ and its Lagrangian density up to $\mathcal{O}(1/M^2)$ reads
\begin{eqnarray}
  \mathcal{L}_{\textrm{NRQED}_{\textrm{DM}}}\!\!&=&\!
        \psi^\dagger \left( i D_0 -M + \frac{\bm{{\rm{D}}}^2}{2 M}
       +c_{{\hbox{\tiny F}}} \frac{\bm{\sigma} \cdot g \bm{{\rm{B}}}}{2 M}
       +c_{{\hbox{\tiny D}}} \frac{\bm{\nabla} \cdot g \bm{{\rm{E}}}}{8 M^2}
       +i c_{{\hbox{\tiny S}}} \frac{\bm{\sigma} \cdot (\bm{{\rm{D}}} \times g \bm{{\rm{E}}}-g \bm{{\rm{E}}} \times \bm{{\rm{D}}} }{8 M^2} \right) \psi 
    \nonumber 
    \\
       \!\!&+&\! \chi^\dagger \left( i D_0 + M - \frac{\bm{{\rm{D}}}^2}{2 M}
       -c_{{\hbox{\tiny F}}} \frac{\bm{\sigma} \cdot g \bm{{\rm{B}}}}{2 M}
       +c_{{\hbox{\tiny D}}} \frac{\bm{\nabla} \cdot g \bm{{\rm{E}}}}{8 M^2} 
       +i c_{{\hbox{\tiny S}}} \frac{\bm{\sigma} \cdot (\bm{{\rm{D}}} \times g \bm{{\rm{E}}}-g \bm{{\rm{E}}} \times \bm{{\rm{D}}} }{8 M^2} \right) \chi 
    \nonumber 
    \\
      \!\!&-&\!\frac{1}{4} F^{\mu \nu}F_{\mu \nu}
       +\frac{d_2}{M^2} F^{\mu \nu} \bm{{\rm{D}}}^2 F_{\mu \nu}
       +\frac{d_s}{M^2} \psi^\dagger \chi \, \chi^\dagger \psi
       +\frac{d_v}{M^2} \psi^\dagger \, \bm{\sigma} \, \chi \cdot \chi^\dagger \, \bm{\sigma} \, \psi \, .
\label{NREFT_lag}
\end{eqnarray}
Here, $\psi$ is the two-component Pauli spinor that annihilates a dark matter fermion,
$\chi^\dagger$ is the Pauli spinor that annihilates an antifermion,
$\bm{{\rm{E}}}$ is the (dark) electric field, $E^i=F^{i0}$, and $\bm{{\rm{B}}}$ is the (dark) magnetic field, $B^i=-\epsilon_{ijk} F^{jk}/2$.
The first two lines in eq.~\eqref{NREFT_lag} describe how non-relativistic dark fermions and antifermions propagate and interact with low-energy dark photons of energy smaller than $M$.
The third line describes the propagation and effective self interaction of the photons; it also contains two four-fermion operators.

To keep track of the thermalization of the physical fields, we do not redefine the fermion and antifermion fields $\psi$ and $\chi$ to reabsorb their mass terms, which we leave explicit. 
In the matching, the thermal scales and any other energy scale below $M$ can be set to zero.
Hence, upon our assumption $M \gg T$, no finite temperature effect enters the EFT  Lagrangian~\eqref{NREFT_lag}.

\begin{figure}[ht]
\centering
\includegraphics[scale=0.52]{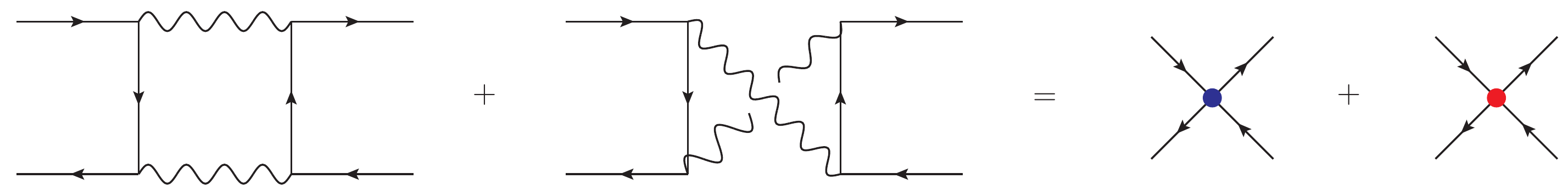}
\caption{\label{fig:DM_0_ann} Matching between annihilation diagrams in the relativistic theory at one loop (left diagrams)
  and the corresponding four-fermion interactions in the non-relativistic EFT (right diagrams).
  For initial and final pairs in an S-wave, 
  the latter correspond to the spin-singlet and spin-triplet four-fermion operators in \eqref{NREFT_lag}.
  The associated matching coefficients are given in \eqref{match_ds_nreft} and \eqref{match_dv_nreft}, where only $d_s$ has an imaginary part at order $\alpha^2$.
  Instead we have $\textrm{Im}(d_v)=\mathcal{O}(\alpha^3)$, reflecting the fact that spin singlets decay into two dark photons, whereas spin triplets decay into three dark photons.
  Explicit expressions at order $\alpha^3$ are given in eqs.~\eqref{Im_ds_NLO} and~\eqref{Im_dv_NLO}.}
\end{figure}

The one-loop expressions of the  matching coefficients $c_{{\hbox{\tiny F}}}$, $c_{{\hbox{\tiny D}}}$, $c_{{\hbox{\tiny S}}}$
and $d_2$ in the $\overline{\hbox{MS}}$ scheme can be found in ref.~\cite{Manohar:1997qy} taking the abelian limit.
The coefficients of the kinetic operators are fixed to be one at all orders in the coupling by reparametrization (Poincar\'e) invariance \cite{Luke:1992cs,Brambilla:2003nt}. 
As for the four-fermion dimension-six operators, the matching coefficients read at order $\alpha^2$~\cite{Pineda:1998kj}
\begin{eqnarray}
&&d_s= \alpha^2 \left( \frac{1}{6} -2 \ln 2 + \ln \frac{M}{\mu} \right) + i \pi \alpha^2 \, ,
\label{match_ds_nreft}
\\
&& d_v= -\pi \alpha + \alpha^2 \left( \frac{91}{18} + \ln \frac{M}{\mu} \right) \, ,
\label{match_dv_nreft}
\end{eqnarray}
where $\mu$ is the renormalization scale associated with factoring hard from low energy scale contributions.
The renormalization of the coupling is discussed at the end of section~\ref{sec:sommerfeld}.
The imaginary parts in the four-fermion operator matching coefficients originate from the particle-antiparticle annihilation diagrams.
Annihilation processes happen at the scale $2M$ and are therefore integrated out in the non-relativistic EFT.
The four-fermion operators shown in eq.~\eqref{NREFT_lag} encode the annihilation of S-wave fermion-antifermion pairs.
Higher-dimensional four-fermion operators encode the annihilation of fermion-antifermion pairs with non-vanishing orbital angular momentum.
For instance, dimension eight four-fermion operators encode the annihilation of P-wave fermion-antifermion pairs.
The leading order contribution to the imaginary part of the dimension-six operators comes from the two-photon annihilation process $X \bar{X} \to \gamma \gamma$,
when the $X \bar{X}$ pair is in an S-wave, see figure~\ref{fig:DM_0_ann};
the imaginary part in eq.~\eqref{match_ds_nreft} may be obtained by cutting the loop diagrams along the photon propagators~\cite{Bodwin:1994jh,Braaten:1996ix}.
For the computation of the annihilation cross section, we refer to section~\ref{sec:nrqed_annihilation}.
In observables, the $\mu$ dependence of the matching coefficients cancels against low-energy matrix elements.

\subsection{pNRQED$_{\textrm{DM}}$}
\label{sec:pNREFT_part}
Consistently with the hierarchy of energy scales~\eqref{scale_arrang},
the next degrees of freedom to integrate out to describe threshold phenomena at the ultrasoft scale are dark photons of energy or momentum of order $M v_{\text{rel}}$,
  which encompasses the scales $M \alpha$ and $\sqrt{MT}$.
At energies much smaller than $M v_{\text{rel}}$ the dynamical degrees of freedom are dark fermions and antifermions with energy of order $M v_{\text{rel}}^2$,
and ultrasoft dark photons with energy and momentum of order $M v_{\text{rel}}^2$, which encompasses the scales $M \alpha^2$ and $T$.\footnote{
We consider the hierarchy of scales to be just $T \leq |E_1^b|$. 
For excited Coulombic states, however, 
further distinctions of scales due to the principal quantum number may turn out to be necessary, 
as $M\alpha^2/n^2 \leq M\alpha^2$ and similarly for the Bohr momentum $M\alpha/n \leq M\alpha$.
Nevertheless, to keep the analysis of the results simple,  
we refrain in this paper to put stronger constraints on the excited Coulombic states.}
The effective field theory that describes them has the form of potential NRQED (pNRQED)~\cite{Pineda:1997bj,Pineda:1997ie} and we denote it as pNRQED$_\textrm{DM}$.
The case of pNRQED at finite temperature has been studied in refs.~\cite{Escobedo:2008sy,Escobedo:2010tu}.
We can proceed as in section~\ref{sec:NREFT_part} and integrate out the soft scale 
by setting to zero all lower (ultrasoft) energy scales, which include the temperature characterizing the thermal distribution of the dark photons.
The matching is done at weak coupling, i.e. order by order in $\alpha$, although the EFT is suited to 
accommodate a non-perturbative framework as well~\cite{Brambilla:1999xf,Brambilla:2004jw}.

Threshold phenomena affect fermion-antifermion pairs, hence it is convenient to project the EFT on the fermion-antifermion space and express it in terms of gauge singlet fermion-antifermion bilocal fields $\phi(t,\bm{r},\bm{R})$,
where $\bm{r} \equiv \bm{x}_1-\bm{x}_2$ is the distance between a fermion located at $\bm{x}_1$ and an antifermion located at $\bm{x}_2$ and $\bm{R}\equiv(\bm{x}_1+\bm{x}_2)/2$ is the center of mass coordinate.\footnote{
The fermion-antifermion Hilbert space is spanned by the vector 
$\displaystyle \sum_{ij}\int d^3x_1\,d^3x_2~\phi_{ij}(\bm{x}_1,\bm{x}_2)$ $\psi^{i\dagger}(\bm{x}_1) $ $\chi^j(\bm{x}_2)|0\rangle$.}
In order to ensure that the photons are ultrasoft, photon fields are \textit{multipole} expanded in $\bm{r}$.
Hence the pNRQED$_\textrm{DM}$ Lagrangian density for the dark matter theory \eqref{lag_mod_0} is organized as an expansion in $1/M$, 
inherited from NRQED$_\textrm{DM}$, $\bm{r}$ and $\alpha$ (at weak coupling):
\begin{eqnarray}
  \mathcal{L}_{\textrm{pNRQED}_{\textrm{DM}}}&=&   \int d^3r \; \phi^\dagger(t,\bm{r},\bm{R})
             \, \left[ i \partial_0 -H(\bm{r},\bm{p},\bm{P},\bm{S}_1,\bm{S}_2)  + g \, \bm{r} \cdot \bm{E}(t,\bm{R})\right] \phi (t,\bm{r},\bm{R})  + \dots  \nonumber \\
&&-\frac{1}{4} F_{\mu \nu} F^{\mu \nu} \, ,
\label{pNREFT_1}
\end{eqnarray}
where 
\begin{eqnarray}
 &&H(\bm{r},\bm{p},\bm{P},\bm{S}_1,\bm{S}_2) = 2M + \frac{\bm{p}^2}{M}+\frac{\bm{P}^2}{4M} - \frac{\bm{p}^4}{4M^3} +  V (\bm{r},\bm{p},\bm{P},\bm{S}_1,\bm{S}_2) + \ldots\, , 
 \label{ham_pNRQED}\\
  &&V (\bm{r},\bm{p},\bm{P},\bm{S}_1,\bm{S}_2)= V^{(0)} + \frac{V^{(1)}}{M} + \frac{V^{(2)}}{M^2} + \ldots \, ,
 \label{pot_pNRQED}    
\end{eqnarray}
and $\bm{S}_1=\bm{\sigma}_1/2$ and $\bm{S}_2=\bm{\sigma}_2/2$ are the spin operators acting on the fermion and antifermion, respectively.
The static potential is the Coulomb potential:
\begin{equation}
  V^{(0)}=-\frac{\alpha}{r}\,.
\label{V0Coul}
\end{equation}
Because $T\lesssim M \alpha^2$, the potential does not get, by construction, thermal contributions at any order. 
The power counting of the EFT goes as follows: the inverse of the relative coordinate $r$ scales like $M v_{\text{rel}}$,
whereas the inverse of the center-of-mass coordinate $R$ can at most change by $M \alpha^2$ or $T$,
if the DM fermion-antifermion pair recoils against ultrasoft dark photons.
The fact that the variation in $R$ is larger than $r$ guarantees the validity of the multipole expansion.
The dots in eq.~\eqref{pNREFT_1} stand for irrelevant operators of higher order in the $1/M$ and multipole expansion.
The relative momentum $\bm{p} = -i\bm{\nabla}_{\bm{r}}$ and the center of mass momentum $\bm{P} = -i\bm{\nabla}_{\bm{R}}$
are the conjugate variables of $\bm{r}$ and $\bm{R}$, respectively.

The expression of the potential $V(\bm{r},\bm{p},\bm{P},\bm{S}_1,\bm{S}_2)$ in the center of mass frame including $V^{(0)}$, $V^{(1)}$ and $V^{(2)}$ can be found in ref.~\cite{Pineda:1998kn}.
Here, besides in the static potential $V^{(0)}$, we are interested in the contributions to $V(\bm{r},\bm{p},\bm{P},\bm{S}_1,\bm{S}_2)$ that are
responsible for the annihilation or creation of dark fermion-antifermion pairs.
Because annihilation happens at the energy scale $2M$, such contributions are inherited from the imaginary parts of the four-fermion operators in $\mathcal{L}_{\textrm{NRQED}_{\textrm{DM}}}$ 
and enter $V$, starting from the $V^{(2)}/M^2$ term, as
\begin{equation}
  \delta V^{\textrm{ann}}(\bm{r})
  = - \frac{i}{M^2} \, \delta^3(\bm{r})\, \left[ {2\rm{Im}}(d_s) - \bm{S}^2 \left( {\rm{Im}}(d_s)- {\rm{Im}}(d_v) \right) \right] + \dots\;,
\label{pNREFT_2}
\end{equation}
where $\bm{S}=\bm{S}_1+\bm{S}_2$ is the total spin of the pair and the dots stand for terms of higher order in the EFT power counting.
At order $r^0$ in the multipole expansion, the equation of motion of the fermion-antifer\-mion pair is a Schr\"odinger equation with potential $V (\bm{r},\bm{p},\bm{P},\bm{S}_1,\bm{S}_2)$.
Hence the leading order fermion-antifermion propagator in  pNRQED$_\textrm{DM}$ automatically accounts
for bound-state effects and multiple Coulomb rescatterings (Sommerfeld enhancement) in physical observables.
We discuss some of the higher-order corrections to eq.~\eqref{pNREFT_2} and multiple Coulomb exchanges in section~\ref{sec:sommerfeld}.

\begin{figure}[ht]
\centering
\includegraphics[scale=0.53]{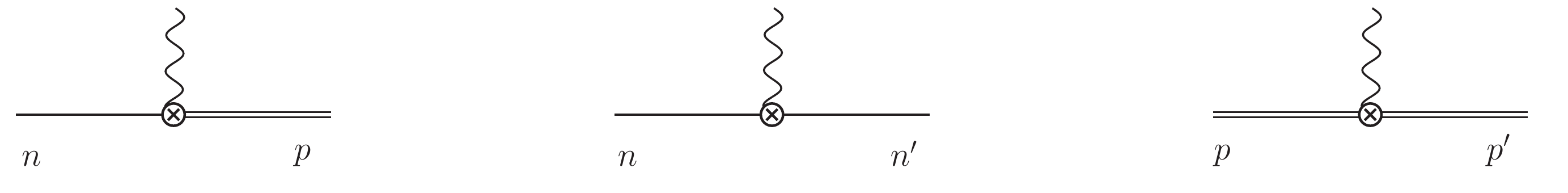}
\caption{\label{fig:pNREFT_DM_fig0}
 Diagrams originating from electric dipole transitions between scattering (double line) and bound states (single line).
 States are labelled by their quantum numbers.
  The electric dipole interaction is represented with a circle-crossed vertex.}
\end{figure}

At order $r$, the term $\phi^\dagger(t,\bm{r},\bm{R}) \, \bm{r} \cdot \bm{E}(t,\bm{R}) \, \phi(t,\bm{r},\bm{R})$ in the pNRQED$_{\textrm{DM}}$ Lagrangian describes
the \emph{electric dipole interaction} of the dark fermion-antifermion pair with ultrasoft dark photons that comprise thermal photons.
The matching coefficient of the electric dipole interaction has been taken equal to one.
Fermion-antifermion pairs above threshold form scattering states of positive energy and
fermion-antifermion pairs below threshold form bound states of negative energy. 
It may be therefore convenient to decompose the fermion-antifermion field $\phi (t,\bm{r},\bm{R})$ into a scattering component and a bound-state component~\cite{Yao:2018nmy},
\begin{align}
  \phi_{ij}(t,\bm{r},\bm{R}) &= \int \frac{d^3P}{(2\pi)^3} \Bigg[ \sum_{n} e^{-iE_nt+i\bm{P}\cdot\bm{R}}\,\Psi_n(\bm{r})\,S_{ij}\,\phi_{n}(\bm{P})
                           \nonumber\\
                         &\hspace{2cm}  + \sum_{\textrm{spin}}\int \frac{d^3p}{(2\pi)^3} e^{-iE_pt+i\bm{P}\cdot\bm{R}}\,\Psi_{\bm{p}}(\bm{r})\,S_{ij}\,\phi_{\bm{p}}(\bm{P})\Bigg] \,,
\label{eq:field}
\end{align}
where $\phi_{n}^\dagger(\bm{P})$ creates a bound state,  $|n,\bm{P}\rangle = \phi_{n}^\dagger(\bm{P})|0\rangle$, 
with center of mass momentum $\bm{P}$,  quantum numbers $n$, energy $E_n$ and wavefunction  $\Psi_n(\bm{r})\,S_{ij}$,
whereas $\phi_{p}^\dagger(\bm{P})$ creates a scattering state, $|\bm{p},\bm{P}\,\rangle = \phi_{\bm{p}}^\dagger(\bm{P})|0\rangle$, 
with center of mass momentum $\bm{P}$,  relative momentum $\bm{p}$, energy $E_p$ and wavefunction  $\Psi_{\bm{p}}(\bm{r})\,S_{ij}$.
The indices $i,j$ are spin indices.
In particular, S-wave dark fermion-antifermion pairs may be either in a spin-singlet state, in which case  $S_{ij} = \delta_{ij}/\sqrt{2}$,
or in a spin-triplet state, in which case $S_{ij} = (\bm{\sigma}\cdot \bm{\epsilon})_{ij}/\sqrt{2}$, 
where $\bm{\sigma}$ are the Pauli matrices and $\bm{\epsilon}$ is the polarization vector of the spin-triplet pair. 
The sum over spin in the second line of eq.~\eqref{eq:field} is a sum over all spin configurations; 
in the first line, this sum is included in the sum over the quantum numbers $n$.
If the dark fermion-antifermion pair is bound we may call it \textit{darkonium}, which, in the S-wave case, we may further distinguish
into a spin-singlet \textit{paradarkonium} state, and a spin-triplet \textit{orthodarkonium} state.
The various transitions between scattering and bound states induced by the electric dipole vertex are shown in figure~\ref{fig:pNREFT_DM_fig0}.
We discuss these transitions in section~\ref{sec:Electric_transitions}.

\section{Annihilation cross section}
\label{sec:sommerfeld}
The dynamical quantity entering the rate equation~\eqref{Boltzmann_1} is the thermal average
of the {\em annihilation} (ann) cross section, $\sigma_{\hbox{\scriptsize ann}}$, times the relative velocity.
The annihilation processes that we consider are
\begin{equation}
X\bar{X}\to \gamma \gamma \quad \textrm{or} \quad X\bar{X}\to \gamma \gamma \gamma\,.
\end{equation}  
The thermal average is defined as the normalized integral of the cross section over the incoming momenta weighted by the Boltzmann distribution~\cite{Gondolo:1990dk,Feng:2010zp}.
This assumes kinetic equilibrium of the dark matter particles at and after the chemical freeze-out,
and that the mass $M$ of the dark matter particle is much larger than the temperature,
so that we can describe the statistical distribution of the dark matter particles by means of the Maxwell--Boltzmann distribution.
In the adopted hierarchy of energy scales, corrections to the annihilation cross section that depend on the center of mass momentum $\bm{P}$, 
which are of relative order $\bm{P}^2/M^2 \sim T/M$, are subleading.
If we neglect them, $\sigma_{\hbox{\scriptsize ann}} v_{\hbox{\scriptsize rel}}$ is independent of the center of mass momentum~$\bm{P}$.
Hence, in the thermal average the integral over $\bm{P}$ factorizes and cancels against the normalization; the same happens to the statistical factor $\displaystyle e^{-2M/T}$.
We are left with
\begin{eqnarray}
  \langle\sigma_{\hbox{\scriptsize ann}} v_{\hbox{\scriptsize rel}}  \rangle   &=&
       \sqrt{\frac{2}{\pi}}   \left( \frac{M}{2T} \right)^{\frac{3}{2}}  \int_0^{\infty} d v_{\hbox{\scriptsize rel}} \, v_{\hbox{\scriptsize rel}}^2 \, e^{-\frac{M}{4T}v_{\textrm{rel}}^2}  \,
    \sigma_{\hbox{\scriptsize ann}} v_{\hbox{\scriptsize rel}}
     \label{thermal_averaged_CR} \, ,
\end{eqnarray}
where $\bm{v}_{\hbox{\scriptsize rel}}$ is related to the relative momentum $\bm{p}$ by $M\bm{v}_{\hbox{\scriptsize rel}}/2 = \bm{p}$.
Dark fermion-antifermion annihilation happens at the energy scale $2M$.
At this energy scale the process is best described by NRQED$_\textrm{DM}$, as we see in section~\ref{sec:nrqed_annihilation}.
Multiple soft dark photon exchanges between the fermion-antifermion pair may however significantly distort the fermion-antifer\-mion wavefunction
either above threshold, when they form a scattering state, or below threshold, when they form a bound state.
Multiple soft photon exchanges are best described by pNRQED$_\textrm{DM}$, as we see in section~\ref{sec:pnrqed_annihilation}.

\subsection{Annihilation cross section in NRQED$_{\textrm{DM}}$}
\label{sec:nrqed_annihilation}
Fermion-antifermion annihilation is encoded in NRQED$_{\textrm{DM}}$ in the imaginary part of the matching coefficients multiplying the four-fermion operators.
Accounting only for di\-men\-sion-six four-fermion operators and describing the fermion and antifer\-mion as free non-interac\-ting spinors
we get from the optical theorem the cross section\footnote{
  Cross sections are computed by summing over the final state polarizations and averaging over the initial state ones.
  In the case of annihilation cross sections and bound state formation cross sections (sec.~\ref{sec:Electric_transitions_form}),
  the initial state polarizations are the $4 = 2\times 2$ spin orientations of the incoming fermion-antifermion pair.
  In the case of ionization cross sections (sec.~\ref{sec:Electric_transitions_diss}),
  the initial state polarizations are the $4 = 2\times 2$ spin orientations of the incoming fermion-antifermion pair and the two polarizations of the incoming photon.
}  
\begin{eqnarray}
  \sigma^{\hbox{\tiny NR}}_{\hbox{\scriptsize ann}} v_{\hbox{\scriptsize rel}}  &=&
   \frac{{\rm{Im}}[\mathcal{M}_{\hbox{\tiny NR}}(\psi\chi \to \psi\chi)]}{2} \, = \frac{  {{\rm{Im}}(d_s)}+3 {{\rm{Im}}(d_v)} }{M^2}\, ,
    \label{NR_hard_cross_section}
\end{eqnarray}
where $\mathcal{M}_{\hbox{\tiny NR}}(\psi\chi \to \psi\chi)$ is the non-relativistic fermion-antifermion scattering amplitude,
and ${\rm{Im}}(d_s)$ and  ${\rm{Im}}(d_v)$ are the imaginary parts of the dimension-six NRQED$_{\textrm{DM}}$ operators introduced in section~\ref{sec:NREFT_part}.
At order $\alpha^3$ they read~\cite{Barbieri:1979be,Hagiwara:1980nv} 
\begin{eqnarray}
  &&  {\rm{Im}} (d_s) = \pi \alpha^2 \left[ 1+\frac{\alpha}{\pi} \left(\frac{\pi^2}{4} -5
     \right)\right] \, ,
    \label{Im_ds_NLO}
      \\
               &&  {\rm{Im}} (d_v) = \frac{4}{9} (\pi^2 -9) \alpha^3 \, .
    \label{Im_dv_NLO}
   \end{eqnarray}
The quantity  $\sigma^{\hbox{\tiny NR}}_{\hbox{\scriptsize ann}} v_{\hbox{\scriptsize rel}}$ does not depend on the momentum of the fermion-antifermion pair.
At leading order (LO), when inserting the explicit expression ${{\rm{Im}}(d_s)}= \pi \alpha^2$, one recovers the well known result~\cite{Dirac:1930bga}
\begin{equation}
 (\sigma^{\hbox{\tiny NR}}_{\hbox{\scriptsize ann}} v_{\hbox{\scriptsize rel}})_{\hbox{\tiny LO}} 
  = \frac{\pi \alpha^2}{M^2}.
  \label{LOexpression}
\end{equation}

At order $\alpha^2$ the annihilation occurs when the fermion-antifermion pair forms an S-wave in a spin-singlet configuration, 
while the annihilation of the fermion-antifermion pair in a spin-triplet S-wave may only occur via an odd number of photons, i.e. starting from order~$\alpha^3$.
Velocity suppressed contributions to the cross section can be systematically included by adding four-fermion operators of dimension higher than six to the Lagrangian~\eqref{NREFT_lag}.
Higher dimensional four-fermion operators also account for the annihilation of fermion-antifermion pairs with higher orbital angular momentum.

The EFT provides a straightforward framework to compute higher-order corrections to the annihilation cross section,
either in terms of $\alpha$ corrections to the matching coefficients of the four-fermion operators,
or in terms of subleading terms in the velocity expansion by including higher dimensional four-fermion operators to the NRQED$_{\textrm{DM}}$ Lagrangian 
(in this respect see also refs.~\cite{Beneke:2012tg,Hellmann:2013} for neutralinos DM in the context of the MSSM and \cite{Bodwin:1994jh} for the case of heavy quarkonium in QCD).
Computing next-to-leading order (NLO) annihilation cross sections has been for long pursued in a variety of models,
e.g.~\cite{Baro:2009na,Akcay:2012db,Harz:2012fz,Hellmann:2013jxa,Ciafaloni:2013hya,Ovanesyan:2016vkk,Schmiemann:2019czm,Banerjee:2019luv}.
Neglecting the effect of velocity suppressed operators and multiple soft photon rescatterings, which is important near threshold as we discuss in the next section,
the expression at NLO in $\alpha$ of the annihilation cross section in the abelian dark matter model \eqref{lag_mod_0} reads
\begin{equation}
  (\sigma^{\hbox{\tiny NR}}_{\hbox{\scriptsize ann}} v_{\hbox{\scriptsize rel}})_{\hbox{\tiny NLO}} =
  \frac{\pi \alpha^2}{M^2} \left[ 1 + \frac{\alpha}{\pi} \left( \frac{19}{12} \pi^2 -17
    \right) \right] \, .
\label{NLO_crosssection}
\end{equation}
The corrections are negative and make the cross section smaller.
Taking $\alpha \approx 0.4$, the NLO cross section is reduced by about 17\% with respect to the LO cross section (even larger couplings have been considered in the literature,
see e.g.~\cite{vonHarling:2014kha}).

\subsection{Annihilation cross section in pNRQED$_{\textrm{DM}}$}
\label{sec:pnrqed_annihilation}
Multiple soft dark photon exchanges between the annihilating dark fermion-antifermion pair modify the fermion-antifermion wavefunction near threshold
and lead to a significant change in the annihilation cross section.
The annihilation process of S-wave fermion-antifermion pairs is described in pNRQED$_{\textrm{DM}}$ by the imaginary local potential \eqref{pNREFT_2}
directly inherited from the dimension six four-fermion operators of NRQED$_{\textrm{DM}}$.
Soft photon exchanges are encoded in pNRQED$_{\textrm{DM}}$ in the potential \eqref{pot_pNRQED}.
The fermion-antifermion wavefunction in pNRQED$_{\textrm{DM}}$, which at leading order in the multipole expansion is the solution of the Schr\"odinger equation
with the potential \eqref{pot_pNRQED}, 
accounts by construction for the effect of multiple soft photon rescattering.

We consider first the effect of multiple soft photon exchanges on fermion-antifermion scattering states.
To compute the annihilation cross section we use the optical theorem as in the first equality of eq.~\eqref{NR_hard_cross_section}, but now we compute the scattering amplitude from
an initial to a final scattering state $|\bm{p}, \bm{0}\,\rangle$.
The annihilation cross section reads
\begin{eqnarray}
(\sigma_{\hbox{\scriptsize ann}} v_{\hbox{\scriptsize rel}})(\bm{p}) 
&=& \frac{1}{2} \langle \,  \bm{p}, \bm{0}| \int d^3r \, \phi^\dagger(\bm{r},\bm{R},t)\, 
\left[-\rm{Im} \,\delta V^{\rm ann}(\bm{r})\right] \, \phi(\bm{r},\bm{R},t)\, |  \bm{p},\bm{0} \rangle
\nonumber 
\\
&=&  \frac{{\rm{Im}}(d_s)+3{\rm{Im}}(d_v)}{M^2} \left|\Psi_{\bm{p} 0}(\bm{0})\right|^2
    =  (\sigma^{\hbox{\tiny NR}}_{\hbox{\scriptsize ann}} v_{\hbox{\scriptsize rel}})  \, S_{\hbox{\scriptsize ann}}(\zeta)\,,
\label{ann_fact_scat}
\end{eqnarray}
where we have set from the beginning the center of mass momentum $\bm{P}=\bm{0}$, 
$\delta V^{\rm ann}(\bm{r})$ has been defined in eq.~\eqref{pNREFT_2},
and ${\rm{Im}}(d_s)$ and ${\rm{Im}}(d_v)$ have been given at order $\alpha^3$ in eqs.~\eqref{Im_ds_NLO} and~\eqref{Im_dv_NLO}.
Differently from eq.~\eqref{NR_hard_cross_section}, the quantity $(\sigma_{\hbox{\scriptsize ann}} v_{\hbox{\scriptsize rel}})(\bm{p})$ depends on the relative momentum $\bm{p}$.
The scattering wavefunction, $\Psi_{\bm{p}}$, has been expanded in partial waves,  $\Psi_{\bm{p}\ell}$, where $\ell$ is the orbital angular momentum quantum number;
only the S-wave component of $\Psi_{\bm{p}}$ contributes at leading order in the non-relativistic expansion,
since dimension six four-fermion operators in NRQED$_{\textrm{DM}}$ and the ensuing annihilation potential \eqref{pNREFT_2} only project on S-waves. 
The factor $S_{\hbox{\scriptsize ann}}(\zeta)$ is called \emph{Sommerfeld factor}~\cite{Sommerfeld} and for an attractive Coulombic potential, like in our case, 
it reads (see e.g.~\cite{Iengo:2009ni,Cassel:2009wt}) 
\begin{equation}
S_{\hbox{\scriptsize ann}}(\zeta)=\frac{2 \pi \zeta}{1-e^{-2 \pi \zeta}} \, , \qquad\qquad \zeta \equiv \frac{\alpha}{v_{\hbox{\scriptsize rel}}} = \frac{1}{2}\frac{M\alpha}{p}\, .
\label{Somme_0}
\end{equation}
Equation~\eqref{ann_fact_scat} shows manifestly the factorization of the different energy scales:
the hard dynamics is contained in the NRQED$_{\textrm{DM}}$ matching coefficients ${\rm{Im}}(d_s)$ and ${\rm{Im}}(d_v)$,
whereas the soft dynamics is contained in the wavefunction squared $|\Psi_{\bm{p} 0}(0)|^2$.
The Sommerfeld factor modifies the cross section from $\sigma_{\hbox{\scriptsize ann}}^{\hbox{\tiny NR}} v_{\hbox{\scriptsize rel}}$, which is the cross section computed
at the level of NRQED$_\textrm{DM}$ without Coulomb rescattering effects, see eq.~\eqref{NR_hard_cross_section},
into $(\sigma_{\hbox{\scriptsize ann}}^{\hbox{\tiny NR}} v_{\hbox{\scriptsize rel}}) S_{\hbox{\scriptsize ann}}(\zeta)$. 
For $v_{\hbox{\scriptsize rel}} \lesssim \alpha$, the annihilation cross section is significantly enhanced by the Sommerfeld factor and the prediction from \eqref{Boltzmann_1} changes accordingly.\footnote{
  A derivation of the Sommerfeld enhancement for S-wave pair annihilation that includes the regime of very small momenta (velocities) for the unbound pair 
  has been presented in~\cite{Hisano:2002fk,Hisano:2004ds,Blum:2016nrz}.
  The main result is a saturation of the Sommerfeld factor and a regular behaviour for $v_\textrm{rel} \to 0$.
  Diagrammatically this amounts to resum the annihilation term, namely the local four-fermion interactions shown in figure~\ref{fig:DM_0_ann} (right).
  In this work, we assume to be away from such regime for unbound states.
  For bound states this resummation is never needed because the momentum of the particle in the pair is constrained to be of order $M \alpha$.
  Finally, it is worth noticing that the thermally averaged cross section~\eqref{thermal_averaged_CR} removes the singularity at vanishing $v_\textrm{rel}$.}

We consider now the effect of multiple soft photon exchanges on fermion-antifermion pairs below threshold.
This leads to the formation of Coulombic bound states, whose masses up to  order $\alpha^2$ are 
\begin{equation}
E_n = 2M - \frac{M \alpha^2}{4 n^2} = 2M - \frac{1}{M a_0^2 n^2}\, , \qquad\qquad a_0 \equiv \frac{2}{M \alpha}\, ,
    \label{Coulomb_en_levels}
\end{equation}
where $a_0$ is the Bohr radius.
To compute the annihilation width, we compute the amplitude in the first equality of eq.~\eqref{NR_hard_cross_section} from an initial to a final bound state $|n,\bm{0}\rangle$.\footnote{
The different normalization of the states leads automatically to a cross section for scattering states and a decay width for bound states. }  
At the order we are working, as we have seen in the case of scattering states, only S-wave bound states contribute.
S-wave bound states may be in a spin-singlet paradarkonium state (para) or in a spin-triplet orthodarkonium state (ortho).
At leading order paradarkonium and orthodarkonium are described by the same radial wavefunction $R_{n0}(r)$.
The paradarkonium and orthodarkonium annihilation widths read
\begin{eqnarray}
\Gamma^{n,\hbox{\scriptsize para}}_{\textrm{ann}} &=& \frac{4 {\rm{Im}}(d_s)}{M^2} \frac{|R_{n0}(0)|^2}{4\pi} \,,
\label{ann_para}\\
\Gamma^{n,\hbox{\scriptsize ortho}}_{\textrm{ann}}&=& \frac{4 {\rm{Im}}(d_v)}{M^2} \frac{|R_{n0}(0)|^2}{4\pi} \,.
\label{ann_ortho}
\end{eqnarray}
For 1S states, $R_{10}(r) = 2 \,e^{-r/a_0} / a_0^{3/2}$, and we can write, keeping order $\alpha^3$ terms in ${\rm{Im}}(d_s)$ and ${\rm{Im}}(d_v)$,
\begin{eqnarray}
  \Gamma^{\text{1S},\hbox{\scriptsize para}}_{\textrm{ann}}
  &=& \frac{M \alpha^5}{2} \left[ 1+\frac{\alpha}{\pi} \left(\frac{\pi^2}{4} - 5
      \right)\right] \, ,      
\label{ann_para1S}\\
  \Gamma^{\text{1S},\hbox{\scriptsize ortho}}_{\textrm{ann}}
          &=& \frac{2 (\pi^2 -9) M  \alpha^6}{9 \pi} \, .
\label{ann_ortho1s}
\end{eqnarray}

We conclude this section with two observations.
First, we remark that the inclusion of the orthodarkonium decay width in the DM evolution equations requires for consistency  
the inclusion of the order $\alpha^3$ corrections to ${\rm{Im}}(d_s)$ in the paradarkonium decay width 
and $(\sigma^{\hbox{\tiny NR}}_{\hbox{\scriptsize ann}} v_{\hbox{\scriptsize rel}})_{\hbox{\tiny NLO}}$ in the computation of the annihilation cross section for scattering states,
since all these terms are of the same order and originate from the same set of four-fermion matching coefficients in  NRQED$_{\textrm{DM}}$.
This has not always been the case in some of the original literature~\cite{vonHarling:2014kha}.

The second observation is on the coupling constant. 
In the fundamental theory \eqref{lag_mod_0}, dark photons couple only with the dark fermions $X$,
hence the coupling runs with one flavor at scales greater than $M$ and is frozen at the value $\alpha \equiv \alpha(M)$ at scales below $M$.
The coupling would still run below the mass scale if the gauge bosons would be coupled to themselves like in non-abelian gauge theories (see section~\ref{sec:non_abelian_model}) or to other light degrees of freedom.
In NRQED$_{\textrm{DM}}$ and pNRQED$_{\textrm{DM}}$, the dark matter fermions have been integrated out, hence the coupling appearing there is $\alpha$, and it does not run.
Because in NRQED$_{\textrm{DM}}$ and pNRQED$_{\textrm{DM}}$ the coupling remains the same at the mass scale and at the scale of the binding,
it is parametrically consistent to include subleading corrections in $\alpha$ to the pair annihilation of the scattering states and to the decay widths of the bound states
originating from higher-order corrections to the matching coefficients of the four-fermion operators,
while at the same time ignoring higher-order corrections in $\alpha$ in the near threshold wavefunctions, which are suppressed by $\alpha^2$, as we have done.
The situation is different in QCD-like non-abelian theories, where corrections to the wavefunction are typically as relevant as or more relevant than corrections to the four-fermion matching coefficients, 
on one hand because there, due to asymptotic freedom, the coupling at the mass scale is smaller than the coupling at the near threshold scales, 
and on the other hand because corrections to the wavefunction are suppressed by just $\alpha$~\cite{Titard:1993nn}.

\section{Bound-state formation and dissociation cross sections}
\label{sec:Electric_transitions}
The pNRQED$_{\textrm{DM}}$ Lagrangian \eqref{pNREFT_1} contains electric dipole terms that are responsible
for dark fermion-antifermion pair transitions between scattering states, $(X\bar{X})_p$, and bound states, $(X\bar{X})_n$, as shown in figure~\ref{fig:pNREFT_DM_fig0}.
In section~\ref{sec:Electric_transitions_form} we address transitions that lead to bound-state formations
and in section~\ref{sec:Electric_transitions_diss} transitions that lead to bound-state dissociations.
We complete the analysis in appendix~\ref{sec:app_B} by providing results for the bound-state to bound-state transitions, which amount to excitation and de-excitation processes, 
and in appendix~\ref{sec:app_C} by considering bremsstrahlung and thermal absorption processes in scattering-state to scattering-state transitions.

Electric dipole transitions involve the emission or absorption of an ultrasoft dark photon.
The medium breaks explicitly Lorentz invariance and we have to choose a reference frame.
A convenient reference frame choice is the one where the medium is at rest.
This choice made, the cross section depends on the center of mass momentum $\bm{P}$.
The center of mass momentum in the thermal average of the cross section times $v_{\textrm{rel}}$ scales like $\sqrt{MT}$,
which is the momentum scale in the Boltzmann distribution.
Although $\sqrt{MT}$ is a soft scale, its effect on electric dipole transitions turns out to be subleading.
  The reason is that the center of mass momentum dependence is carried by the kinetic energy of the fermion-antifermion pair,
which changes by an amount $\bm{P}^2/(4M) - (\bm{P-k})^2/(4M) \sim \bm{P}\cdot\bm{k}/(2M)$ when a dark photon of momentum $k \sim T$ or $k \sim M \alpha^2$ is emitted.
The quantity $\bm{P}\cdot\bm{k}/(2M)$ is subleading with respect to $T$ or $M\alpha^2$ even in the momentum region $P \sim \sqrt{MT}$.
We may therefore systematically expand in the center of mass momentum $\bm{P}$.
If we retain the leading order term, this amounts to set $\bm{P}=\bm{0}$ in the cross section,\footnote{
  We come to the same conclusion if we choose the reference frame of the center of mass of the dark fermion-antifermion pair.
  In this case, the velocity of the medium is about $\sqrt{T/M}$, which is much smaller than one, the velocity of light.
  Expanding in it, the thermal distribution of the dark photons reduces at leading order to the thermal distribution of the medium at rest~\cite{Escobedo:2011ie}.
}  
which is our choice in the following sections.

\subsection{Bound-state formation}
\label{sec:Electric_transitions_form}
In this section, we compute at leading order in the non-relativistic and coupling expansion the cross section for the process
\begin{equation}
(X\bar{X})_p\to \gamma + (X\bar{X})_n\,,
\label{ptogamman}
\end{equation}
where a bound state $(X\bar{X})_n$ is formed from a scattering state $(X\bar{X})_p$ via the emission of an ultrasoft dark photon.
The transition is an electric dipole transition.
In  pNRQED$_{\textrm{DM}}$, it corresponds to the inverse process of the one depicted in the first diagram of figure~\ref{fig:pNREFT_DM_fig0}.
The energy of the incoming pair is $E_p = 2M + \bm{p}^2/M + \dots = 2M + M v_{\text{rel}}^2/4 + \dots$,
and the energy of the resulting bound state is $E_n$, defined in eq.~\eqref{Coulomb_en_levels}.
Since $E_p > E_n$, the process \eqref{ptogamman} is allowed in vacuum.
Scattering-state to scattering-state transitions of the type $(X\bar{X})_p\to \gamma + (X\bar{X})_{p'}$ are discussed in appendix~\ref{sec:app_C}.

\begin{figure}[ht]
    \centering
    \includegraphics[scale=0.55]{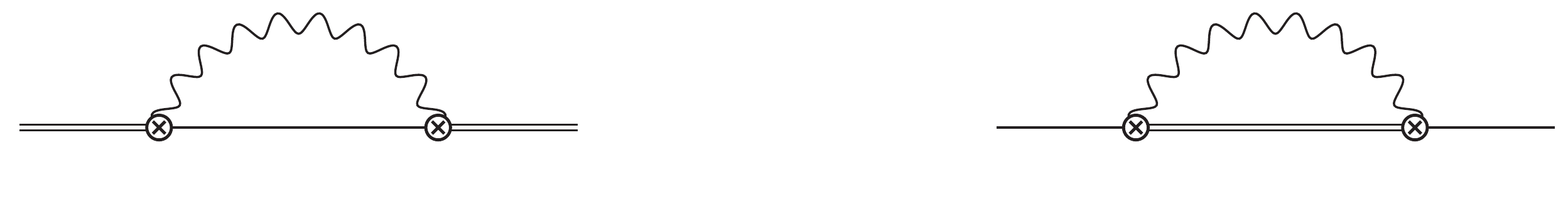}
    \caption{Self-energy diagrams in  pNRQED$_{\textrm{DM}}$ with an initial scattering state and an intermediate bound state on the left,
      and with an initial bound state and an intermediate scattering state on the right.
      The imaginary part of the left diagram is responsible for the bound-state formation process $(X\bar{X})_p\to \gamma + (X\bar{X})_n$,
      whereas the imaginary part of the right diagram is responsible for the bound-state dissociation process $\gamma + (X\bar{X})_n\to (X\bar{X})_p$.}
    \label{fig:pnEFT_DM_self}
\end{figure}

The cross section can be computed at leading order from the imaginary part of the one-loop self energy in pNRQED$_{\textrm{DM}}$
shown in the left diagram of figure~\ref{fig:pnEFT_DM_self}.
The photon could be a thermal photon and therefore the computation needs to be performed in the \emph{thermal field theory} version of pNRQED$_{\textrm{DM}}$.
We use the real-time Schwinger--Keldysh formalism~\cite{Bellac:2011kqa,Laine:2016hma,Ghiglieri:2020dpq}.
The real-time formalism necessarily leads to a doubling of the degrees of freedom called of type 1 and 2.
The type 1 fields are the physical ones, i.e. those that appear in the initial and final states.
Propagators are represented by $2\times 2$ matrices, as they may involve fields of both types.
The thermal dark photon propagator in Coulomb gauge reads
\begin{equation}
    \begin{aligned}
    &D_{00}(|\bm{k}|) 
    = \begin{pmatrix} \displaystyle \frac{i}{|\bm{k}|^2 +i\epsilon} && 0 \\ 0 && \displaystyle \frac{-i}{|\bm{k}|^2 -i\epsilon} \end{pmatrix}\, , \\
    &D_{ij}(k)= \left(\delta_{ij} - \frac{k_ik_j}{|\bm{k}|^2}\right)
    \left[\begin{pmatrix} \displaystyle \frac{i}{k^2 +i\epsilon} && \theta(-k_0)2\pi \delta(k^2) \\ \theta(k_0)2\pi \delta(k^2) && \displaystyle \frac{-i}{k^2 -i\epsilon} \end{pmatrix}
      + 2\pi \delta(k^2)\,n_{\text{B}}(|k_0|)\begin{pmatrix} 1 && 1 \\ 1 && 1 \end{pmatrix}\right] \, ,
  \end{aligned}
\label{XXpropGT}  
\end{equation}
where $n_{\text{B}}(E) = 1/(e^{E/T}-1)$ is the Bose--Einstein distribution.
The choice of the gauge is irrelevant for our computation of the electric dipole transition, since the electric field is gauge invariant.
The thermal propagator of the fermion-antifermion field $\phi$ in  pNRQED$_{\textrm{DM}}$ reads
\begin{equation}
    G^{\phi}(p_0) 
    = \begin{pmatrix} \displaystyle \frac{i}{p_0 - H +i\epsilon} && 0 \\ 2\pi \delta(p_0 - H) && \displaystyle \frac{-i}{p_0 - H -i\epsilon} \end{pmatrix}
    + 2\pi \delta(p_0 - H)\,n_{\text{B}}(p_0)\begin{pmatrix} 1 && 1 \\ 1 && 1 \end{pmatrix} \, ,
\label{XXpropT}
  \end{equation}
where $p_0$ is the energy of the dark fermion-antifermion pair.
Note that the fermion-antifermion pair behaves like a boson.
In the heavy mass limit, recalling that $H = 2M + \dots$, $n_{\text{B}}(H)$ is exponentially suppressed as $e^{-2M/T}$, so that it holds 
\begin{equation}
    G^{\phi}(p_0) 
    \approx \begin{pmatrix} \displaystyle \frac{i}{p_0 - H +i\epsilon} && 0 \\ 2\pi \delta(p_0 - H) && \displaystyle \frac{-i}{p_0 - H -i\epsilon} \end{pmatrix}\,.
\label{XXpropTnr}
\end{equation}
It has been remarked in~\cite{Brambilla:2008cx} that, because the 12 component of a heavy-field propagator vanishes in the heavy-mass limit,
the physical heavy fields do not propagate into type 2 fields.
Therefore, to a large extent the type 2 fermion-antifermion fields decouple and may be ignored in the heavy-mass limit,
which makes the real-time formalism convenient when dealing with heavy fields.

From eqs.~\eqref{XXpropGT} and~\eqref{XXpropTnr} it follows that the 11 component of the self-energy diagram
shown in the left or right panel of figure~\ref{fig:pnEFT_DM_self} reads 
\begin{align}
\Sigma^{11}(p_0) = -ig^{2}\frac{d-2}{d-1}\mu^{4-d}r^{i}\int \frac{\textrm{d}^{d}k}{(2\pi)^{d}}~&\frac{i}{p^{0}-k^{0}-H + i\epsilon}\nonumber\\
&\times k_{0}^{2}\left[\frac{i}{k_{0}^{2} -|\bm{k}|^{2}+i\epsilon} + 2\pi \delta(k_{0}^{2} - |\bm{k}|^{2})n_{B}(|k_{0}|) \right]r^{i} \,,
\label{self_00_11}
\end{align}
where $p_0$ is the energy of the incoming pair.
We have regularized in $d$ dimensions to get rid of scaleless integrals.
In eq.~\eqref{self_00_11} one can distinguish the in-vacuum and thermal contributions originating from the photon propagator,
while the heavy fermion-antifermion pair has been taken to propagate in vacuum.
In order to extract the imaginary part of \eqref{self_00_11}, we can use the identity
\begin{equation}
\frac{1}{p^0-k^{0}-H+i\epsilon} = \textrm{P}\left(\frac{1}{p^0-k^{0}-H}\right) -i\pi\delta(p^0-k^{0}-H)\,;
\end{equation}
P stands for the principal value prescription.
Following refs.~\cite{Brambilla:2008cx,Brambilla:2011sg}, we obtain
\begin{equation}
  {\rm{Im}}[\Sigma^{11}(p_0)] = -\frac{g^2}{6 \pi} r^i
  \left[\theta(\Delta E) \, (\Delta E)^3 \left( 1 + n_{\text{B}}(\Delta E) \right) + \theta(-\Delta E) \, (-\Delta E)^3 n_{\text{B}}(-\Delta E) \right]
  r^i \,,
\label{SigmaDeltaE}
\end{equation}
where  $\Delta E \equiv p^0-H$  and we have set $d=4$ since the imaginary part of $\Sigma^{11}$ is finite. 
Finally, the imaginary part of the self energy is related to the cross section of the process $(X\bar{X})_p \to \gamma + X\bar{X}$ by the optical theorem 
\begin{align}
&(\sigma_{(X\bar{X})_p \to \gamma + X\bar{X}} \, v_{\hbox{\scriptsize rel}})(\bm{p})
= -2 \langle \,\bm{p}|  {\rm{Im}}[\Sigma^{11}(E_p)] | \bm{p} \, \rangle                
    \nonumber\\
  &= \frac{g^2}{3 \pi}
    \langle \,\bm{p}|
    r^i \left[\theta(\Delta E) \, (\Delta E)^3 \left( 1 + n_{\text{B}}(\Delta E) \right) + \theta(-\Delta E) \, (-\Delta E)^3 n_{\text{B}}(-\Delta E) \right] r^i
    | \bm{p}\, \rangle   \,.
  \label{bsf_projected}
\end{align}
We have projected on the scattering state $\langle \bm{r}|\bm{p}\,\rangle = \Psi_{\bm{p}}(\bm{r})$; $p_0 = E_p$ is the energy of the incoming fermion-antifermion pair.
The difference $\Delta E$ is positive on any intermediate state that is a bound state, and in that case the process may happen both in vacuum and in the medium.
If $\Delta E <0$ the process may only happen through the absorption of a dark photon from the medium.

We can now proceed and compute the \emph{bound-state formation} (bsf) cross section
by projecting eq.~\eqref{bsf_projected} on intermediate bound states $\langle \bm{r}|n\,\rangle = \Psi_n(\bm{r})$, 
\begin{align}
  (\sigma_{\hbox{\scriptsize bsf}} \, v_{\hbox{\scriptsize rel}})(\bm{p}) \equiv  \sum \limits_n   (\sigma^n_{\hbox{\scriptsize bsf}} \, v_{\hbox{\scriptsize rel}})(\bm{p}) 
  = \frac{g^2}{3\pi} \sum_{n} \left[ 1 + n_{\text{B}}(\Delta E_{n}^{p}) \right] |\langle n | \bm{r} | \bm{p}\,\rangle|^2  (\Delta E_{n}^{p})^3 \, .
\label{bsfn}  
\end{align}
We have defined 
\begin{equation}
\Delta E_{n}^{p} \equiv E_p-E_n = \frac{M}{4}v_{\textrm{rel}}^2\left(1+\frac{\alpha^2}{n^2v_{\textrm{rel}}^2}\right) \,,
\label{energy_photon_bsf}
\end{equation}
which holds at LO;
$\sigma^n_{\hbox{\scriptsize bsf}}$ is a shorthand notation for the cross section $\sigma_{(X\bar{X})_p \to \gamma + (X\bar{X})_n}$.

An alternative way of computing the bound-state formation cross section is from the 21 component of the self energy,
\begin{equation}
  \Sigma^{21}(p_0) = i\frac{g^2}{3 \pi}
  r^i \left[\theta(\Delta E) \, (\Delta E)^3 \left( 1 + n_{\text{B}}(\Delta E) \right) + \theta(-\Delta E) \, (-\Delta E)^3 n_{\text{B}}(-\Delta E) \right] r^i
  \, ,
\label{Sigma12}
\end{equation}
in terms of which
\begin{equation}
(\sigma_{(X\bar{X})_p \to \gamma + X\bar{X}} \, v_{\hbox{\scriptsize rel}})(\bm{p}) = \langle\,\bm{p}|  [-i\Sigma^{21}(E_p)] | \bm{p} \, \rangle   \,.      
\end{equation}
Once projected on intermediate bound states, this leads to the bound-state formation cross section given in eq.~\eqref{bsfn}.

If we would keep in our derivations the thermal distribution of the heavy fermion-antifermion pair in the self-energy loop,
which amounts to keep the temperature dependent part of \eqref{XXpropT},
then the bound-state formation cross section would modify into
\begin{equation} 
  (\sigma_{\hbox{\scriptsize bsf}} v_{\hbox{\scriptsize rel}})(\bm{p})
  = \frac{g^2}{3\pi}
  \sum_{n}
    \left[ 1 + n_{\text{B}}(\Delta E_{n}^{p}) \right]
    \left[ 1 + n_{\text{B}}(E_n) \right]
    |\langle \, n | \bm{r} | \bm{p} \,\rangle|^2
  (\Delta E_{n}^{p})^3 \, .
\label{sigmabsf12full}
\end{equation}
Clearly, the thermal distribution of the heavy fermion-antifermion pair vanishes exponentially for $T \ll M \sim E_n/2$, in which case the above expression reduces to~\eqref{bsfn}.

In order to compare with existing literature and to provide an application of the above expression,
we consider the formation of the lowest-lying $1\textrm{S}$ bound state,
whose wavefunction is $\langle \bm{r}|1\textrm{S}\rangle = R_{10}(r)/(4\pi)$.
In this case, only scattering states in the partial wave $\ell=1$ contribute, whose wavefunction is 
$\langle \bm{r}|\bm{p}1\,\rangle = \Psi_{\bm{p}1}(\bm{r})$.
The bound-state formation cross section reads
\begin{eqnarray}
  (\sigma^{1\textrm{S}}_{\hbox{\scriptsize bsf}} v_{\hbox{\scriptsize rel}})(\bm{p})
  &=& \frac{g^2}{3\pi} \left[ 1 + n_{\text{B}}(\Delta E_{1}^{p}) \right] |\langle 1 \textrm{S} | \bm{r} | \bm{p}  1\rangle|^2  (\Delta E_{1}^{p})^3
\nonumber 
     \\
  &=& \frac{\alpha^7 \pi^2 \, 2^{10}}{3 \, M^2 \, v_{\hbox{\scriptsize rel}}^5 \left( 1+ \frac{\alpha^2}{v_{\textrm{rel}}^2}\right)^2}
      \frac{e^{-4 \frac{\alpha}{v_{\textrm{rel}}} \hbox{\scriptsize arccot} \frac{\alpha}{v_{\textrm{rel}}} }}{1-e^{-2 \pi \frac{\alpha}{v_{\textrm{rel}}}}}
      \, \left[ 1 + n_{\text{B}}(\Delta E_{1}^{p}) \right] \, ,
     \label{our_gamma_1p}
\end{eqnarray}
with $p=M v_{\hbox{\scriptsize rel}}/2$ and 
\begin{equation}
  \Delta E_{1}^{p}
  =\frac{Mv_{\textrm{rel}}^2}{4} \left( 1+ \frac{\alpha^2}{v_{\textrm{rel}}^2}\right)\,.
    \label{final_fromation}
\end{equation} 
If we select the spin of the final state,
the bound-state formation cross section for paradarkonium is 
$\sigma^{1\textrm{S,para}}_{\hbox{\scriptsize bsf}} = \sigma^{1\textrm{S}}_{\hbox{\scriptsize bsf}}/4$ and for orthodarkonium is $\sigma^{1\textrm{S,ortho}}_{\hbox{\scriptsize bsf}} = 3\,\sigma^{1\textrm{S}}_{\hbox{\scriptsize bsf}}/4$.
The evaluation of the dipole  matrix element squared $|\langle n| \bm{r} | \bm{p}\rangle|^2$ can be found
in appendix~\ref{sec:app_A} for a generic scattering-state to bound-state transition (see also refs.~\cite{Petraki:2015hla,Petraki:2016cnz,Garny:2021qsr}).

By rewriting the result in \eqref{our_gamma_1p} in terms of $\zeta = \alpha/v_{\textrm{rel}} $,
we recover in the zero temperature limit, i.e. by setting to zero the Bose--Einstein distribution in eq.~\eqref{our_gamma_1p},
the expression derived in refs.~\cite{vonHarling:2014kha,Petraki:2015hla},
and also the abelian limit of the QCD expression derived in ref.~\cite{Brambilla:2011sg}.
We also agree with the finite temperature expression presented in ref.~\cite{Binder:2020efn}.

\begin{figure}[ht]
    \centering
    \includegraphics[scale=0.75]{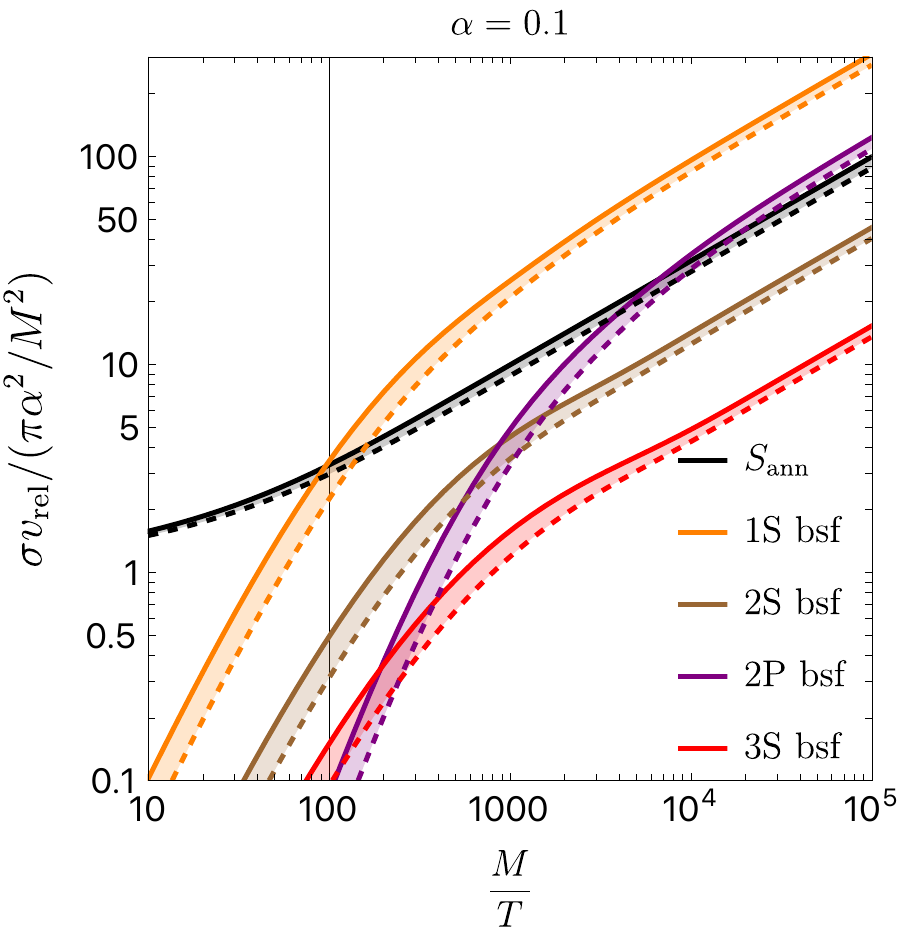}
    \hspace{0.5 cm}
     \includegraphics[scale=0.75]{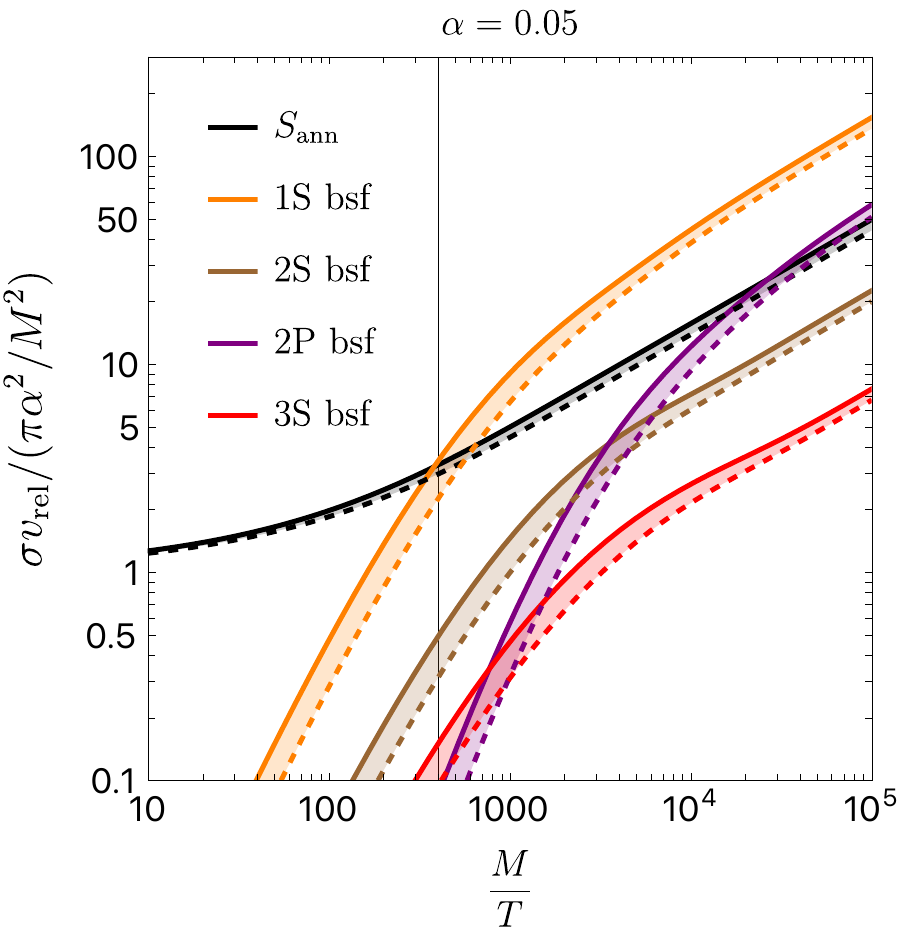}
     \caption{Annihilation cross section for scattering states and the 1S ($n=1$), 2S, 2P ($n=2$) and 3S ($n=3$) bound-state formation cross section normalized
       by the free LO annihilation cross section, $(\sigma^{\hbox{\tiny NR}}_{\hbox{\scriptsize ann}} v_{\hbox{\scriptsize rel}})_{\hbox{\tiny LO}}$.
       We take two bookmark values $\alpha=0.1$ and $\alpha=0.05$;
       solid lines stand for $v_{\hbox{\scriptsize rel}}=v_p$, whereas dashed lines for $v_{\hbox{\scriptsize rel}}=v_m$.
       In the 2P bsf cross section we sum over the energy-degenerate contributions $m=0,\pm 1$.
       The vertical line marks the position where $T=M\alpha^2$.} 
    \label{fig:bound_state_formation_ann}
\end{figure}

From the Maxwell--Boltzmann distribution
\begin{equation}
f_{\hbox{\tiny MB}}(v_{\hbox{\scriptsize rel}})=\sqrt{\frac{2}{\pi}} \left(\frac{M}{2T} \right)^{3/2}
v_{\hbox{\scriptsize rel }}^2  e^{-\frac{Mv^2_{\textrm{rel}}}{4T}},
\label{MBdistr}
\end{equation}
we can define a \emph{most-probable velocity} $v_p=\sqrt{4T/M}$, which is the velocity that maximizes $f_{\hbox{\tiny MB}}(v_{\hbox{\scriptsize rel}})$,
and a \emph{mean velocity}  $\displaystyle v_m = \int_{0}^\infty d v_{\hbox{\scriptsize rel }} v_{\hbox{\scriptsize rel }}   f_{\hbox{\tiny MB}}(v_{\hbox{\scriptsize rel}}) = 2 v_p/\sqrt{\pi}$.
Both velocities depend on the ratio $T/M$ and are much smaller than one, consistently with the non-relativistic assumption.
In figure~\ref{fig:bound_state_formation_ann}, we give the particle-antiparticle annihilation cross section for scattering states and the bound-state formation cross section
normalized to that of free pairs at LO, $\pi \alpha^2/M^2$, for $\alpha=0.1$ and $\alpha=0.05$.
Solid lines stand for cross sections computed at $v_{\hbox{\scriptsize rel}}=v_p$, dashed lines for cross sections computed at $v_{\hbox{\scriptsize rel}}=v_m$,
and the shadowing for cross sections at $v_m<v_{\hbox{\scriptsize rel}}<v_p$.
Both cross sections increase as the temperature (and the typical particle velocity) decreases.
For the same $M/T$, smaller coupling constants imply less enhancement and the bound-state formation takes longer to overcome the pair-annihilation cross section.
We show the bound-state formation cross section for the 1S, 2S, 2P and 3S state;
explicit expressions for the cross section of states above the ground state can be found in appendix~\ref{sec:app_A}.
Our results agree with ref.~\cite{Petraki:2016cnz}. 

\begin{figure}[ht]
    \centering
    \includegraphics[scale=0.75]{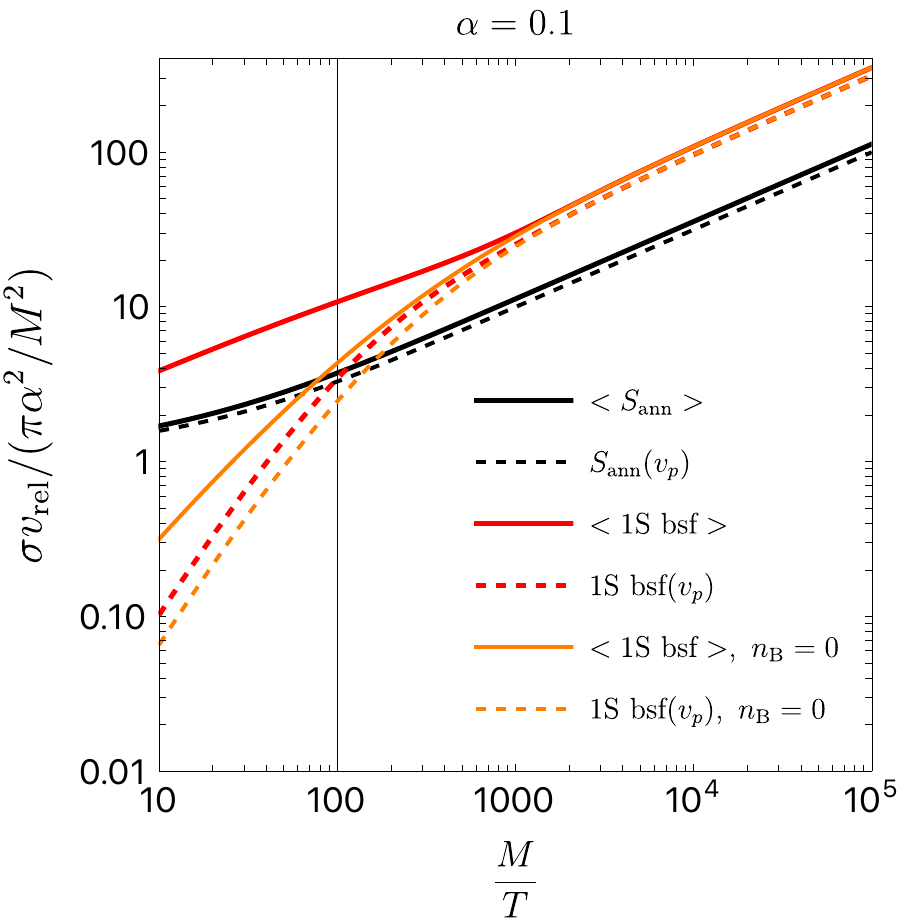}
    \hspace{0.5cm}
    \includegraphics[scale=0.75]{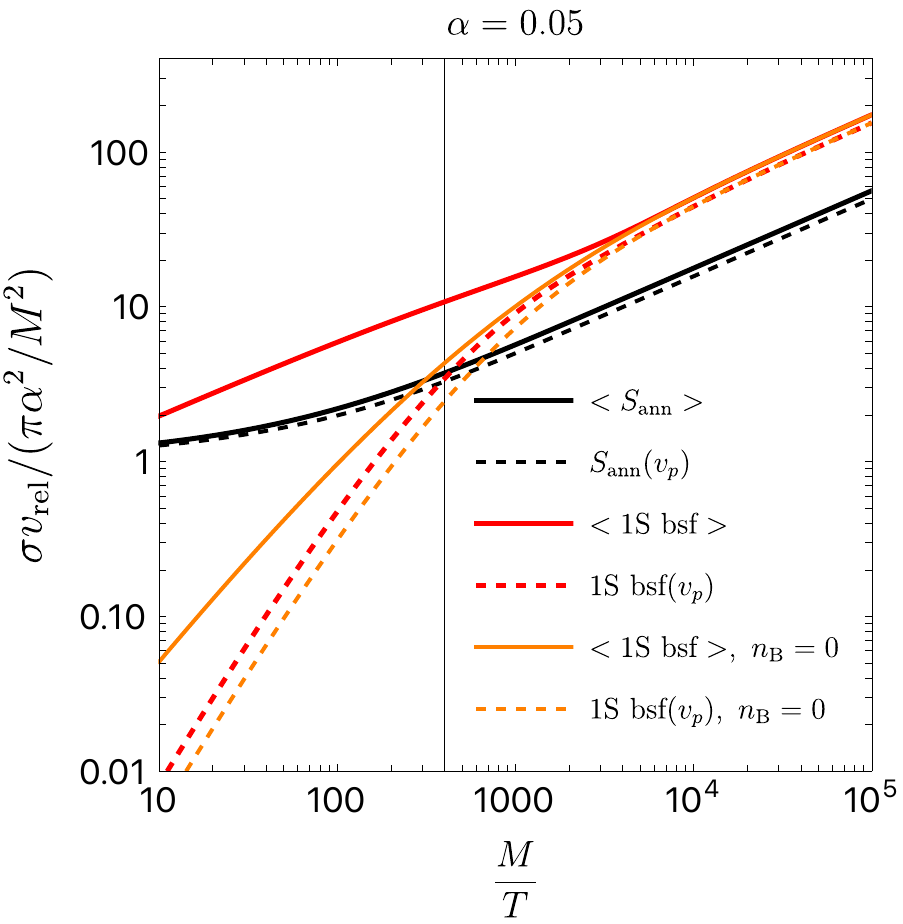}
    \caption{Annihilation cross section for scattering states and the 1S bound-state formation cross section normalized by the free LO annihilation cross section,
      $(\sigma^{\hbox{\tiny NR}}_{\hbox{\scriptsize ann}} v_{\hbox{\scriptsize rel}})_{\hbox{\tiny LO}}$.
      The solid black (red) line stands for the thermally averaged annihilation (bound-state formation) cross section.
      The corresponding dashed curves stand for the cross sections computed at $v_{\hbox{\scriptsize rel}}=v_p$.
      The orange lines stand for the averaged (solid line) and at $v_{\hbox{\scriptsize rel}}=v_p$ (dashed line) bound-state formation cross sections 
      computed setting $n_{\text{B}}=0$ in eq.~\eqref{our_gamma_1p}.
      The vertical line marks the position where $T=M\alpha^2$.
          }
    \label{fig:sigma_bsf_01_TA}
\end{figure}

We can also define, similarly to what done for the annihilation cross section, cf. eq.~\eqref{thermal_averaged_CR}, 
an average bound-state formation cross section  $\langle \sigma_{\hbox{\scriptsize bsf}} \, v_{\hbox{\scriptsize rel}}\rangle$.
In figure~\ref{fig:sigma_bsf_01_TA}, we give the thermally averaged annihilation cross section (solid black curve) and bound-state formation cross section (solid red curve).
In order to show the impact of the thermal average on the different quantities,
we also include as dashed lines the corresponding cross sections with a fixed velocity, namely $v_p$.
We notice that for the annihilation cross section the difference is very small, whereas it is larger for the bound-state formation cross section.
Finally, the effect of the Bose--Einstein distribution for the emitted photon is way more important in the thermally averaged cross section
than in the cross section at fixed-velocity (cf. orange curves).
This is due to the fact that for $T\to \infty$ (or $M/T\to 0$) the Bose enhancement, i.e. the fact that $n_{\text{B}}(\Delta E_{1}^{p}) \to \infty$ for $T\to\infty$,
is lost when $v_{\hbox{\scriptsize rel}}$ is fixed to be $v_p =\sqrt{4T/M}$, in which case  $n_{\text{B}}(\Delta E_{1}^{p}) \to 1/(e-1) \approx 0.58$ for $T\to\infty$.

\subsection{Bound-state dissociation}
\label{sec:Electric_transitions_diss}
Dark photons from the plasma may trigger {\em bound-state dissociation} (bsd) through the reaction 
\begin{equation}
\gamma + (X\bar{X})_n\to (X\bar{X})_p\,.
\label{thermaldiss}
\end{equation}
The process is also called \textit{photo-dissociation}.
As the plasma temperature decreases, fewer darkonium states can be ionized.
In the literature, the dissociation width has been often derived from the in-vacuum bound-state formation cross section through the Milne relation~\cite{vonHarling:2014kha, Mitridate:2017izz},
while the thermal character of the process has been recovered by thermally averaging over the Bose--Einstein distribution of the photon at a later stage.
In the following, we compute the dissociation width of a bound state starting from  $\textrm{pNRQED}_{\textrm{DM}}$  at finite temperature.
In this way, the thermal photon distribution appears from the beginning in the self energy.
Bound-state to bound-state transitions of the type $\gamma + (X\bar{X})_n\to (X\bar{X})_{n'}$ and $(X\bar{X})_n\to \gamma + (X\bar{X})_{n'}$ are discussed in appendix~\ref{sec:app_B}.

At one loop, we can obtain the photo-dissociation width of a bound state with quantum numbers $n$
from the self-energy diagram shown in the right panel of figure~\ref{fig:pnEFT_DM_self}.
The optical theorem relates the bound-state width to the imaginary part of the self energy
either in the form $\Gamma^n = -2\,\langle n| {\rm{Im}}[\Sigma^{11}(E_n)] |n\rangle $ or  $\Gamma^n = \langle n|[-i\Sigma^{21}(E_n)]|n\rangle$
with $E_n$ being the energy of the incoming bound state.
The expressions of ${\rm{Im}}\,\Sigma^{11}$ and $\Sigma^{21}$ have been given in eq.~\eqref{self_00_11} and eq.~\eqref{Sigma12}, respectively.
Projecting on intermediate unbound fermion-antifermion pairs of relative momentum $\bm{p}$ selects the part of ${\rm{Im}}[\Sigma^{11}(E_n)]$ or $\Sigma^{21}(E_n)$ 
with negative $\Delta E = E_n - E_p = - \Delta E_n^p$, i.e. the one proportional to $n_{\text{B}}(-\Delta E) = n_{\text{B}}(\Delta E_n^p)$.
The Bose--Einstein distribution  $n_{\text{B}}(\Delta E_n^p)$ vanishes in the $T=0$ limit,
which reflects the fact that the decay of a bound state into an unbound pair is kinematically forbidden in vacuum.
The photo-dissociation width is therefore a purely {\em thermal width}, which reads
\begin{equation}
  \Gamma^{n}_{\textrm{bsd}} =
    \frac{g^2}{3 \pi} \int \frac{d^3p}{(2\pi)^3} \,  n_{\text{B}}(\Delta E_n^p) \, \left| \sum_\ell \langle n | \bm{r} | \bm{p} \ell \rangle\right|^2 (\Delta E_n^p)^3\,,
 \label{gamma_diss_n_state}
\end{equation}
where we have expanded the scattering states into partial waves of orbital angular momentum $\ell$;
note that the state $|\bm{p} \ell \rangle$ is an eigenstate of the orbital angular momentum, but no more of the momentum $\bm{p}$.
Keeping the thermal distribution of the heavy fermion-antifermion pair in the self-energy diagram would modify the photo-dissociation width into
\begin{equation}
  \Gamma^{n}_{\textrm{bsd}}  =
  \frac{g^2}{3 \pi}\,\int \frac{d^3p}{(2\pi)^3} \,  n_{\text{B}}(\Delta E_n^p) \,
  \left[ 1 + n_{\text{B}}(E_p) \right]\,
  \left| \sum_\ell \langle n | \bm{r} | \bm{p} \ell \rangle\right|^2 (\Delta E_n^p)^3\,.
 \label{gamma_diss_n_statefull}
\end{equation}
Taking into account that the thermal distribution of the heavy fermion-antifermion pair vanishes exponentially for $T \ll M \sim E_p/2$, 
the above expression reduces to the one in eq.~\eqref{gamma_diss_n_state}.

For photo-dissociation of the lowest-lying darkonium, eq.~\eqref{gamma_diss_n_state} becomes\footnote{
  Due to the selection rule of the electric dipole matrix element, only a transition into a scattering state with orbital angular momentum quantum number $\ell=1$ is possible.}
\begin{eqnarray}
  \Gamma^{1 \textrm{S}}_{\textrm{bsd}}  =
  \int_{|\bm{k}|\geq |E^b_{1}|} \frac{d^3k}{(2\pi)^{3}}\,n_{B}(|\bm{k}|)\,\frac{g^{2}}{3\pi}\,
  \frac{M^{\frac{3}{2}}}{2}\,|\bm{k}|\,\sqrt{|\bm{k}|+E_{1}^b}\;
  |\langle \text{1S}|\bm{r}|\bm{p}1\rangle|^2\bigg \vert_{|\bm{p}| = \sqrt{M(|\bm{k}|+E^b_{1})}} \,,
  \nonumber\\
  \label{dis_width}
\end{eqnarray}
where $E^b_n \equiv E_n - 2M$; in particular, $E_1^b = -M\alpha^2/4$ is the binding energy of the 1S state.
Note that the photon needs to have a threshold momentum to trigger the breaking of the bound state.
The result agrees in the abelian limit with the gluo-dissociation width of a color singlet quark-antiquark bound state
in the temperature regime $T \sim M \alpha_s^2$~\cite{Brambilla:2011sg,Brezinski:2011ju}.
The gluo-dissociation width of a heavy quarkonium in the static limit was obtained in ref.~\cite{Brambilla:2008cx} and for a hydrogen atom in QED in ref.~\cite{Escobedo:2008sy}.

The thermal width can be also understood as the convolution of the in-vacuum {\em ionization} (ion) cross section of the bound-state for the process $\gamma + (X\bar{X})_n\to (X\bar{X})_p$,
which we denote $\sigma^{n}_{\hbox{\scriptsize ion}}$, with the thermal distribution of the incoming photon,
\begin{equation}
  \Gamma^{n}_{\textrm{bsd}}  = 2 \int_{|\bm{k}|\geq |E^b_{n}|} \frac{d^3k}{(2\pi)^3} \,  n_{\text{B}}(|\bm{k}|)\, \sigma^{n}_{\hbox{\scriptsize ion}}(|\bm{k}|)\,,
\label{photogammacrosssection}
\end{equation}
where 2 is the number of final state photon polarizations and the relative velocity between the darkonium and the photon from the bath has been set equal to one.
Comparing the above equation with eq.~\eqref{dis_width} and using the expression of the dipole matrix element given in appendix~\ref{sec:app_A}, we obtain
\begin{equation}
\sigma^{1 \textrm{S}}_{\hbox{\scriptsize ion}}(|\bm{k}|)= \alpha \frac{ 2^{9} \pi^2}{3} \frac{|E_1^b|^3}{M  |\bm{k}|^4}  \frac{e^{-\frac{4}{w_1(|\bm{k}|)}\arctan(w_1(|\bm{k}|))} }{1-e^{-\frac{2\pi}{w_1(|\bm{k}|)}}} \, ,
\label{our_disso_cross}
\end{equation}
with $w_1(|\bm{k}|) \equiv \sqrt{|\bm{k}|/|E_1^b|-1}$. 
The result in eq.~\eqref{our_disso_cross} agrees with the ionization cross section given in ref.~\cite{vonHarling:2014kha},
where it was obtained through the Milne relation.\footnote{
  One has to express eq.~\eqref{dis_width} in terms of the momentum of the scattering state, $|\bm{p}|=M v_{\hbox{\scriptsize rel}}/2$, 
  i.e. $|\bm{k}| = M (v_{\textrm{rel}}^2 + \alpha^2)/4 = \Delta E_{1}^{p}$ and $w_1(|\bm{k}|)=v_{\textrm{rel}}/\alpha$, cf. eq.~\eqref{dis_width_2}.}
Here, we did not rely on an explicit use of the Milne relation, but on thermal field theory alone.
Thermal field theory provides, by construction, the dissociation width with the correct temperature dependence.

\begin{figure}[ht]
         \centering
         \includegraphics[scale=0.76]{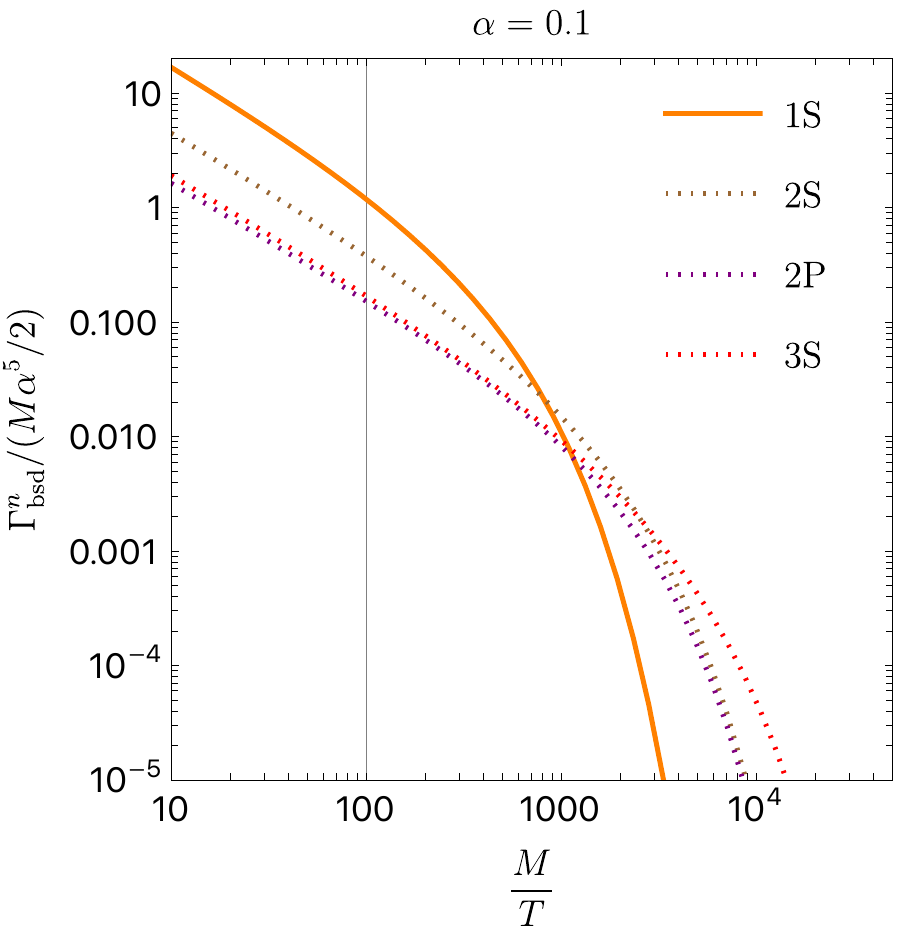}
         \hspace{0.5 cm}
           \includegraphics[scale=0.75]{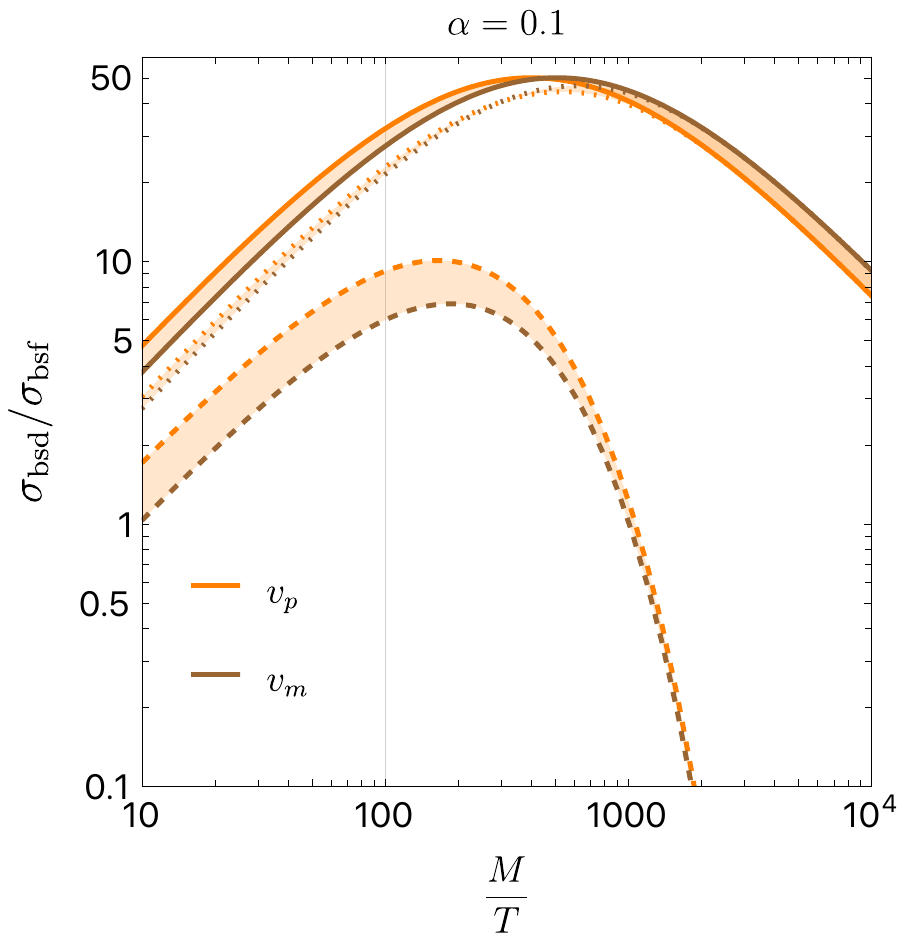}
           \caption{(Left) Thermal widths from photo-dissociation, eq.~\eqref{photogammacrosssection},
             for the 1S bound state  (solid line), the 2S, 2P and 3S excited states (dotted lines).
             The width for the 2P bound state has been averaged over the polarizations $m=0,\pm 1$.
             (Right) Milne relation for the most-probable velocity $v_p$ and average velocity $v_m$ for the 1S bound state.
             Dotted lines stand for the Milne relation given in eq.~\eqref{our_milne}, whereas dashed lines have been obtained by replacing the ionization cross section with the left-hand side of eq.~\eqref{milne1}.
             The solid lines stand for the Milne relation~\eqref{milne}, also given in ref.~\cite{vonHarling:2014kha}.
             The vertical line marks the position where $T= M \alpha^2$.
             Results are given for $\alpha=0.1$.
           }
         \label{fig:Diss_1_and_2}
       \end{figure}

It is straightforward to compute the dissociation width of excited bound states.
We consider 2S, 2P and 3S states, with binding energies $E^b_2=-M\alpha^2/16$ and $E_3^b=-M\alpha^2/36$, respectively, 
and compute the corresponding in-vacuum photo-dissociation cross sections in appendix~\ref{sec:app_A_1}. 
In the left panel of figure~\ref{fig:Diss_1_and_2}, we plot the corresponding photo-dissociation widths, accordingly to eq.~\eqref{photogammacrosssection}.
At very small temperatures the thermal width for the 1S state vanishes faster than the one for the 2S state, at higher temperatures the 1S thermal width is larger than the 2S one.

\subsubsection{Milne relation at finite temperature}
\label{sec:milne}
The processes of bound-state dissociation and formation have a rather different behaviour depending on the temperature of the plasma.
Typically, dissociation gets suppressed at low temperatures, whereas formation becomes larger.
Nevertheless dissociation and formation cross section are related by the Milne relation.

Let us consider, as an example, the 1S bound-state case.
In terms of $v_{\textrm{rel}} = 2p/M$, the ionization cross section \eqref{our_disso_cross} reads
\begin{equation}
  \sigma^{1 \textrm{S}}_{\textrm{ion}}(\bm{p})   =  \frac{\alpha^7 \pi^2 \, 2^{11}}{3 \, M^2 \, v_{\textrm{rel}}^8 \left( 1+ \frac{\alpha^2}{v_{\textrm{rel}}^2}\right)^4}
       \frac{e^{-4 \frac{\alpha}{v_{\textrm{rel}}} \hbox{\scriptsize arccot} \frac{\alpha}{v_{\textrm{rel}}} }}{1-e^{-2 \pi \frac{\alpha}{v_{\textrm{rel}}}}}\,
         \label{dis_width_2}\, .
\end{equation}
The ratio of  $\sigma^{1 \textrm{S}}_{\textrm{ion}}$ and $\sigma^{1\textrm{S}}_{\textrm{bsf}}$ from  eq.~\eqref{our_gamma_1p} is then
\begin{equation}
  \frac{\sigma^{1 \textrm{S}}_{\textrm{ion}}(\bm{p}) }{\sigma^{1 \textrm{S}}_{\textrm{bsf}}(\bm{p})}   =
  \frac{M^2v_{\textrm{rel}}^2}{8(\Delta E_{1}^{p})^2}\frac{1}{1+n_{\text{B}}(\Delta E_{1}^{p})}\, .
\label{our_milne}
\end{equation}
In the $T = 0$ limit at fixed $v_{\textrm{rel}}$ one recovers the in-vacuum Milne relation as given in ref.~\cite{vonHarling:2014kha},
\begin{equation}
  \left.\frac{\sigma^{1 \textrm{S}}_{\textrm{ion}}(\bm{p})}{\sigma^{1 \textrm{S}}_{\textrm{bsf}}(\bm{p})}\right|_{T=0} = \frac{M^2v_{\textrm{rel}}^2}{8(\Delta E_{1}^{p})^2}\, .
\label{milne}
\end{equation}
If the initial state dark photons in the process \eqref{thermaldiss} are thermally distributed then it could make sense to define a temperature-dependent bound-state dissociation cross section
\begin{equation}
\sigma^{1 \textrm{S}}_{\textrm{bsd}}(\bm{p}) \equiv \sigma^{1 \textrm{S}}_{\textrm{ion}}(\bm{p}) \, n_{\text{B}}(\Delta E_{1}^{p})\,.
\label{milne1}
\end{equation}
It is precisely this quantity that enters the width \eqref{photogammacrosssection} for the ground state.
Differently from $\sigma^{1 \textrm{S}}_{\textrm{ion}}$, $\sigma^{1 \textrm{S}}_{\textrm{bsd}}$ vanishes in vacuum, i.e. for  $T = 0$.
The ratios $\left. \sigma^{1 \textrm{S}}_{\textrm{ion}}/\sigma^{1 \textrm{S}}_{\textrm{bsf}}\right|_{T=0}$ (at fixed $v_{\textrm{rel}}$),
$\sigma^{1 \textrm{S}}_{\textrm{ion}}/\sigma^{1 \textrm{S}}_{\textrm{bsf}}$ and $\sigma^{1 \textrm{S}}_{\textrm{bsd}}/\sigma^{1 \textrm{S}}_{\textrm{bsf}}$
have been plotted in the right panel of figure~\ref{fig:Diss_1_and_2} as a function of the temperature for the choices 
$v_{\textrm{rel}} = v_p$ and $v_{\textrm{rel}} = v_m$.\footnote{
  The temperature dependence of  $\left. \sigma^{1 \textrm{S}}_{\textrm{ion}}/\sigma^{1 \textrm{S}}_{\textrm{bsf}}\right|_{T=0}$ comes
  from the temperature dependence of the velocities $v_p$ and $v_m$.
}  
Note that $\sigma^{1 \textrm{S}}_{\hbox{\scriptsize bsd}}(\bm{p}) /\sigma^{1 \textrm{S}}_{\textrm{bsf}}(\bm{p})$ at small $T$ goes like 
$\left. \sigma^{1 \textrm{S}}_{\hbox{\scriptsize ion}}(\bm{p}) /\sigma^{1 \textrm{S}}_{\textrm{bsf}}(\bm{p}) \right|_{T=0} e^{-\Delta E_{1}^{p} /T}$, i.e. it is exponentially suppressed.

\section{Dark matter energy density}
\label{sec:numerics}
In this section, we collect numerical results for the dark matter energy density.
The main motivation for a precise calculation of the DM energy density relies on the accurate determination of such observable,
$\Omega_{\textrm{DM}} h^2=0.1200 \pm 0.0012$ \cite{Planck:2018nkj}, which features a 1\% uncertainty. 

As previously mentioned, in this work we rely on Boltzmann equations as far as the rate equations are concerned.
Upon including bound states in the network of Boltzmann equations, their solution can be cumbersome.
We use an approximation, first introduced in ref.~\cite{Ellis:2015vaa} and commonly adopted in the literature, that is based on an effective treatment of dark matter bound states.
In a typical cosmological setting, the annihilation rate of bound states is pretty efficient, $\Gamma_{\textrm{ann}} \gg H$, so that they quickly adjust to their equilibrium number densities.
Upon neglecting the (de-)excitations between bound states, one obtains a single Boltzmann equation that depends only on the density of scattering states $n$, and is governed by an effective cross section (see footnote~\ref{footnote_coupled_Boltzmann_eqs}).
The latter comprises the effects of DM annihilation via unbound pairs and bound states, and bound-state formation cross sections and dissociation widths.
Recent studies have further investigated the validity and generalization of this approach
when transitions between different bound states are kept in the network of Boltzmann equations \cite{Garny:2021qsr,Binder:2021vfo}. 

\begin{figure}[ht]
    \centering
    \includegraphics[scale=0.75]{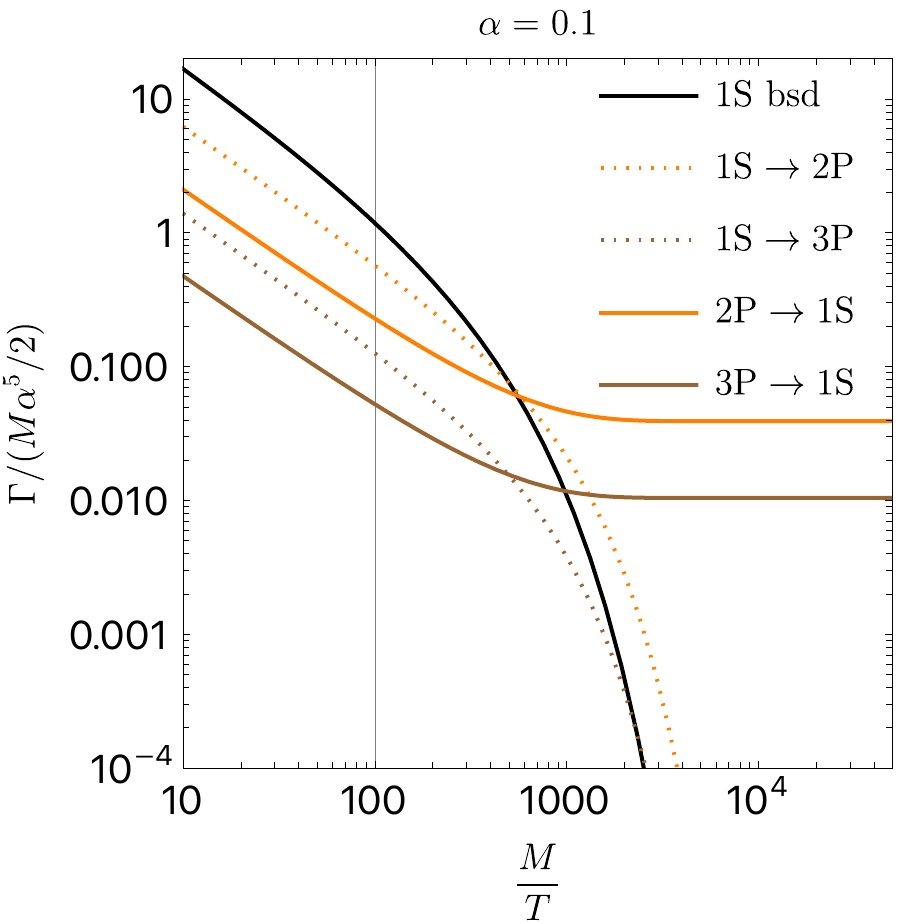}
    \hspace{0.5cm}
    \includegraphics[scale=0.75]{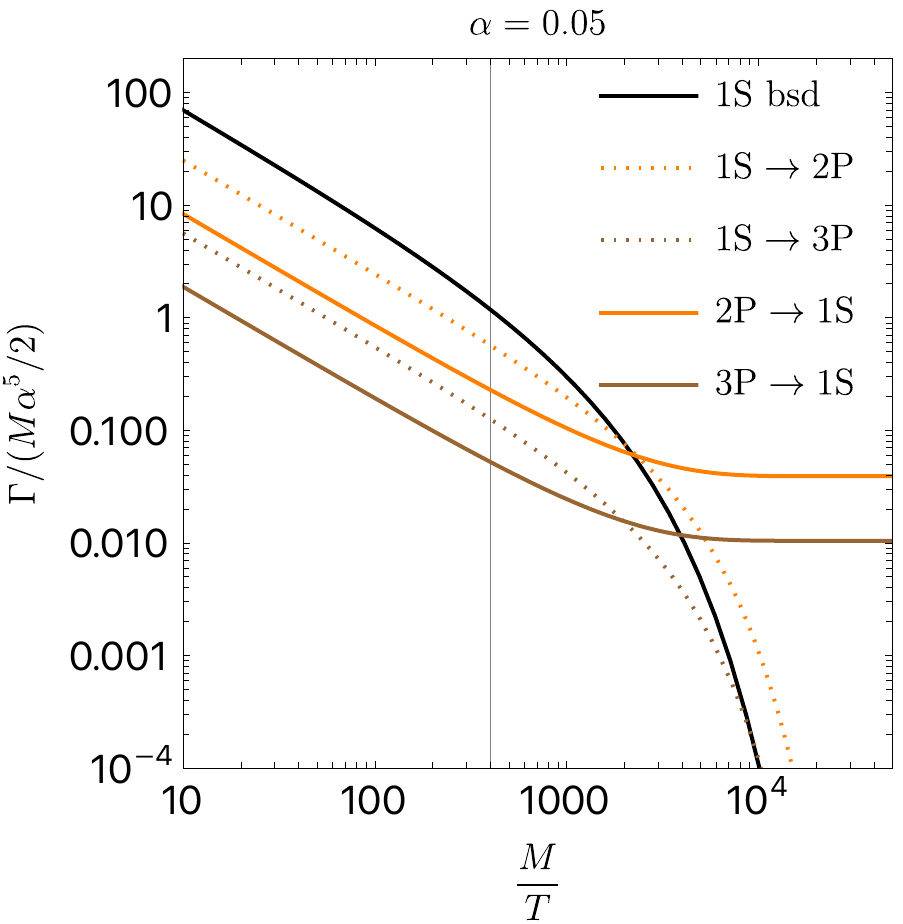}
    \caption{Bound state dissociation and bound-state to bound-state transition widths relevant for the 1S state normalized with respect to the leading order 1S-state annihilation width
      as a function of $M/T$.
      In the left plot $\alpha = 0.1$, in the right plot $\alpha=0.05$.
      The vertical line marks the position where $T=M\alpha^2$.}
    \label{fig:differentwidths}
  \end{figure}
  
The Boltzmann equation that we use in the following numerical analyses reads 
\begin{equation}
    (\partial_t + 3H) n = - \frac{1}{2}\langle \sigma_{\textrm{eff}} \, v_{\textrm{rel}} \rangle (n^2-n^2_{\textrm{eq}}) \,.
    \label{Boltzmann_eq_eff}
\end{equation}
The dissociation width and different transition widths contributing to the effective cross section $ \sigma_{\textrm{eff}}$
entering eq.~\eqref{Boltzmann_eq_eff} are shown as a function of the temperature in figure~\ref{fig:differentwidths} for the case of the 1S state.
When bound-state to bound-state transitions are much smaller than $\Gamma_{\textrm{ann}}^n$ and $\Gamma_{\textrm{bsd}}^n$,
the thermally averaged effective cross section can be written as 
 \begin{equation}
   \langle  \sigma_{\textrm{eff}} \, v_{\textrm{rel}} \rangle  =
   \langle \sigma_{\textrm{ann}} v_{\textrm{rel}} \rangle
   + \sum_{n} \langle   \sigma^n_{\textrm{bsf}} \, v_{\textrm{rel}} \rangle \, \frac{\Gamma_{\textrm{ann}}^n}{\Gamma_{\textrm{ann}}^n+\Gamma_{\textrm{bsd}}^n} \, .
    \label{Cross_section_eff}
 \end{equation}
The first term stems from the annihilation of a pair in a scattering state, yielding eq.~\eqref{Boltzmann_1},
whereas the second term encodes the reprocessing of an unbound pair into a bound state.
The sum over $n$ extends over all bound states: the 1S spin singlet paradarkonium state, the 1S spin triplet orthodarkonium states, and so on.
The combination $\Gamma^{n}_{\textrm{ann}}/(\Gamma_{\textrm{ann}}^n+\Gamma_{\textrm{bsd}}^n)$ includes the information about the dissociation width of the bound state.
It is small at large temperatures, but it is close to one at temperatures of the order of the binding energy or smaller,
when bound state dissociation becomes less efficient than annihilation, i.e. $\Gamma_{\textrm{bsd}}^n \ll \Gamma_{\textrm{ann}}^n$.
At sufficiently small temperatures, the bound-state to bound-state transitions become as large as or larger than the dissociation widths
and cannot be neglected with respect to them (see figures~\ref{fig:differentwidths} and~\ref{bound-to-bound_pic}).
In this situation, we replace eq.~\eqref{Cross_section_eff} with the expression that can be found in ref.~\cite{Garny:2021qsr}, which takes into account transition rates.
Finally, we remark that late in the evolution of the universe, i.e. at temperatures such that the DM particles are very diluted,
DM particle interactions become negligible with respect to the universe expansion rate that dominates the evolution equation \eqref{Boltzmann_eq_eff}.

It is convenient to recast the Boltzmann equation \eqref{Boltzmann_eq_eff} in terms of the yield $Y\equiv n/s$, with $s$ being the entropy density.
The present-day DM relic density is then $\Omega_{\textrm{DM}} = M s_0 Y_0/\rho_{\textrm{crit},0}$,
where $Y_0, s_0$ and $\rho_{\textrm{crit},0}$  denote respectively the present yield, entropy density and critical density.
The values for $s_0$ and  $\rho_{\textrm{crit},0}$ can be taken from  e.g.~\cite{ParticleDataGroup:2022pth},
and one obtains $\Omega_{\textrm{DM}} h^2=(M/\textrm{GeV)}\, Y_0 /(3.645 \times 10^{-9})$, where $h$ is the reduced Hubble constant.
The temperature-dependent relativistic degrees of freedom entering the Hubble rate in eq.~\eqref{Boltzmann_eq_eff}
are assumed to be those of the SM with the addition of the dark photon~\cite{vonHarling:2014kha}.

\begin{figure}[t!]
    \centering
    \includegraphics[scale=0.6]{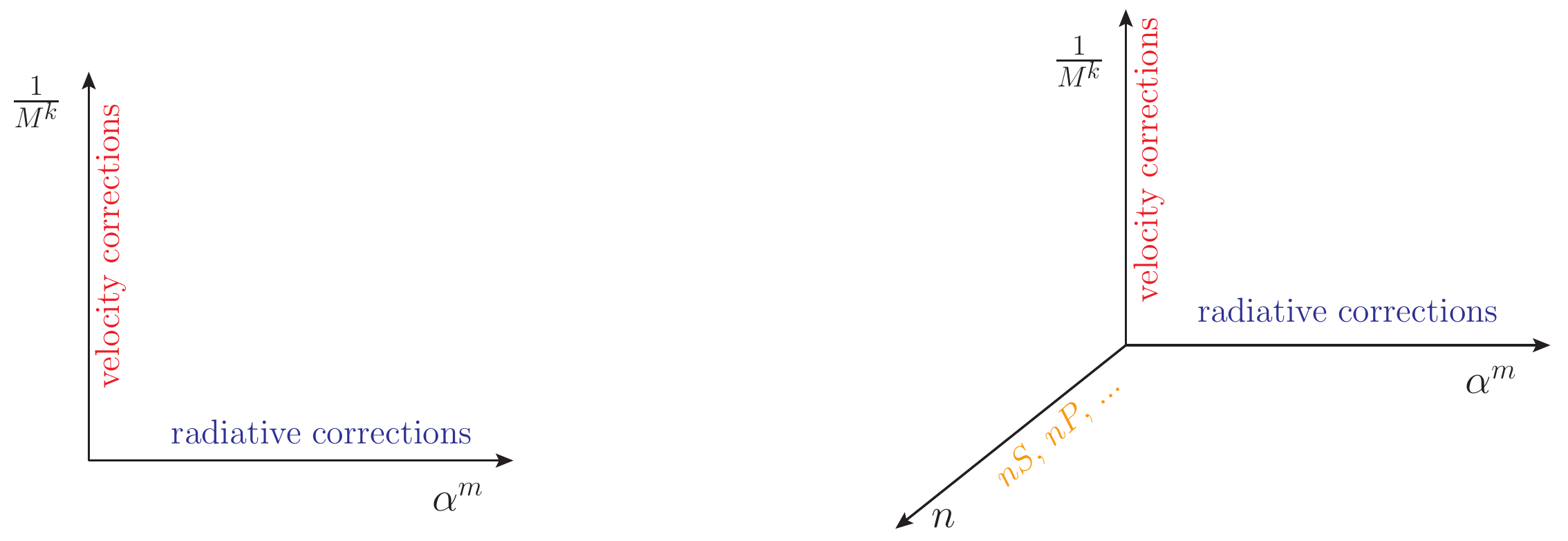}
    \caption{Schematic representation of the corrections to cross sections and widths.
      The left plot is for scattering states, whereas the right plot is for bound states. }
    \label{fig:EFT_scheme_corr}
\end{figure}

The main advantage of the EFT framework is to allow for a rigorous derivation and a systematic inclusion of corrections to the relevant observables.
In the problem at hand, we are interested in the cross sections and widths entering the thermally averaged effective cross section.
We refer to the scheme in figure~\ref{fig:EFT_scheme_corr} to pictorially illustrate the different corrections.
As for the scattering states, which are integrated over all momenta, we can improve cross sections and widths by adding relativistic corrections.
This amounts at including higher-dimensional operators in both $\textrm{NRQED}_{\textrm{DM}}$ and $\textrm{pNRQED}_{\textrm{DM}}$.
Furthermore, we can add radiative corrections to the matching coefficients of the effective field theories (see, for instance, section~\ref{sec:nrqed_annihilation}).
As for the bound states, the situation is similar.
Higher-dimensional operators account for higher-order relativistic corrections, which at the same time may open new decay channels.
Radiative corrections improve the matching coefficients.
Moreover, we add a third dimension to the scheme that accounts for the number of bound states to be included in the analyses consistently with the other corrections.
The inclusion of excited states can be seen as a further improvement towards an accurate description of the actual physical system,
and hence a correction to the simplest possible situation, when only the ground state is considered.
We provide below some examples that may help to assess the relative importance of the different corrections. 

\begin{figure}[ht]
    \centering
    \includegraphics[scale=0.8]{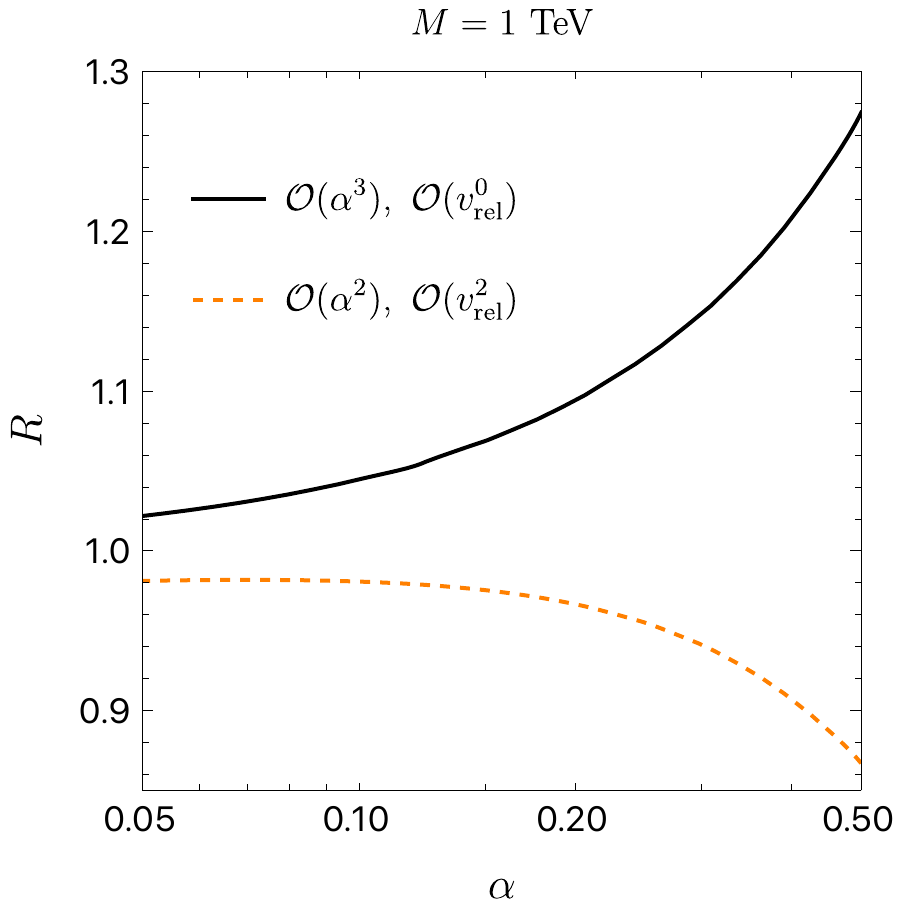}
    \caption{Ratio of the DM energy densities extracted from effective cross sections comprising scattering states only.
      The black solid line is the ratio between the energy density $\Omega_{\textrm{DM}} h^2$ as obtained
      from $(\sigma^{\hbox{\tiny NR}}_{\hbox{\scriptsize ann}} v_{\hbox{\scriptsize rel}})_{\textrm{NLO}} S(\zeta)$
      and the one obtained from $(\sigma^{\hbox{\tiny NR}}_{\hbox{\scriptsize ann}} v_{\hbox{\scriptsize rel}})_{\textrm{LO}} S(\zeta)$.
      The dashed orange line is the ratio between the energy density $\Omega_{\textrm{DM}} h^2$ from the cross section in eq.~\eqref{ann_fact_scat_v2}
      and the one from $(\sigma^{\hbox{\tiny NR}}_{\hbox{\scriptsize ann}} v_{\hbox{\scriptsize rel}})_{\textrm{LO}} S(\zeta)$.}
    \label{fig:NLO_versus_V2}
  \end{figure}

In the first case that we examine, we want to investigate the relative importance of NLO radiative corrections to the hard matching coefficients,
which are corrections of higher order in $\alpha$, 
with respect to relativistic corrections, which are corrections of higher order in the relative velocity of the dark fermion-antifermion pair.
For illustration, we consider a particular subset of higher-order corrections in $v_{\textrm{rel}}$, those stemming from four-fermion operators of dimension 8,
which show up at order $1/M^4$.\footnote{
Relativistic corrections also affect the dark fermion-antifermion pair wavefunctions. We do not consider these corrections here.}
Those operators can be read off from ref.~\cite{Bodwin:1994jh}, and the corresponding matching coefficients at $\mathcal{O}(\alpha^2)$ from refs.~\cite{Bodwin:1994jh,Vairo:2003gh}.
The resulting annihilation cross section is 
\begin{eqnarray}
  (\sigma_{\hbox{\scriptsize ann}} v_{\hbox{\scriptsize rel}})(\bm{p}) 
  =(\sigma^{\hbox{\tiny NR}}_{\hbox{\scriptsize ann}} v_{\hbox{\scriptsize rel}})_{\textrm{LO}}  \,
  \left[ \left( 1 - \frac{v^2_\textrm{rel}}{3}\right)S_{\hbox{\scriptsize ann}}(\zeta) +\frac{7}{12} v^2_\textrm{rel} S_{\textrm{ann}}^{\ell=1}(\zeta) \right]\,,
\label{ann_fact_scat_v2} 
\end{eqnarray}
where $S_{\textrm{ann}}^{\ell=1}(\zeta)=(1+\zeta^2)S_{\textrm{ann}}(\zeta)$ is the Sommerfeld enhancement for a scattering state in a P-wave \cite{Iengo:2009ni,Cassel:2009wt},
and $(\sigma^{\hbox{\tiny NR}}_{\hbox{\scriptsize ann}} v_{\hbox{\scriptsize rel}})_{\textrm{LO}}$ is given in eq.~\eqref{LOexpression}.
We solve the Boltzmann equation  \eqref{Boltzmann_eq_eff} without bound-state effects, i.e. we neglect the second term in the right-hand side of eq.~\eqref{Cross_section_eff}. 
The result is shown in figure~\ref{fig:NLO_versus_V2}.
The black solid line is the ratio, $R$, between the energy density $\Omega_{\textrm{DM}}h^2$ as obtained with $(\sigma^{\hbox{\tiny NR}}_{\hbox{\scriptsize ann}} v_{\hbox{\scriptsize rel}})_{\textrm{NLO}} S(\zeta)$,
where $(\sigma^{\hbox{\tiny NR}}_{\hbox{\scriptsize ann}} v_{\hbox{\scriptsize rel}})_{\textrm{NLO}}$ is given in eq.~\eqref{NLO_crosssection}, 
and $(\sigma^{\hbox{\tiny NR}}_{\hbox{\scriptsize ann}} v_{\hbox{\scriptsize rel}})_{\textrm{LO}} S(\zeta)$ as input for $\langle \sigma_{\textrm{ann}} v_{\textrm{rel}} \rangle$, respectively.
As elaborated in section \ref{sec:nrqed_annihilation}, the NLO corrections to the matching coefficients make the cross section smaller,
and hence a more abundant dark matter population is found for each value of $\alpha$.
Accordingly, we find $R>1$, as shown in the plot.
The dashed orange line stems for the ratio of the energy density as obtained from the cross section in eq.~\eqref{ann_fact_scat_v2}
and $(\sigma^{\hbox{\tiny NR}}_{\hbox{\scriptsize ann}} v_{\hbox{\scriptsize rel}})_{\textrm{LO}} S(\zeta)$.
The trend here is different.
The corrections to the cross section make it larger for each $\alpha$, and accordingly we find a smaller DM energy density that results in $R<1$.
The P-wave contribution overcomes the negative correction of the velocity dependent S-wave correction, especially at large values of $\alpha$.
The result has a rather mild dependence on the specific value of the DM mass (set to $M=1$~TeV in the plot). 

\begin{figure}[ht]
    \centering
    \includegraphics[scale=0.75]
    {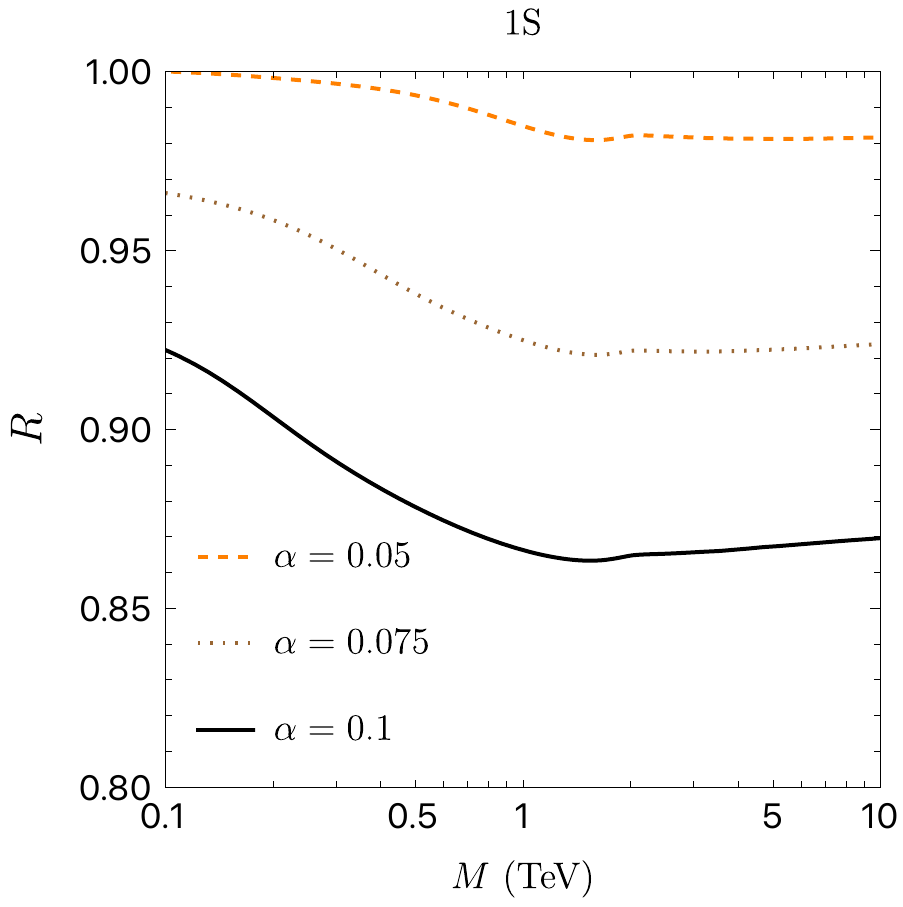}
    \hspace{0.3 cm} \includegraphics[scale=0.75]{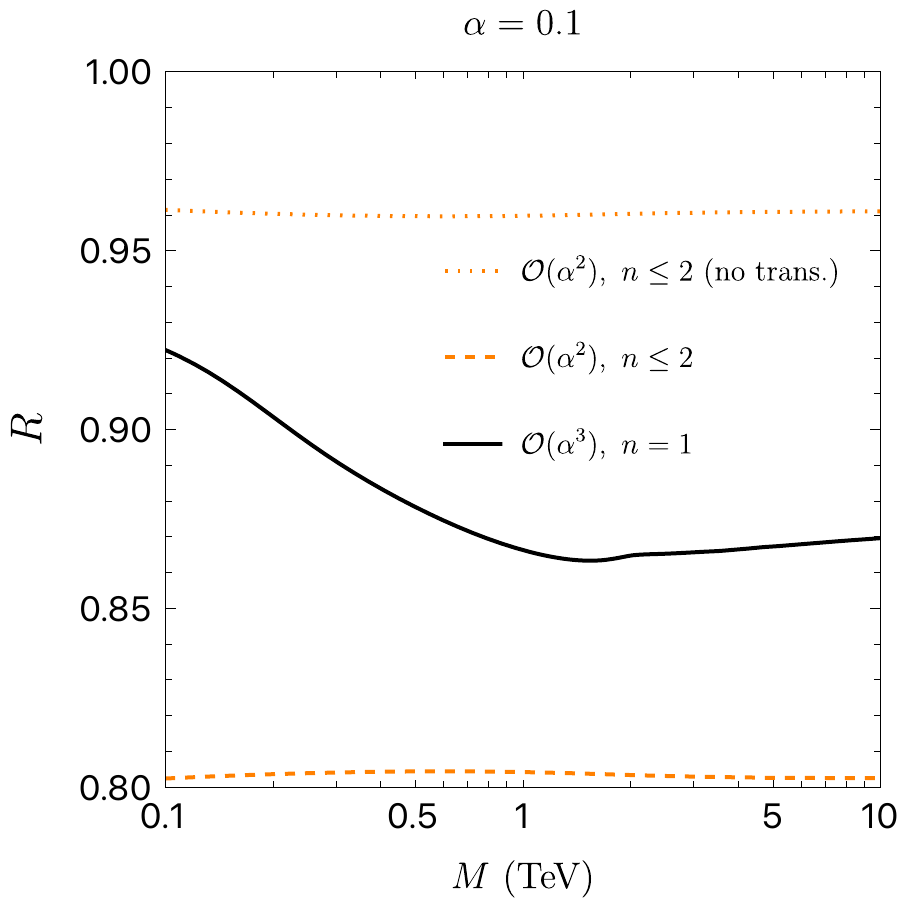}
     \caption{(Left) Ratio of the DM energy density obtained from the effective cross section \eqref{Cross_section_eff} at $\mathcal{O}(\alpha^3)$ and
      the one at $\mathcal{O}(\alpha^2)$,
      including both scattering states and the bound state 1S, for different values of $\alpha$.
      (Right) The solid-black line is the same as in the left panel and is reproduced here for ease of comparison.
      The orange-dotted (dashed) line stands for the DM energy density computed from the effective cross section including 
      the excited states 2S and 2P$_{m=0,\pm 1}$ within  (without) the no-transition approximation. 
       }
    \label{fig:NLO_versus_nS}
\end{figure}

In the second case that we consider, we include bound-state effects, i.e. we reinstate the second term in the effective cross section in eq.~\eqref{Cross_section_eff}.
We work at leading order in the $1/M$ expansion.
Therefore, we do not include dimension-8 operators neither in the annihilation cross section, nor in the annihilation widths of the bound states that can only be S-wave states,
since P-wave states decay through dimension-8 operators at order $1/M^4$.
We solve the Boltzmann equation with both contributions from scattering and bound states at order $\alpha^3$ and $\alpha^2$.
We restrict the sum over $n$ just to 1S states, which means that at order $\alpha^2$ we consider only paradarkonium
and at order $\alpha^3$ we include also the formation and decay of orthodarkonium.
In the left panel of figure~\ref{fig:NLO_versus_nS}, we show the ratio of DM energy densities for the $\alpha^3$ case over the  $\alpha^2$ case.
As one may see, the ratio is $R<1$ because the effective cross section \eqref{Cross_section_eff} is larger whenever we add the spin-triplet bound state.
The formation and decay of orthodarkonium induces an effect that is much larger than the few-per-cent decrease in the annihilation cross section due to $\mathcal{O}(\alpha^3)$ terms.
The overall effect is larger than the uncertainty on the relic density, which is about 1\%, already for $\alpha=0.05$, and is  larger than 13\% for $\alpha=0.1$.
We remark that the conditions required in order to use eq.~\eqref{Cross_section_eff} are satisfied in the orthodarkonium case.
Indeed, the annihilation rate of the orthodarkonium \eqref{ann_ortho1s} is much larger than the Hubble rate,
and the transition rate from the spin-triplet to the spin-singlet 1S state satisfies $\Gamma^{\textrm{1S,ortho} \to \textrm{1S,para}}/\Gamma^{\text{1S},\textrm{ortho}} \ll 1$.\footnote{
  The transition between ortho- and paradarkonium is a magnetic transition, i.e. it is triggered by a magnetic-dipole operator.
  The width is proportional to $\alpha$ times the ratio of the hyperfine splitting to the third power over $M^2$, which gives 
  $\Gamma^{\textrm{1S,ortho} \to \textrm{1S,para}} \sim M \alpha^{13}$; see, e.g.~ref.~\cite{Brambilla:2005zw}. }

In our last case, we compare the effect of adding excited states ($n>1$) to the second term in the right-hand side of eq.~\eqref{Cross_section_eff} at order $\alpha^2$
with having only $\langle \sigma_{\textrm{ann}} v_{\textrm{rel}} \rangle$ plus the effect of the 1S state at order $\alpha^3$.
As for the excited states, we consider four additional states, namely the states 2S and 2P$_{m}$ with $m=-1,0,1$.
Moreover, for $n>1$ we also compare the no-transition case, i.e. the energy density obtained from the effective cross section given in eq.~\eqref{Cross_section_eff},
with the energy density obtained from the effective cross section derived in ref.~\cite{Garny:2021qsr} that accounts for transitions between bound states.
For these three approximations, we compute the DM energy densities normalized, like in the previous case,
with respect to the DM energy computed from the effective cross section at order $\alpha^2$ including the 1S bound state and scattering states.   
In the right panel of figure~\ref{fig:NLO_versus_nS}, we show the result for $\alpha=0.1$.
The formation and decay of orthodarkonium for the 1S state (black-solid line) gives a larger contribution than including excited states with $n=2$
in the spin-singlet configuration, when neglecting transitions among them (orange-dotted line).
The reason for this is that the bound-state formation cross section for orthodarkonium is significantly larger than the bound-state formation cross sections for excited states.
Upon including the bound-state to bound-state transitions (see appendix~\ref{sec:app_B} for the corresponding widths),
2P$_{m=0,\pm 1}$ states may decay into the 1S ground state providing an additional contribution to the depletion of DM.
This results in the orange-dashed line lying below the orange-dotted line, where transitions are neglected, and the black line.  

In table~\ref{table}, we provide a summary of the different effects 
%contributions and its radiative or velocity corrections 
on the DM relic density 
%as with respect to the DM relic density 
obtained accounting for 
%obtained from 
the S-wave annihilation process with LO cross section, $\sigma_{\textrm{ann}}v_{\textrm{rel}}=\pi \alpha^2/M^2$, only.
We consider different DM masses $M$ and couplings $\alpha$. 
For the specific case of $M=1$~TeV and $\alpha=0.1$, the Sommerfeld enhancement decreases the relic density by 73\%, and including bound-state effects reduces it even more: 
including the 1S state with annihilation matching coefficients at LO by 80\%, 
and including excited states up to $n\leq 2$ with bound-to-bound transitions together with annihilation matching coefficients at NLO by 91\%.

\begin{table}
\begin{center}
\begin{tabular}{ |c|c|c|c|c|c|c| } 
 \hline
  contributions & $1$,~$0.05$ & $1$,~$0.1$ & $1$,~$0.2$ & $10$,~$0.05$ & $10$,~$0.1$ & $10$,~$0.2$ \\ \hline
 Sommerfeld enhancement (SE) & 54\% & 73\% & 86\% & 52\% & 72\% & 85\% \\ \hline
 SE + $\mathcal{O}(\alpha^3),\mathcal{O}(v_{\textrm{rel}}^0)$ correction & 53\% & 71\% & 84\% & 51\% & 71\% & 84\% \\ \hline 
 SE + $\mathcal{O}(\alpha^2),\mathcal{O}(v_{\textrm{rel}}^2)$ correction & 54\% & 73\% & 86\% & 53\% & 72\% & 86\% \\ \hline
SE + bsf + $\mathcal{O}(\alpha^2),\mathcal{O}(v_{\textrm{rel}}^0),n=1$ & 60\% & 80\% & 91\% & 58\% & 79\% & 91\%  \\ \hline
SE + bsf + $\mathcal{O}(\alpha^3),\mathcal{O}(v_{\textrm{rel}}^0),n=1$ & 60\% & 83\% & 94\% & 59\% & 82\% & 93\% \\ \hline

SE + bsf + $\mathcal{O}(\alpha^2),\mathcal{O}(v_{\textrm{rel}}^0),n\leq 2$ & 65\% & 84\% & 93\% & 64\% & 83\% & 92\% \\ \hline
SE + bsf + $\mathcal{O}(\alpha^3),\mathcal{O}(v_{\textrm{rel}}^0),n\leq 2$ & 75\% & 91\% & 97\% & 74\% & 91\% & 97\% \\
\hline
\end{tabular}
\caption{\label{table} 
Effects of Sommerfeld enhancement, radiative corrections and/or velocity corrections on the present DM energy density 
relative to accounting for only the free annihilation of S-waves at LO. 
We consider DM masses $M=1$~TeV and 10~TeV, and couplings $\alpha=0.05,~0.1 ~\textrm{and}~0.2$, specified in the first row of each column as $(M, \alpha)$. 
The percentages denote the fractions of decrease in the relic density.
}
\end{center}
\end{table}

\begin{figure}[ht]
    \centering
\includegraphics[scale=0.79]{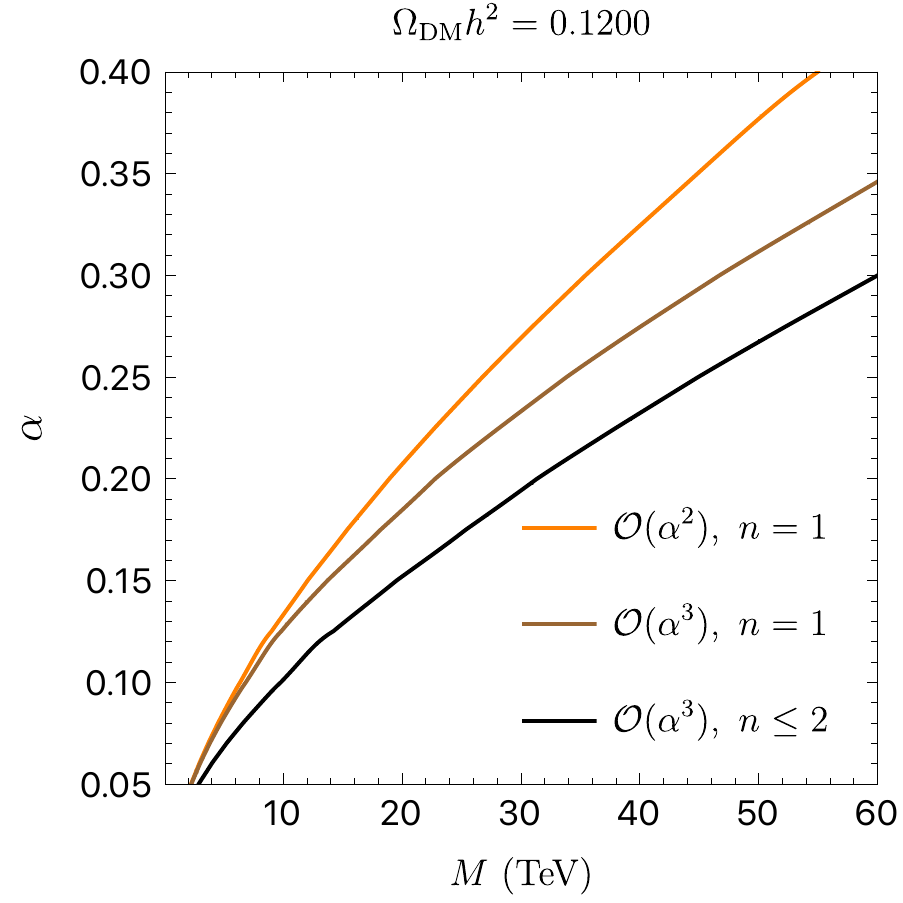}
\caption{Contours in the parameter space $(M,\alpha)$ that correspond to the observed DM energy density obtained from different approximations of the effective cross section (see text).  }
    \label{fig:yields_DM}
\end{figure}

In figure~\ref{fig:yields_DM}, we show the energy density contours for $\Omega_{\textrm{DM}}h^2=0.1200$ in the model parameter space $(M, \alpha)$.
The orange line corresponds to the inclusion of paradarkonium and the annihilation cross section at order $\alpha^2$,
whilst the brown line also includes effects at order $\alpha^3$.
The brown line is systematically below the orange curve because the additional annihilation channel
via orthodarkonium makes the effective cross section in eq.~\eqref{Cross_section_eff} larger, so that the same energy density is reproduced for smaller $\alpha$. 
The black line includes the effects of order $\alpha^3$ corrections as well as the excited states 2S and 2P$_{m=0,\pm 1}$ with transitions among them. 
This setting allows even smaller values of $\alpha$ to reproduce the observed energy density because of the additional channels for DM annihilations. 

\begin{figure}[ht]
    \centering
    \includegraphics[scale=0.79]{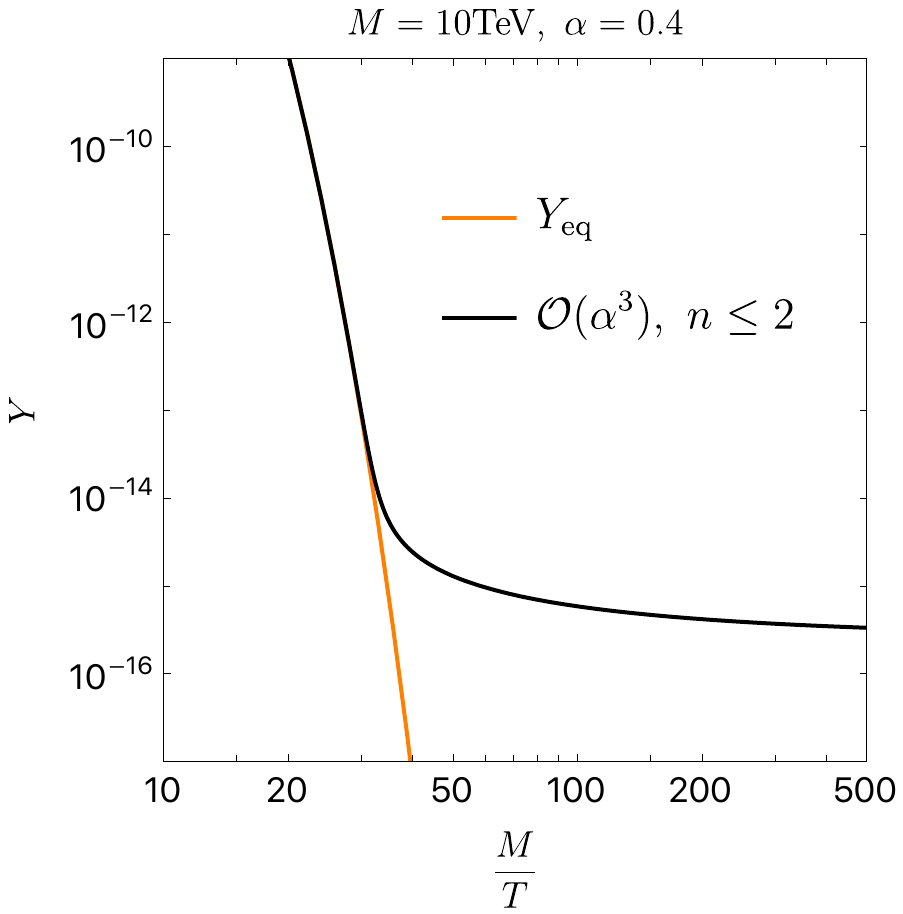}
    \caption{Dark matter yield for $\alpha=0.4$ and $M=10$~TeV obtained from the effective cross section at order $\mathcal{O}(\alpha^3)$ in the annihilation channels,
      and with $n \leq 2$ bound states and their electric transitions included (black line).
      For the whole temperature range after freeze-out it holds that $T \leq |E_1^b|$.
      The orange line is the equilibrium yield.
      }
    \label{fig:yield_alpha_04}
  \end{figure}

  In this section, we have solved the Boltzmann equation over a wide range of temperatures;  
  as mentioned in the introduction, the freeze-out temperature can be estimated to be about $M/25$. 
  The condition $T \lesssim  M\alpha^2$ in eq.~\eqref{scale_arrang}, or more conservatively $T \le |E_1^b|$,
  is satisfied for all times after freeze-out only if $\alpha \gtrsim 0.4$, see figure~\ref{fig:yield_alpha_04}.
  For smaller couplings, there exists a period of time after freeze-out when $T$ is larger than $M\alpha^2$.
  In general, when $T > M\alpha^2$, thermal effects modify the dark fermion-antifermion potential and therefore the wavefunction.
  In the model \eqref{lag_mod_0} with $\mathcal{L}_{\textrm{portal}} = 0$, it has been shown that these modifications are subleading
  with respect to the Coulomb potential as long as $T \ll M$~\cite{Escobedo:2008sy}.\footnote{
    In Coulomb gauge, the photon propagator is given by eqs.~\eqref{XXpropGT}.
      Temporal photon propagators do not depend on the temperature at leading order 
      and do not get screened by light fermion loops, which are absent, 
      but get thermal corrections starting from two insertions of the $F^{\mu \nu} \bm{{\rm{D}}}^2 F_{\mu \nu}/M^2$ operator appearing in the third line of eq.~\eqref{NREFT_lag},
      whose contribution is $T^4/M^4$ suppressed. 
      Spatial photons, on the other hand, couple to the heavy DM fermions and antifermions through $1/M$ suppressed couplings, cf. the first two lines of eq.~\eqref{NREFT_lag}.
      Hence thermal corrections do not modify the Coulomb potential~\eqref{V0Coul}, but only $1/M$ suppressed potentials,
      whose contribution to the wavefunction is subleading as long as $T \ll M$.
    }
  This is why we have ignored them here and computed widths and cross sections on Coulombic wavefunctions.
  Another important observation is that when $T$ approaches $M\alpha$, the multipole expansion for thermal dark photons breaks down (and one has to treat them at the level of NRQED$_{\textrm{DM}}$).
  However, for all the couplings considered in this section to compute the DM energy density,
  it holds that $T \ll M\alpha$ after freeze-out.

The numerical results for the DM energy density presented in this section are based on the single effective Boltzmann equation~\eqref{Boltzmann_eq_eff} or its modification
that includes transitions among bound states.
If instead we solve the coupled Boltzmann equations for the ground state and the unbound pairs numerically
either by taking $\textrm{Im}(d_s)$ at order $\alpha^2$ and $\textrm{Im}(d_v)=0$ or $\textrm{Im}(d_s)$ and $\textrm{Im}(d_v)$ at order $\alpha^3$,
in which case an additional Boltzmann equation due to orthodarkonium has to be added,
we get results that at most differ by 1\% from the ones presented here, and so are within the uncertainty of the measured relic density.\footnote{
 The coupled Boltzmann equations for the 1S bound-state number densities $n_{1\textrm{S}}^{\textrm{para}}$, $n_{1\textrm{S}}^{\textrm{ortho}}$
  and the sum of the dark matter particle and antiparticle number densities $n = n_X + n_{\bar{X}} = 2 n_X$ are~\cite{Gondolo:1990dk,vonHarling:2014kha}
\begin{equation}
\begin{aligned}
  (\partial_t + 3H) n &= -
  \frac{1}{2}\langle \sigma_{\textrm{ann}} \, v_{\textrm{rel}} \rangle (n^2-n^2_{\textrm{eq}})
  - \frac{1}{2}\langle \sigma^{1\textrm{S}}_{\hbox{\scriptsize bsf}} v_{\hbox{\scriptsize rel}} \rangle n^2
  + 2\Gamma^{1\textrm{S}}_{\textrm{bsd}} (n_{1\textrm{S}}^{\textrm{para}} + n_{1\textrm{S}}^{\textrm{ortho}}) \, ,\nonumber\\
  (\partial_t + 3H)n_{1\textrm{S}}^{\textrm{para}} &=
  -\Gamma^{\text{1S},\hbox{\scriptsize para}}_{\textrm{ann}}(n_{1\textrm{S}}^{\textrm{para}} - n_{1\textrm{S},\textrm{eq}}^{\textrm{para}})
  - \Gamma^{1\textrm{S}}_{\textrm{bsd}}n_{1\textrm{S}}^{\textrm{para}}
  + \frac{1}{16}\langle \sigma^{1\textrm{S}}_{\hbox{\scriptsize bsf}} v_{\hbox{\scriptsize rel}} \rangle n^2 \, ,\nonumber\\
(\partial_t + 3H)n_{1\textrm{S}}^{\textrm{ortho}} &= -
\Gamma^{1\textrm{S},\hbox{\scriptsize ortho}}_{\textrm{ann}}(n_{1\textrm{S}}^{\textrm{ortho}} - n_{1\textrm{S},\textrm{eq}}^{\textrm{ortho}})
-  \Gamma^{1\textrm{S}}_{\textrm{bsd}}n_{1\textrm{S}}^{\textrm{ortho}}
+ \frac{3}{16}\langle \sigma^{1\textrm{S}}_{\hbox{\scriptsize bsf}} v_{\hbox{\scriptsize rel}} \rangle n^2 \, ,\nonumber
\end{aligned}
\end{equation}
where $n_{1\textrm{S},\textrm{eq}}^{\textrm{ortho}} = 3n_{1\textrm{S},\textrm{eq}}^{\textrm{para}}$,
and $n_{1\textrm{S},\textrm{eq}}^{\textrm{para}} = (MT/\pi)^{3/2}e^{-E_1/T}$ and $n_\textrm{eq}/2 = n_{X,\textrm{eq}}=2(MT/2 \pi)^{3/2}e^{-M/T}$
(the factor of 2 for $n_{X,\textrm{eq}}$ is due to the spin polarization of the fermion).
The coupled equations reduce to equation \eqref{Boltzmann_eq_eff} with the effective cross section given by \eqref{Cross_section_eff} through the following steps and approximations.
{\it i)} From the {\it detailed balance} condition at equilibrium between bound-state formation and dissociation it follows that 
$$
\frac{1}{16} \langle \sigma^{1\textrm{S}}_{\hbox{\scriptsize bsf}} v_{\hbox{\scriptsize rel}} \rangle \, n^2_{\textrm{eq}} =
\Gamma^{1\textrm{S}}_{\textrm{bsd}} \, n_{1\textrm{S,eq}}^{\textrm{para}} \qquad \hbox{or} \qquad 
\frac{3}{16}\langle \sigma^{1\textrm{S}}_{\hbox{\scriptsize bsf}} v_{\hbox{\scriptsize rel}} \rangle \, n^2_{\textrm{eq}} =
\Gamma^{1\textrm{S}}_{\textrm{bsd}} \, n_{1\textrm{S,eq}}^{\textrm{ortho}}\,.
$$
The detailed balance condition is equivalent to requiring $(\partial_t + 3H) n_\textrm{eq} \approx 0$, i.e. that at equilibrium the number density changes only because of the universe expansion.
{\it ii)} By using the detailed balance condition in the right-hand sides of the equations for $n_{1\textrm{S}}^{\textrm{para}}$ and $n_{1\textrm{S}}^{\textrm{ortho}}$,
we may express them in terms of the annihilation and bound-state dissociation widths.
One may explicitly verify that because $H \ll \Gamma^{1\textrm{S},\hbox{\scriptsize para\,(ortho)}}_{\textrm{ann}},  \Gamma^{1\textrm{S}}_{\textrm{bsd}}$, 
the left-hand sides of these equations can be neglected, i.e. $(\partial_t + 3H)n_{1\textrm{S}}^{\textrm{para\,(ortho)}} \approx 0$.
The equations become then algebraic equalities that fix the 1S bound-state number densities as functions of $n$:
$$
\frac{n_{1\textrm{S}}^{\textrm{para\,(ortho)}}}{n_{1\textrm{S},\textrm{eq}}^{\textrm{para\,(ortho)}}} =
\frac{\Gamma^{1\textrm{S}}_{\textrm{bsd}}\, n^2/n^2_{\textrm{eq}} + \Gamma^{1\textrm{S},\hbox{\scriptsize para\,(ortho)}}_{\textrm{ann}}}
{\Gamma^{1\textrm{S}}_{\textrm{bsd}} + \Gamma^{1\textrm{S},\hbox{\scriptsize para\,(ortho)}}_{\textrm{ann}}}\,.
$$
{\it iii)} Finally, using in the differential equation for $n$ the detailed balance condition {\it i)}, the 1S bound-state number densities determined in {\it ii)},
and recalling that 
$\sigma^{1\textrm{S,para}}_{\hbox{\scriptsize bsf}} = \sigma^{1\textrm{S}}_{\hbox{\scriptsize bsf}}/4$
and $\sigma^{1\textrm{S,ortho}}_{\hbox{\scriptsize bsf}} = 3\,\sigma^{1\textrm{S}}_{\hbox{\scriptsize bsf}}/4$,
we end up with the single Boltzmann equation \eqref{Boltzmann_eq_eff} and the effective cross section \eqref{Cross_section_eff}.
\label{footnote_coupled_Boltzmann_eqs}
}

\section{Non-abelian SU($N$) model}
\label{sec:non_abelian_model}
In this section, we consider the DM fermion to be in the fundamental representation of a non-abelian SU($N$) gauge group, $N\ge 2$. 
We go through the same step-by-step construction as for the abelian model studied in the previous sections. 
We discuss pair annihilation in section~\ref{sec:non_abelian_pair_ann}, and bound-state formation and dissociation in section~\ref{sec:non_abelian_bsf_bsd}. 
Numerical applications for the DM energy density are presented in section~\ref{sec:non_abelian_energy_density}.

The model Lagrangian for a DM fermion coupled to SU($N$) gauge bosons reads
\begin{equation}
    \mathcal{L}=\bar{X} (i \slashed {D} -M) X -\frac{1}{4} G^a_{\mu \nu} G^{a\,\mu \nu} + \mathcal{L}_{\textrm{portal}} \, ,
    \label{non_ab_model}
\end{equation}
where $D_\mu=\partial_\mu+i g A_\mu^a T^a$, $T^a$ are the group generators, $A_\mu^a$ the SU($N$) dark gauge fields and $G^a_{\mu \nu}$ the corresponding field strength tensor. 
As in the abelian case, we set $\mathcal{L}_{\textrm{portal}}=0$.

We are interested in the energy regime $M \gg M \alpha \gtrsim \sqrt{MT} \gg M \alpha^2 \gtrsim T \gg \Lambda$,
where $\Lambda$ denotes the non perturbative scale where a weak-coupling expansion in $\alpha = g^2/(4\pi)$ breaks down.\footnote{
  In a non-abelian gauge theory, even in the absence of light fermions, a Debye mass, $m_D$, is generated by the self interaction of the gauge fields.
    In a weakly-coupled plasma, we take $m_D \sim gT \ll T$ and ignore its effects at leading order. 
   
}  
This choice of hierarchy for the energy scales allows us, like in the abelian case, to integrate out modes associated with the hard and soft scale
  by setting to zero the temperature characterizing the thermal distribution of the dark gauge fields.
  Moreover, computations at the hard, soft and ultrasoft energy scale may be done in weak-coupling perturbation theory.
  The ultimate EFT, made just of dark fermion-antifermion pairs of energy $M v_{\textrm{rel}}^2$ and dark gauge bosons of energy and momentum $M v_{\textrm{rel}}^2$,
  which encompasses the scales $M \alpha^2$ and $T$, or $\Lambda$, is a pNRQCD-like EFT~\cite{Pineda:1997bj,Brambilla:1999xf} for DM fields and an SU($N$) gauge group.
  We dub it pNREFT$_{\textrm{DM}}$.
The Lagrangian at order $r$ in the multipole expansion reads
\begin{equation}
    \begin{aligned}
    &\mathcal{L}_{\textrm{pNREFT}_{\textrm{DM}}} = \int d^3r\;\Bigg\{ \textrm{Tr}\left[\textrm{S}^\dagger \left( i \partial_0 -H_s\right) \textrm{S}
          + \textrm{O}^\dagger\left( i D_0 -H_o\right) \textrm{O} \right] \\
            &~~~~~~~~~~+\textrm{Tr}\left[V_A(r)g(\textrm{S}^\dagger \boldsymbol{r} \cdot \boldsymbol{E}\textrm{O} + \textrm{O}^\dagger \boldsymbol{r} \cdot \boldsymbol{E}\textrm{S})
              +\frac{V_B(r)}{2}g(\textrm{O}^\dagger \boldsymbol{r} \cdot \boldsymbol{E}\textrm{O} + \textrm{O}^\dagger \textrm{O}\boldsymbol{r} \cdot \boldsymbol{E})\right]\Bigg\}\\
            &~~~~~~~~~~~~~~~~~~ -\frac{1}{4} G^a_{\mu \nu} G^{a\, \mu \nu} \, .
    \label{su(N)_case}
    \end{aligned}
\end{equation}
The field $\textrm{S}=S\,\mathds{1}_{N\times N}/\sqrt{N}$ is an SU($N$) singlet field made of a dark fermion and antifermion 
and $\textrm{O}=O^a T^a/\sqrt{T_F}$ is the corresponding SU($N$) adjoint field; they depend on time, the relative coordinate $\bm{r}$ and the center of mass coordinate $\bm{R}$.
The Hamiltonians, $H_s$ and $H_o$, 
can be written as in~\eqref{ham_pNRQED} and~\eqref{pot_pNRQED} with the leading order potentials given respectively by
\begin{equation}
    V^{(0)}_{s} = -C_{F}\frac{\alpha}{r} \,, \qquad\qquad V^{(0)}_{o} = \frac{\alpha}{2 N r} \, .
\label{eq:SUNpot}
\end{equation}
The Casimir of the fundamental representation is $C_F=(N^2-1)/(2N)$ and the Dynkin index is $T_F=1/2$.
The adjoint field $\textrm{O}$ is a color octet field for $N=3$, which is the QCD case.

The electric dipole terms in the second line of eq.~\eqref{su(N)_case}, with $V_A(r)=V_B(r)=1$ at leading order,
allow for transitions between an unbound adjoint dark fermion-antifermion pair and a bound or unbound singlet pair.
Such transitions are responsible for bound-state formation and dissociation.
Moreover, they allow for scattering-state to scattering-state transitions among dark fermion-antifermion pairs in the SU($N$) adjoint representation.
At variance with the abelian case, SU($N$) singlet transitions, either involving bound states or scattering states, cannot happen in this model because of the SU($N$) charge conservation.

Another crucial difference with the abelian model \eqref{lag_mod_0} is in the running of the coupling.
For the non-abelian model \eqref{non_ab_model}, despite the fact that we do not consider additional light fermions or scalars,
the interactions among gauge bosons are enough to make the renormalized coupling run also at scales below $M$, 
the first coefficient of the beta function being $\beta_0=11 N/3$.
Therefore, the coupling constant assumes different values when computing processes occurring at the hard, soft and ultrasoft scale.
We typically renormalize $\alpha$ at the scale $2M$ in annihilation processes, at a soft scale of order $M\alpha$ in the wavefunction\footnote{
The coupling $\alpha$ in the potentials~\eqref{eq:SUNpot} typically runs at a scale of order $1/r$.
}  
and at a scale of order $M \alpha^2$ or $T$ when considering gauge bosons at the ultrasoft scale.
A further scale, $|\bm{p}|=M v_{\textrm{rel}}/2$, can be associated to the coupling of gauge bosons with non-relativistic scattering states. 
Because of asymptotic freedom: $\alpha(2M) \ll \alpha(M\alpha), \alpha(M v_{\textrm{rel}}) \ll \alpha(M\alpha^2)$.

\subsection{Pair annihilation}
\label{sec:non_abelian_pair_ann}
In the non-abelian theory, there are two more four-fermion dimension six operators contributing to fermion-antifermion annihilation 
than in the abelian case due to the SU($N$) adjoint configurations. 
The contribution of the four-fermion dimension six operators to the part of the pNREFT$_{\textrm{DM}}$ Lagrangian relevant for annihilation is~\cite{Brambilla:2002nu}
\begin{equation}
    \begin{aligned}
    &\delta \mathcal{L}^{\textrm{ann}}_{\textrm{pNREFT}_{\textrm{DM}}} =\\
    &~~ \frac{i}{M^2} \, \int d^3r \, \textrm{Tr}\Bigg\{\textrm{S}^\dagger \delta^3(\boldsymbol{r}) \, N
  \left[ {2\rm{Im}}(f_{[\bm{1}]}(^1 S_0)) - \boldsymbol{S}^2 \left( {\rm{Im}}(f_{[\bm{1}]}(^1 S_0))- {\rm{Im}}(f_{[\bm{1}]}(^3 S_1)) \right) \right] \textrm{S} \\
  &~~~~~~~~~+\textrm{O}^\dagger \delta^3(\boldsymbol{r}) \, T_F
  \left[ {2\rm{Im}}(f_{[\textbf{adj}]}(^1 S_0)) - \boldsymbol{S}^2 \left( {\rm{Im}}(f_{[\textbf{adj}]}(^1 S_0))- {\rm{Im}}(f_{[\textbf{adj}]}(^3 S_1)) \right) \right]\textrm{O}\Bigg\} \,.
    \end{aligned}
\end{equation}
At order $\alpha^3$, the imaginary parts of the singlet matching coefficients read~\cite{Bodwin:1994jh,Petrelli:1997ge}:\footnote{
  The matching coefficients $\rm{Im}(f_{[\bm{1}]}(^3 S_1))$ reported in~\cite{Bodwin:1994jh} and ~\cite{Petrelli:1997ge} agree for $N=3$,
  however they do not for $N\neq 3$, as the factor $1/54$ in~\cite{Bodwin:1994jh} reads $1/(18 N)$ 
in~\cite{Petrelli:1997ge}. Equation~\eqref{Imf13S1} is the one that can be found in~\cite{Petrelli:1997ge}.}
\begin{align}
  \rm{Im}(f_{[\bm{1}]}(^1 S_0)) &= \frac{N^2-1}{4N^2}\pi\, \alpha(2M)^2
  \left\{1 + \frac{\alpha}{\pi}\left[ \frac{1}{2N}\left(5-\frac{\pi^2}{4}\right) + N \left( \frac{77}{9} - \frac{5}{12}\pi^2\right)\right] \right\}\,,
  \label{Imf11S0}
  \\
  \rm{Im}(f_{[\bm{1}]}(^3 S_1)) &= \frac{2}{9}(\pi^2-9)\frac{(N^2-1)(N^2-4)}{8N^3}\alpha^3\,,
  \label{Imf13S1}
\end{align}
and the imaginary parts of the SU($N$) adjoint matching coefficients are~\cite{Bodwin:1994jh,Petrelli:1997ge,Maltoni1999}:\footnote{
  The matching coefficient $\rm{Im}(f_{[\textbf{adj}]}(^3 S_1))$ vanishes at order $\alpha^2$ because the gauge bosons in \eqref{non_ab_model} do not couple to light fermions.
  At order $\alpha^3$ the expression in \eqref{Imf83S1} is the one reported in ref.~\cite{Maltoni1999}, while the expression in ref.~\cite{Petrelli:1997ge} is three times larger.
}
\begin{align}
  \rm{Im}(f_{[\textbf{adj}]}(^1 S_0)) &= \frac{N^2-4}{4N}\pi\,\alpha(2M)^2 \left\{1 + \frac{\alpha}{\pi}\left[ \frac{1}{2N}\left(5-\frac{\pi^2}{4}\right) + N \left( \frac{199}{18} - \frac{7}{12}\pi^2\right)\right] \right\} \,,
  \label{Imf81S0}\\
  \rm{Im}(f_{[\textbf{adj}]}(^3 S_1)) &= \frac{5}{6} \left(-\frac{73}{4} + \frac{67}{36}\pi^2\right) \alpha^3\,.
  \label{Imf83S1}
\end{align}  

From eqs.~\eqref{Imf11S0} and~\eqref{Imf13S1}, and the SU($N$) equivalent of eq.~\eqref{ann_fact_scat}, it follows that the Sommerfeld-enhanced SU($N$)-singlet annihilation cross section reads 
\begin{equation}
  (\sigma^{\hbox{\tiny NR}}_{\hbox{\scriptsize ann}} v_{\hbox{\scriptsize rel}})^{[\mathds{1}]}_{\hbox{\tiny NLO$^*$}}(\boldsymbol{p}) = 
  (\sigma^{\hbox{\tiny NR}}_{\hbox{\scriptsize ann}} v_{\hbox{\scriptsize rel}})^{[\mathds{1}]}_{\hbox{\tiny NLO}} \, |\Psi^{[\mathds{1}]}_{\bm{p}}(\boldsymbol{0})|^2 \, ,
    \label{singlet_non_abelian_xsection_Som}
\end{equation}
with the free annihilation cross section at NLO given by 
\begin{equation}
  (\sigma^{\hbox{\tiny NR}}_{\hbox{\scriptsize ann}} v_{\hbox{\scriptsize rel}})^{[\mathds{1}]}_{\hbox{\tiny NLO}} =
  \frac{C_F}{2}\frac{\pi \alpha(2M)^2}{M^2} \left[ 1 + \frac{\alpha}{\pi} \left( \frac{400N^2+1044-3\pi^2(2N^2+35)}{72N}  
  \right) \right] \, .
    \label{singlet_non_abelian_xsection}
\end{equation}
The Sommerfeld factor for the singlet wavefunction at leading order is\footnote{
  The SU($N$)-singlet scattering (bound-state) wavefunction in the non-abelian model can be inferred from the abelian version~\eqref{coulomb_wave0} (\eqref{boundwave})
  by replacing $\alpha \rightarrow C_F \alpha$, while the SU($N$)-adjoint wavefunction follows from replacing $\alpha \rightarrow -\alpha/(2N)$ in~\eqref{coulomb_wave0}.}
\begin{equation}
   |\Psi^{[\mathds{1}]}_{\bm{p}}(\boldsymbol{0})|^2 = \frac{2 \pi C_F\zeta}{1-e^{-2 \pi C_F\zeta}} \, , \qquad\qquad \zeta = \frac{\alpha}{v_{\hbox{\scriptsize rel}}} \, .
   \label{sommerfeld_color_singlet}
\end{equation}
The subscript NLO$^*$ reminds that our expression for the cross section is complete at NLO only in the hard part,
the one encoded in the four-fermion matching coefficients, while it lacks NLO corrections in the wavefunction, i.e. in the Sommerfeld factor.

Similarly, for the annihilation cross section of a pair in the adjoint representation we find 
\begin{equation}
  (\sigma^{\hbox{\tiny NR}}_{\hbox{\scriptsize ann}} v_{\hbox{\scriptsize rel}})^{[\textbf{adj}]}_{\hbox{\tiny NLO$^*$}}(\boldsymbol{p}) =
  (\sigma^{\hbox{\tiny NR}}_{\hbox{\scriptsize ann}} v_{\hbox{\scriptsize rel}})^{[\textbf{adj}]}_{\hbox{\tiny NLO}} \,  |\Psi^{[\textbf{adj}]}_{\bm{p}}(\boldsymbol{0})|^2 \, ,
\label{octet_non_abelian_xsection_Som}
\end{equation}
where
\begin{equation}
\begin{aligned}
   &  (\sigma^{\hbox{\tiny NR}}_{\hbox{\scriptsize ann}} v_{\hbox{\scriptsize rel}})^{[\textbf{adj}]}_{\hbox{\tiny NLO}} = \frac{N^2-4}{8 N}\frac{\pi\alpha(2M)^2}{M^2}
    \\
    &~~~~~~\times \left[ 1 + \frac{\alpha}{\pi} \left( \frac{796 N^4-16144 N^2 -720 +\pi^2 (-42 N^4+1499 N^2+36)}{72 N(N^2-4)} 
    \right)\right] \, ,
\label{octet_non_abelian_xsection}
\end{aligned}
\end{equation}
and we have averaged over the $N^2-1$ configurations of the incoming dark fermion-antifermion pair in the adjoint representation of SU($N$).
The Sommerfeld factor is
\begin{equation}
    |\Psi^{[\textbf{adj}]}_{\bm{p}}(\boldsymbol{0})|^2 = \frac{\pi \zeta/N}{e^{\pi \zeta/N}-1} \, ,
    \label{sommerfeld_adjoint}
\end{equation}
with $\zeta$ defined as in \eqref{sommerfeld_color_singlet}.
The total cross section reads
\begin{equation}
  (\sigma^{\hbox{\tiny NR}}_{\hbox{\scriptsize ann}} v_{\hbox{\scriptsize rel}})_{\hbox{\tiny NLO$^*$}}(\boldsymbol{p}) =
  \frac{ (\sigma^{\hbox{\tiny NR}}_{\hbox{\scriptsize ann}} v_{\hbox{\scriptsize rel}})^{[\mathds{1}]}_{\hbox{\tiny NLO$^*$}}(\boldsymbol{p})
    + (N^2-1) (\sigma^{\hbox{\tiny NR}}_{\hbox{\scriptsize ann}} v_{\hbox{\scriptsize rel}})^{[\textbf{adj}]}_{\hbox{\tiny NLO$^*$}}(\boldsymbol{p})}{N^2}\,.
    \label{total_non_abelian_ann_colored_av}
\end{equation}
In contrast to the abelian case, the NLO corrections increase the annihilation cross section both for SU($N$) singlet and adjoint pairs.
However, the Sommerfeld factor has a different impact on the cross sections. 
For the attractive singlet channel, the Sommerfeld factor \eqref{sommerfeld_color_singlet} is larger than one and increases the cross section,
whereas for the adjoint repulsive channel, the Sommerfeld factor \eqref{sommerfeld_adjoint} is smaller than one and consequently decreases the cross section.
At $\mathcal{O}(\alpha^2)$ our results in eqs.~\eqref{singlet_non_abelian_xsection} and \eqref{octet_non_abelian_xsection} agree with those in ref.~\cite{ElHedri:2016onc},
once the cross sections in ref.~\cite{ElHedri:2016onc} are expanded in $v_\textrm{rel}$, and only annihilations into gauge bosons are taken into account (we have no light fermions in the model).

As mentioned above, the expressions~\eqref{sommerfeld_color_singlet} and~\eqref{sommerfeld_adjoint} for the Sommerfeld factors do not include relativistic corrections to the wavefunctions.
It is important to realize that in a non-abelian theory these corrections may eventually turn out to be more important than the NLO corrections to the free annihilation cross section,
as the first ones are governed by the coupling at the soft scale, whereas the latter ones are governed by the coupling at the hard scale.

Bound states are sustained only by the SU($N$)-singlet configuration.
The annihilation width of a spin- and SU($N$)-singlet pair in an $n\textrm{S}$ wave reads at NLO in $\rm{Im}(f_{[\bm{1}]}(^1 S_0))$
\begin{equation}
  \Gamma^{n\textrm{S},\textrm{para}}_{\hbox{\scriptsize ann} } =
  C^4_F\frac{M \alpha(\mu_{\textrm{s}})^3 \alpha(2M)^2}{4n^3} 
  \left\{ 1 + \frac{\alpha}{\pi}
    \left[ \frac{1}{2N}\left(5-\frac{\pi^2}{4}\right) + N \left( \frac{77}{9} - \frac{5}{12}\pi^2\right)
\right] \right\},
\label{widthparaSUN}
\end{equation}
where at leading order we have distinguished between the coupling coming from the four-fermion matching coefficient, which is renormalized at the scale $2M$, 
and the coupling coming from the wavefunction, which is renormalized at a scale $\mu_{\textrm{s}}$ of the order of the soft scale.
At next-to-leading order the natural scale of $\alpha$ appearing in \eqref{widthparaSUN} is $2M$, since it originates from $\rm{Im}(f_{[\bm{1}]}(^1 S_0))$.
The decay width \eqref{widthparaSUN} does not include relativistic corrections to the wavefunction, which, as we have remarked,
may be as much important as or more important than the included NLO corrections to the four-fermion matching coefficient.\footnote{
A simple analytic expression that approximates the NLO correction to the wavefunction can be found in ref.~\cite{Titard:1993nn}.
Including the wavefunction correction of~\cite{Titard:1993nn} modifies eq.~\eqref{widthparaSUN} into 
$$
  \Gamma^{n\textrm{S},\textrm{para}}_{\hbox{\scriptsize ann} } \approx
  C^4_F\frac{M \alpha(\mu_{\textrm{s}})^3 \alpha(2M)^2}{4n^3} 
  \left\{ 1 + \frac{\alpha(2M)}{\pi}
    \left[ \frac{1}{2N}\left(5-\frac{\pi^2}{4}\right) + N \left( \frac{77}{9} - \frac{5}{12}\pi^2\right)\right]
    + \frac{\alpha(\mu_{\textrm{s}})}{\pi}\, \frac{31}{12}\,N
 \right\},
$$
where we have indicated the natural scale of the coupling at NLO, in this way keeping distinguished NLO corrections coming from $\rm{Im}(f_{[\bm{1}]}(^1 S_0))$ and the wavefunction.
}
At leading order, the decay width of a spin-triplet SU($N$)-singlet pair in an $n\textrm{S}$ wave reads
\begin{equation}
  \Gamma^{n\textrm{S},\textrm{ortho}}_{\hbox{\scriptsize ann}} = \frac{\pi^2 -9}{36\pi n^3}C_F^4\frac{N^2-4}{N} M  \alpha(\mu_{\textrm{s}})^3 \alpha(2M)^3 \, .
\label{widthorthoSUN}
\end{equation}

\subsection{Bound-state formation and dissociation}
\label{sec:non_abelian_bsf_bsd}
In this section, we start by considering the formation of bound states out of unbound adjoint states. 
The bsf cross section may be computed analogously to the abelian case, resulting in
\begin{equation}
  (\sigma_{\hbox{\scriptsize bsf}} v_{\hbox{\scriptsize rel}})^{[\textbf{adj}]}(\boldsymbol{p}) = \frac{2}{3N}\,\alpha(\mu_{\textrm{us}})
  \sum \limits_{n,\ell,m} \left[ 1 + n_{\text{B}}(\Delta E_{n}^{p}) \right] \left|\langle n\ell m | \boldsymbol{r} | \boldsymbol{p} \rangle^{[\textbf{adj}]}\right|^2
  (\Delta E_{n}^{p})^3 \, ,  
  \label{bsf_adjoint_to_bound}
\end{equation}
where $|n\ell m\rangle$ is the eigenstate of the Hamiltonian $H_s$ describing a bound state with quantum numbers $n$, $\ell$ and $m$, 
$|\boldsymbol{p} \rangle^{[\textbf{adj}]}$ is the eigenstate of the Hamiltonian $H_o$ labeled by the relative momentum $\bm{p} = M\bm{v}_{\textrm{rel}}/2$
of the unbound dark fermion-antifermion pair, 
and we have averaged over the $N^2-1$ configurations of the incoming dark fermion-antifermion pair in the adjoint representation of SU($N$).
Although we work at LO in $\alpha$, we have made explicit in eq.~\eqref{bsf_adjoint_to_bound} that the natural renormalization scale of the electric-dipole coupling in such a process is a scale $\mu_{\textrm{us}}$
of the order of the ultrasoft scale. 
The formation of a bound state out of an unbound singlet state by emission of a gauge boson is not possible in a non-abelian theory due to the SU($N$) charge conservation, i.e. $(\sigma_{\hbox{\scriptsize bsf}} v_{\hbox{\scriptsize rel}})^{[\mathds{1}]}(\boldsymbol{p}) = 0$.
The total bound-state formation cross section, defined similarly to eq.~\eqref{total_non_abelian_ann_colored_av}, is then just $(N^2-1)/N^2$ times 
$(\sigma_{\hbox{\scriptsize bsf}} v_{\hbox{\scriptsize rel}})^{[\textbf{adj}]}(\boldsymbol{p})$.
It is the total bound-state formation cross section that can be found in the literature~\cite{Harz:2018csl,Garny:2021qsr}.

For the reverse process, namely bound-state dissociation, the thermal width is given by
\begin{equation}
    \Gamma^{n\ell m}_{\textrm{bsd}} =  2(N^2-1)\int _{|\boldsymbol{k}|\geq |E^b_{n}|} \frac{d^{3}k}{(2\pi)^{3}}\,n_{B}(|\boldsymbol{k}|)\,\sigma^{n\ell m}_{\textrm{ion}}(\boldsymbol{k}) \, ,
\end{equation}
where $E^b_n = -M (C_F \alpha)^2/(4n^2)$ is the binding energy and the in-vacuum ionization cross section is 
\begin{equation}
  \sigma^{n\ell m}_{\textrm{ion}}(\boldsymbol{k}) =
\frac{1}{3N}
\,\alpha(\mu_{\textrm{us}})\,
  \frac{M^{\frac{3}{2}}}{2}|\boldsymbol{k}|\sqrt{|\boldsymbol{k}|+E^b_{n}}\,
  \left|\langle n\ell m|\boldsymbol{r}|\boldsymbol{p}\rangle^{[\textbf{adj}]}\right|^{2}\bigg \vert_{|\boldsymbol{p}| = \sqrt{M(|\boldsymbol{k}|+E^b_{n})}} \, .
\end{equation}
In the ionization cross section we have averaged over the $N^2-1$ degrees of freedom of the incoming dark gauge field.

As for the abelian model, the remaining computational effort is in the electric dipole matrix element. 
The general result is collected in appendix~\ref{sec:app_D}.
In the following, we specify the expression of the electric dipole matrix element for the ground state, 
and use it to compute the formation of the ground state 
and its dissociation into an unbound SU($N$) adjoint pair.
From the general expression \eqref{matrixelement_su(N)}, the matrix element squared for the ground state reads
\begin{equation}
\begin{aligned}
  |\langle 1  \textrm{S}|\bm{r}|\bm{p}\rangle^{[\textbf{adj}]}|^{2} =
  \frac{\pi^2}{M^5}\frac{2^{11}C_F^3\alpha^6}{N v_{\textrm{rel}}^{11}}\frac{(2C_F+N)^2}{\left[1+\left(\frac{C_F\alpha}{v_{\textrm{rel}}}\right)^2\right]^6}
  \left[1+\left(\frac{1}{2N}\frac{\alpha}{v_{\textrm{rel}}}\right)^2\right]\frac{e^{\frac{2\alpha}{Nv_{\textrm{rel}}}
        \arccot\left(\frac{C_F\alpha}{v_{\textrm{rel}}}\right)}}{e^{\frac{\pi}{N}\frac{\alpha}{v_{\textrm{rel}}}}-1} \, ,
\end{aligned}
\end{equation}
which is in agreement with the result in ref.~\cite{Brambilla:2011sg}. 

The total bsf cross section for the ground state is
\begin{equation}
\begin{aligned}
  &(\sigma^{1\textrm{S}}_{\hbox{\scriptsize bsf}} v_{\hbox{\scriptsize rel}})(\boldsymbol{p}) = \frac{4C_F}{3N^2}\,\alpha(\mu_{\textrm{us}})
  \left[ 1 + n_{\text{B}}(\Delta E_{1}^{p}) \right] \left|\langle 1\textrm{S} | \boldsymbol{r} | \boldsymbol{p} \rangle^{[\textbf{adj}]}\right|^2
  (\Delta E_{1}^{p})^3 \\
  &= \alpha(\mu_{\textrm{us}})\frac{2^7 \pi^2 \alpha^6}{3 \, M^2 \, v_{\hbox{\scriptsize rel}}^5}\frac{C_F^4}{N^3}\left(2C_F+N\right)^2
  \frac{1+\left(\frac{\alpha}{2Nv_{\textrm{rel}}}\right)^2}{\left[ 1+ \left(\frac{C_F\alpha}{v_{\textrm{rel}}}\right)^2\right]^3}
      \frac{e^{\frac{2\alpha}{Nv_{\textrm{rel}}} \hbox{\scriptsize arccot} \frac{C_F\alpha}{v_{\textrm{rel}}} }}{e^{\frac{\pi}{N} \frac{\alpha}{v_{\textrm{rel}}}}-1}
      \, \left[ 1 + n_{\text{B}}(\Delta E_{1}^{p}) \right] ,
  \label{bsf_adjoint_to_bound_1S}
  \end{aligned}
\end{equation}
with $p=M v_{\hbox{\scriptsize rel}}/2$ the incoming relative momentum of the adjoint pair, and 
\begin{equation}
  \Delta E_{1}^{p}
  =\frac{Mv_{\textrm{rel}}^2}{4} \left[ 1+ \left(\frac{C_F \alpha}{v_{\textrm{rel}}}\right)^2\right]\,.
\end{equation} 
Our result agrees with the outcome of ref.~\cite{Harz:2018csl}.\footnote{
In ref.~\cite{Harz:2018csl}, the authors distinguish between the coupling coming from the bound-state wavefunction, renormalized at a soft scale of order $M\alpha$, and 
the coupling coming from the scattering-state wavefunction, renormalized at a soft scale of order $M v_{\hbox{\scriptsize rel}}$. 
Although it may be useful to keep track of the two different couplings,
the distinction between them is a higher-order effect that goes beyond the accuracy of the bsf and ionization cross section formulas that we use in this work. 
In those formulas, we only keep track of the ultrasoft scale in the electric dipole coupling.}

\begin{figure}[ht]
    \centering
    \includegraphics[scale=0.75]{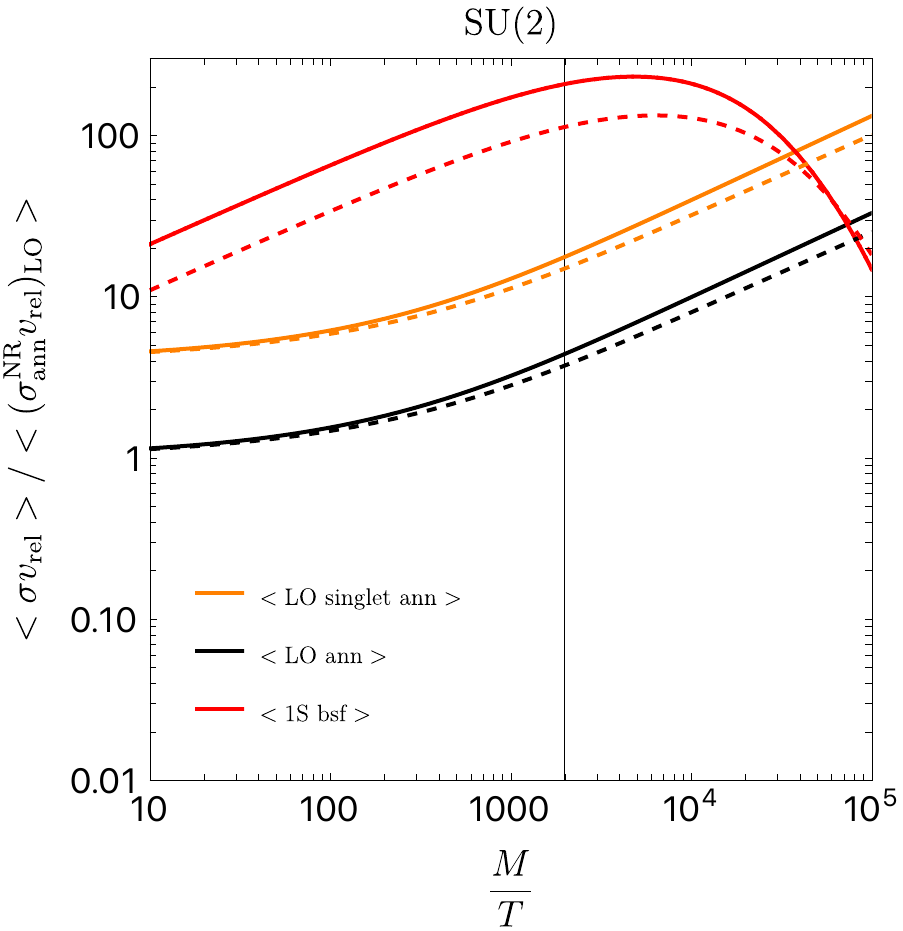}
    \hspace{0.5 cm}
     \includegraphics[scale=0.75]{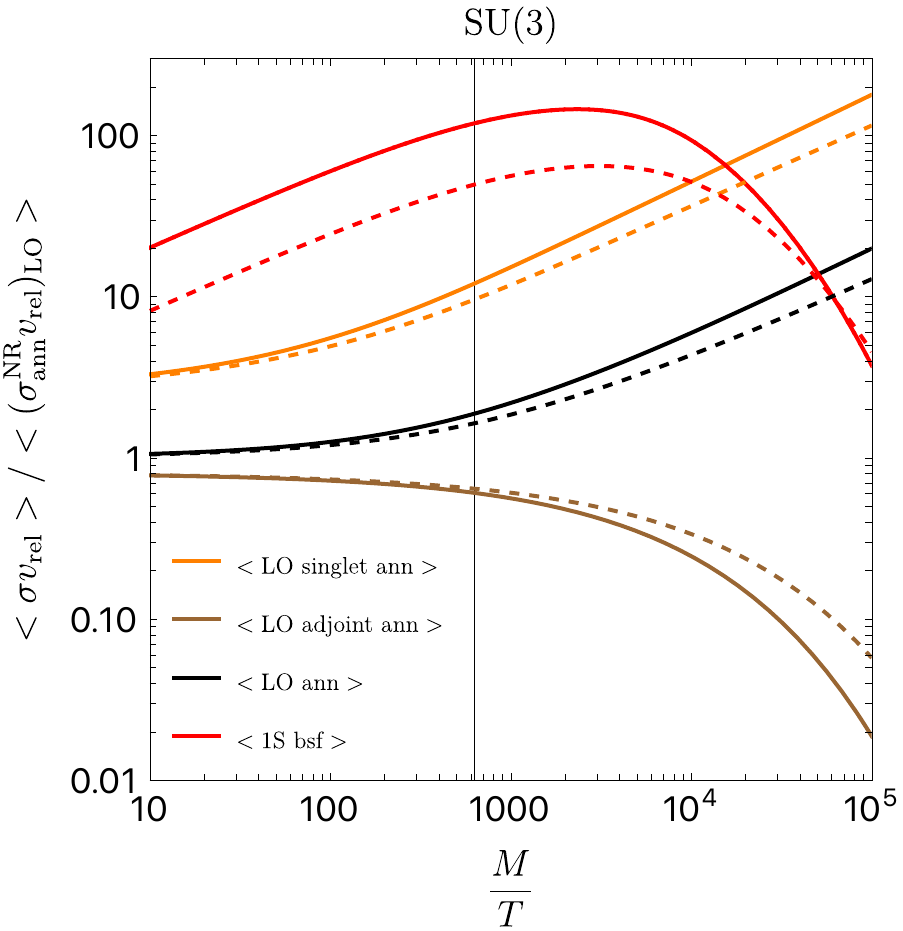}
    \caption{Ratios of the thermally averaged cross sections in the non-abelian model SU(2) (left) and SU(3) (right). 
    The orange lines denote the ratio when taking the thermal average of~\eqref{singlet_non_abelian_xsection_Som} but at LO in the free annihilation cross section,
    the brown lines when taking the thermal average of~\eqref{octet_non_abelian_xsection_Som} but at LO in the free annihilation cross section,
    the black lines when taking the thermal average of~\eqref{total_non_abelian_ann_colored_av} but at LO in the free annihilation cross section,
    and the red lines when taking the thermal average of~\eqref{bsf_adjoint_to_bound_1S}. 
    We take the ratio with respect to the thermally averaged free annihilation cross section~\eqref{LOxsection_nonabelian}.
    Solid lines depict the ratios in the case of running coupling $\alpha$, starting from $\alpha(2M)=0.03$,
    dashed lines for constant $\alpha = 0.03$. 
    The vertical lines mark $T=MC_F^2\alpha^2$.
      }
    \label{fig:plotannBSFSU2}
  \end{figure}

Similarly to the abelian case, we thermally average the bsf cross section over the incoming momenta of the adjoint pair following the Maxwell--Boltzmann distribution~\eqref{MBdistr}. 
In figure~\ref{fig:plotannBSFSU2}, we plot the thermally averaged annihilation and bsf cross section, normalized by the thermally averaged free annihilation cross section at LO of an unbound pair,
\begin{equation}
 \langle (\sigma^{\hbox{\tiny NR}}_{\hbox{\scriptsize ann}} v_{\hbox{\scriptsize rel}})_{\hbox{\tiny LO}} \rangle =
  \frac{1}{N^2} \, \left\langle (\sigma^{\hbox{\tiny NR}}_{\hbox{\scriptsize ann}} v_{\hbox{\scriptsize rel}})^{[\mathds{1}]}_{\hbox{\tiny LO}}
    + (N^2-1) (\sigma^{\hbox{\tiny NR}}_{\hbox{\scriptsize ann}} v_{\hbox{\scriptsize rel}})^{[\textbf{adj}]}_{\hbox{\tiny LO}}\right\rangle
    = \frac{C_F}{4N^2}(N^2-2)
    \frac{\pi \alpha(2M)^2}{M^2} 
    \, ,
\label{LOxsection_nonabelian}
\end{equation}
as a function of $M/T$. 
We show the behaviour of the cross sections for a constant (dashed lines) and running coupling (solid lines) in an SU(2) and SU(3) dark matter model,
respectively in the left and right panel of figure~\ref{fig:plotannBSFSU2}. 
In the case of a constant coupling, we take as a benchmark value $\alpha=0.03$. 
For the running coupling, the same value is taken at the hard annihilation scale, $\alpha(2M)=0.03$, and run down to the smaller soft and ultrasoft scales,\footnote{
Through this section, we evolve $\alpha$ at one loop.}
where one finds $\alpha(2M)<\alpha(\mu_{\textrm{s}})<\alpha(\mu_{\textrm{us}})$ for typical non-relativistic velocities. 
We take $\mu_{\textrm{s}} = M v_{\hbox{\scriptsize rel}}/2$ and  $\mu_{\textrm{us}} = M v_{\hbox{\scriptsize rel}}^2/4$.
This gives more prominent near-threshold effects, as one may see in the annihilations of the singlet pair (orange dashed lines below the solid lines) and in the total annihilation cross section (black dashed lines below the solid lines). 
Conversely, for the adjoint SU(3) unbound pairs, the running coupling implies a stronger repulsion (brown solid curve below the dashed curve). 
We remark that the annihilation cross section of the pair in the adjoint representation vanishes for the SU(2) model at LO:
$(\sigma^{\hbox{\tiny NR}}_{\hbox{\scriptsize ann}} v_{\hbox{\scriptsize rel}})^{[\textbf{adj}],\text{SU(2)}}_{\hbox{\tiny LO}}=0$.   
Comparing with the abelian case in figure~\ref{fig:sigma_bsf_01_TA}, we see that the annihilation processes for the pair in a color singlet show a similar behaviour,
namely a Sommerfeld enhancement that increases for smaller temperatures. 
However, in the SU(3) case, the contribution for the adjoint pair annihilations is suppressed by a Sommerfeld factor smaller than unity. 
This is due to the repulsive potential experienced by the adjoint pair, which becomes more relevant for smaller $T$, and thus lower velocities. 
The effect of a repulsive potential appears also clearly in the bsf process (red solid and dashed curves). 
At variance with the abelian case, the rising of the bsf cross section is saturated by the repulsive potential at small temperatures, and the bsf process becomes progressively less likely. 
This is signaled by the fact that for small $v_{\hbox{\scriptsize rel}}$, whereas the right-hand side of eq.~\eqref{our_gamma_1p} goes like $1/v_{\hbox{\scriptsize rel}}$,
the right-hand side of eq.~\eqref{bsf_adjoint_to_bound_1S} is exponentially suppressed.
The running coupling enhances the bsf cross section with respect to a frozen coupling for the range of $M/T$ shown in the plot. 

\begin{figure}[ht]
    \centering
    \includegraphics[scale=0.76]{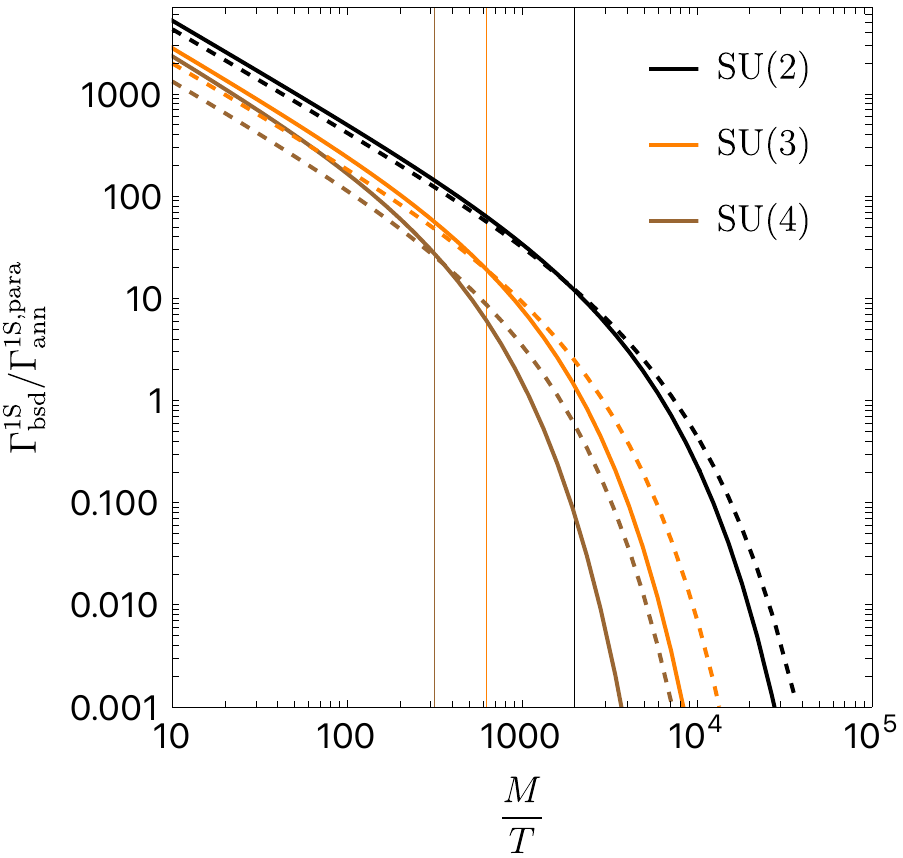}
        \hspace{0.5 cm}
     \includegraphics[scale=0.73]{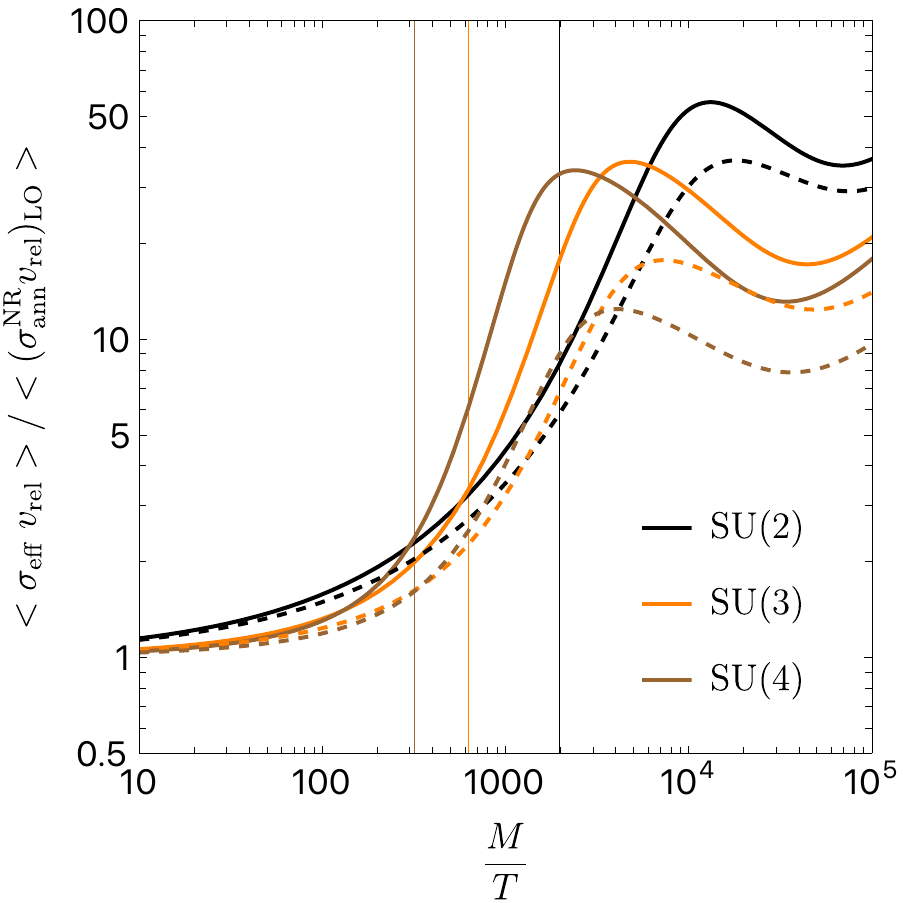}
    \caption{(Left) Thermal dissociation widths over the 1S paradarkonium decay width at LO in the SU(2) (black lines), SU(3) (orange lines) and SU(4) (brown lines) theory.
    Solid lines depict the ratios in the case of a running coupling $\alpha$, dashed lines in the case of a constant $\alpha =0.03$. 
    (Right) Thermally averaged effective cross sections over~\eqref{LOxsection_nonabelian} in the SU(2), SU(3) and SU(4) theory. 
    Dashed lines are for a constant $\alpha$, solid lines for a running $\alpha$. 
    In both plots, the vertical lines, from left to right, mark the position where $T=MC_F^2\alpha^2$ in the SU(4), SU(3) and SU(2) theory, respectively.
      }
    \label{fig:plotBSD}
  \end{figure}

The dissociation width of an SU($N$)-singlet ground state into an unbound adjoint state is
\begin{equation}
  \Gamma^{1 \textrm{S}}_{\textrm{bsd}}  = 2(N^2-1) \int_{|\bm{k}|\geq |E^b_{1}|} \frac{d^3k}{(2\pi)^3} \,  n_{\text{B}}(|\bm{k}|)\, \sigma^{1 \textrm{S}}_{\hbox{\scriptsize ion}}(|\bm{k}|)\,,
\label{nonabelian_bsd_width_groundstate}
\end{equation}
where the ionization cross section, averaged over the incoming SU($N$) gauge fields, is
\begin{equation}
  \sigma^{1 \textrm{S}}_{\hbox{\scriptsize ion}}(|\bm{k}|)= \alpha(\mu_{\textrm{us}})\frac{ 2^{3} \pi^2}{3} \frac{|E_1^b|^4}{M  |\bm{k}|^5}
  \frac{(2C_F+N)^2}{N^4 C_F^5}
  \left[1+(2NC_F w_1(|\bm{k}|))^2\right] 
  \frac{e^{\frac{2\arctan(w_1(|\bm{k}|))}{NC_F w_1(|\bm{k}|)}}}{e^{\frac{\pi}{NC_F w_1(|\bm{k}|)}}-1} \, ,
\label{nonabelian_bsd_xsection_groundstate}
\end{equation}
with $w_1(|\bm{k}|) \equiv \sqrt{|\bm{k}|/|E_1^b|-1}$. 
The expression is in agreement with the result of ref.~\cite{Brambilla:2011sg}. 
The left panel of figure~\ref{fig:plotBSD} shows the behaviour of the dissociation width, normalized to the paradarkonium decay width at LO for the ground state,
$\Gamma^{1\textrm{S},\textrm{para}}_{\hbox{\scriptsize ann} } = C^4_F M \alpha(\mu_{\textrm{s}})^3 \alpha(2M)^2/4$ with $\mu_{\textrm{s}}=M\alpha/2$ and $\mu_{\textrm{us}}=M\alpha^2/4$, 
as a function of $M/T$ for different non-abelian models, with and without the running of the coupling. 

\subsection{Energy density}
\label{sec:non_abelian_energy_density}
At last we solve the coupled Boltzmann equations.
Differently from the abelian case, the number density of scattering states comprises now the unbound pairs in both the SU($N$) singlet and adjoint configurations. 
We consider the simple case where we include the ground state only, see footnote~\ref{footnote_coupled_Boltzmann_eqs}.
The coupled equations can be traded with a single effective Boltzmann equation in the form of eq.~\eqref{Boltzmann_eq_eff} for unbound pairs.
Since bound-state to bound-state transitions are zero in the SU($N$) models under study, the thermally averaged effective cross section~\eqref{Cross_section_eff} does not get modified by them. 
We use for the annihilation cross section the thermal average of eq.~\eqref{total_non_abelian_ann_colored_av}, 
for the bsf cross section the thermal average of eq.~\eqref{bsf_adjoint_to_bound_1S}, 
for the bsd width eqs.~\eqref{nonabelian_bsd_width_groundstate} and~\eqref{nonabelian_bsd_xsection_groundstate}, and for the annihilation widths eqs.~\eqref{widthparaSUN} and~\eqref{widthorthoSUN}.
In the right panel of figure~\ref{fig:plotBSD},
we plot the thermally averaged effective cross section normalized to~\eqref{LOxsection_nonabelian} for the SU(2) (black lines), SU(3) (orange lines) and SU(4) (brown lines) theory. 
Dashed lines correspond to a constant $\alpha = 0.03$, whereas solid lines stand for a running coupling that has been evolved from the initial value $\alpha(2M)=0.03$.
The bell shape originates from the bound-state formation and decay contribution,
$\langle   \sigma^{\textrm{1S}}_{\textrm{bsf}} \, v_{\textrm{rel}} \rangle \, \Gamma_{\textrm{ann}}^{\textrm{1S}}/(\Gamma_{\textrm{ann}}^{\textrm{1S}}+\Gamma_{\textrm{bsd}}^{\textrm{1S}})$,
being dominant with respect to the annihilation term, $\langle \sigma_{\textrm{ann}} v_{\textrm{rel}} \rangle$, for some $N$-dependent temperature regions. 
Note that the solid curves are systematically higher than the dashed curves, i.e. accounting for the running of the coupling increases the effective cross section.  

\begin{figure}[ht]
    \centering
    \includegraphics[scale=0.80]{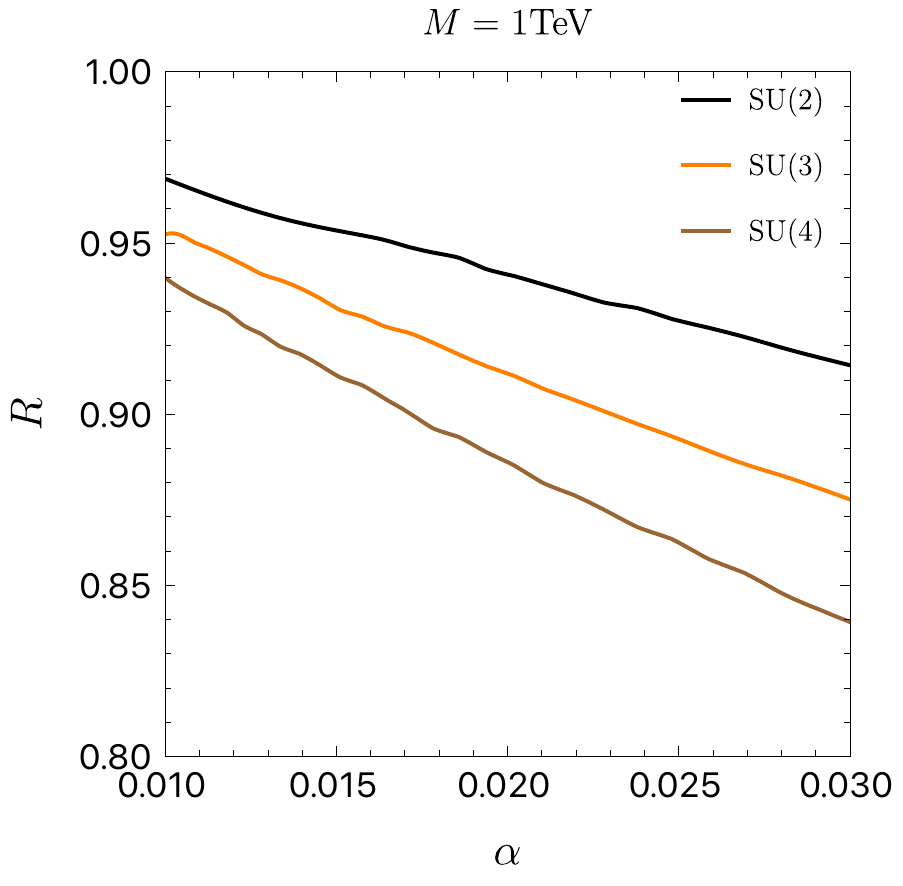}
        \hspace{0.3 cm} \includegraphics[scale=0.78]{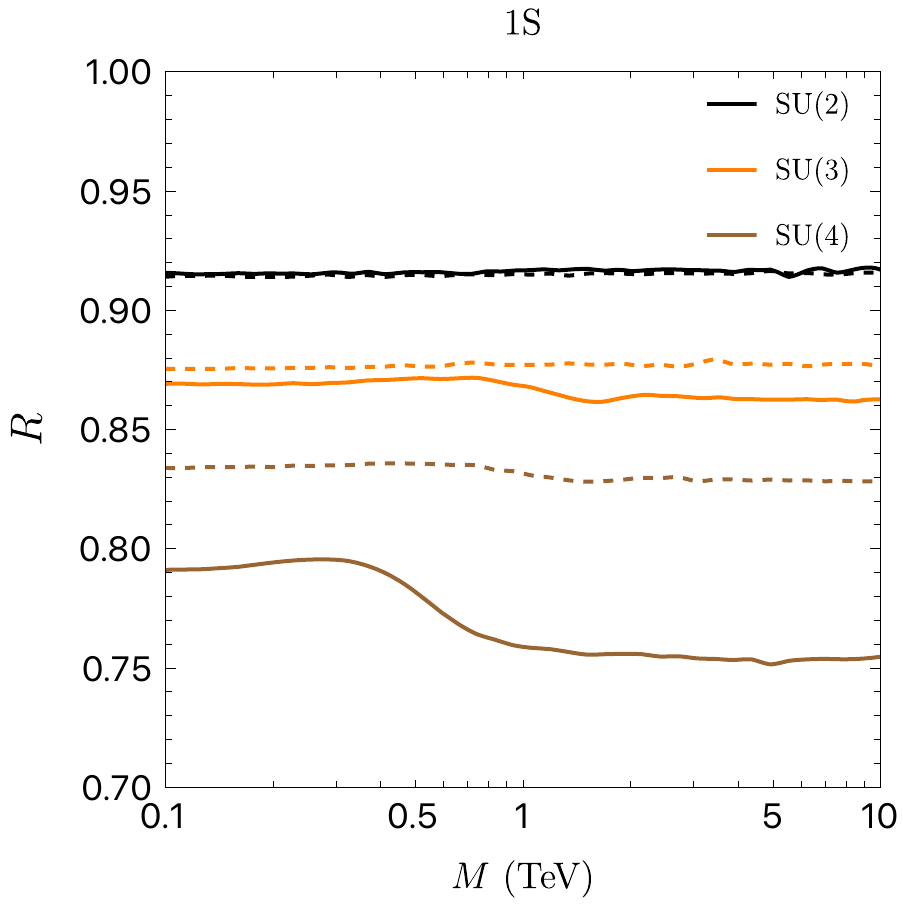}
        \caption{(Left) Ratios of the DM energy densities obtained from the annihilation cross section at NLO$^*$ given in eq.~\eqref{total_non_abelian_ann_colored_av}
          and the annihilation cross section at LO, 
    without bound-state effects, for the non-abelian models SU(2), SU(3) and SU(4). 
    The dark matter mass is fixed at $M=1$~TeV. 
    The curves are obtained with a running coupling evolved from the values of $\alpha(2M)$ given on the horizontal axis.
    (Right) Ratio of the DM energy density obtained from the effective cross section \eqref{Cross_section_eff} at $\mathcal{O}(\alpha^3)$ and at $\mathcal{O}(\alpha^2)$,
    including both scattering states and the 1S ground state, as a function of the DM mass $M$ in the SU(2), SU(3) and SU(4) model. 
    Solid lines are from the running coupling $\alpha$ evolved from $\alpha(2M)=0.03$, dashed lines are from the constant $\alpha=0.03$.
      }
    \label{fig:NLO_versus_V2_nonabelian}
  \end{figure}

In analogy with the numerical results presented for the abelian model, 
we first solve the effective Boltzmann equation without bound-state effects, while retaining the Sommerfeld factors in the annihilation cross section. 
In the left panel of figure~\ref{fig:NLO_versus_V2_nonabelian},
we show the ratios of the DM energy density $\Omega_{\textrm{DM}}h^2$ as obtained from the cross section in eq.~\eqref{total_non_abelian_ann_colored_av}
and the one extracted by replacing $(\sigma^{\hbox{\tiny NR}}_{\hbox{\scriptsize ann}} v_{\hbox{\scriptsize rel}})^{[\mathds{1}]}_{\hbox{\tiny NLO$^*$}}$
with $(\sigma^{\hbox{\tiny NR}}_{\hbox{\scriptsize ann}} v_{\hbox{\scriptsize rel}})^{[\mathds{1}]}_{\hbox{\tiny LO}}$,
and $(\sigma^{\hbox{\tiny NR}}_{\hbox{\scriptsize ann}} v_{\hbox{\scriptsize rel}})^{[\textbf{adj}]}_{\hbox{\tiny NLO$^*$}}$
with $(\sigma^{\hbox{\tiny NR}}_{\hbox{\scriptsize ann}} v_{\hbox{\scriptsize rel}})^{[\textbf{adj}]}_{\hbox{\tiny LO}}$.
Differently from the abelian case, NLO corrections increase the free annihilation cross sections \eqref{singlet_non_abelian_xsection} and~\eqref{octet_non_abelian_xsection}
and, therefore, dark matter is more effectively depleted for increasing values of the coupling
(compare with the different behaviour of the black solid line in figure~\ref{fig:NLO_versus_V2} for the abelian model). 

We then reinstate 1S bound-state effects in the effective cross section and numerically solve the effective Boltzmann equation~\eqref{Boltzmann_eq_eff}. 
In the right panel of figure~\ref{fig:NLO_versus_V2_nonabelian}, we show the ratio of the DM energy density, $\Omega_{\textrm{DM}} h^2$,
obtained from annihilation cross sections and decay  widths at NLO and the energy density obtained from annihilation cross sections and decay widths at LO. 
We observe that for the SU(3) and SU(4) models the NLO corrections are more significant, as there is an additional annihilation/decay channel due to the orthodarkonium,  
than for the SU(2) model, where $\Gamma^{n\textrm{S},\textrm{ortho},\textrm{SU(2)}}_{\hbox{\scriptsize ann}}=0$ at order $\alpha^6$.

As a cross check for the validity of the numerical results obtained by solving the single effective Boltzmann equation, we numerically evaluate also the coupled evolution equations. 
Considering the running of the coupling $\alpha$ and annihilations at NLO, we observe a $\lesssim 1\%$ difference to the obtained energy density in the SU(2), SU(3) and SU(4) model, 
which is within the uncertainty of the measured relic density.

As a final remark, we comment on the value chosen for the coupling in this section 
when treating the running coupling case, namely $\alpha(2M)=0.03$. 
This is a fairly smaller value than the ones that have been used in the abelian case. 
The reason for this choice is that larger values of the coupling would endanger the weak-coupling expansion in the 
lowest temperature region, $T \approx 10^{-5}M$, considered in this work.
The temperature sets the magnitude of the ultrasoft scale. 
If we require that the coupling at the ultrasoft scale is weak, then we need to require that, in particular, $\alpha(10^{-5}M) < 1$.
This condition is fulfilled for the SU(4) theory at one-loop running if we choose $\alpha(2M) \lesssim 0.03$, and by somewhat larger values of $\alpha(2M)$ for smaller gauge groups:
 $\alpha(2M) \lesssim 0.04$ for the SU(3) theory and $\alpha(2M) \lesssim 0.06$ for the SU(2) theory. 
For these values of the couplings, the absolute value of the binding
energy, $|E_1^b|$, is smaller than the freeze-out temperature $M/25$.
Differently from the abelian case (see figure~\ref{fig:yield_alpha_04}), 
it is not possible to accommodate weak coupling with having $|E_1^b|$ of the order of the freeze-out temperature, 
i.e. in the weakly-coupled non-abelian theory, the evolution after freeze-out proceeds necessarily for some time
at a temperature that is larger than the binding energy of the dark fermion-antifermion bound states.
In dependence of the coupling and the theory, also the soft scale may turn out to be smaller than or of the same order as the freeze-out temperature,
  which signals the break down of the multipole expansion for thermal gauge fields at freeze out.
  This happens for the SU(2), SU(3) and SU(4) theories considered here at $\alpha(2M) \lesssim 0.03$.
  We have not accounted for this in figure~\ref{fig:NLO_versus_V2_nonabelian}, where we assume that the multipole expansion holds also at freeze-out.

\section{Conclusions and outlook}
\label{sec:concl}
In recent years, there has been a considerable effort in transferring and extending known techniques adopted in atomic and heavy-quarkonium physics to dark matter models, 
both at zero and finite temperature.
Indeed, whenever the dark particles experience self-interactions via vector (or scalar) mediators, and the DM abundance is fixed through the freeze-out mechanism, 
the dark particle dynamics shares some similarities with the one of heavy quarkonium in a quark gluon plasma.

In the context of the freeze-out mechanism, the relevant processes that determine the non-equilibrium dynamics of dark matter single particles and pairs,
namely annihilations and decays, bound-state formation and dissociation, occur in a non-relativistic regime, i.e. when $T \ll M$.
Dark matter particles may therefore form non-relativistic bound states, whose energy scales together with the thermal scales characterize the dark matter dynamics.
The computation of physical observables becomes cumbersome and not very transparent when all the scales are intertwined as in the full relativistic field theory.
Instead, one can take advantage of the fact that the energy scales are hierarchically ordered by systematically replacing the relativistic field theory
by a tower of simpler non-relativistic effective field theories.
The advantage in using non-relativistic effective field theories is that one deals with the degrees of freedom that describe the physics of interest at the Lagrangian level,
and that long- and short-range contributions are factorized for any observable.
Moreover, it is straightforward to systematically improve the determination of physical quantities by
including higher-order corrections in the coupling and in the relative velocity, guided by the power counting of the effective theory.
In this paper, we have scrutinized the implementation of non-relativistic effective field theories
to address the dynamics of heavy dark matter particles in the thermal environment provided by the early universe. 

The main aim of this work is to provide a detailed step-by-step construction of non-relativistic effective field theories for some simple dark matter models.
In sections \ref{sec:NREFTs} to \ref{sec:numerics}, 
we consider a dark version of QED whose Lagrangian is given in eq.~\eqref{lag_mod_0}. 
We complement the existing literature with some theoretical and technical aspects that, in our perception, have not been sufficiently highlighted.
The hierarchy of scales that we assume is given in eq.~\eqref{scale_arrang}.
We show how the relevant reactions, which comprise pair annihilations of scattering states, decays of bound states, their formation, dissociation and transition rates in a thermal environment, 
can be accounted for in the effective field theories NRQED$_{\textrm{DM}}$ and pNRQED$_{\textrm{DM}}$, 
which are NRQED and pNRQED applied to dark fermions and photons, respectively.
In particular, pNRQED$_{\textrm{DM}}$ is well suited to describe particle-antiparticle pairs in scattering states and in bound states, as well as transitions among them. 
In section \ref{sec:non_abelian_model}, we consider an SU($N$) extension of the previous model 
whose  Lagrangian is given in eq.~\eqref{non_ab_model}.  
The same processes studied in the abelian case are also studied 
in the non-abelian case within a similar set of non-relativistic effective field theories.
We emphasise the differences between the two models.

For pairs in scattering states, we derive the annihilation cross section in pNRQED$_{\textrm{DM}}$, see eq.~\eqref{ann_fact_scat}.
It accounts for the effect of multiple soft photon exchanges that ultimately leads to the Sommerfeld enhancement.
Our expression makes manifest the factorization between contributions from hard modes, which are encoded in the four-fermion operator matching coefficients of NRQED$_{\textrm{DM}}$, 
and the soft dynamics captured by the wavefunction of the pair.
For pairs in bound states, we write the paradarkonium and orthodarkonium decay widths in eqs.~\eqref{ann_para1S} and \eqref{ann_ortho1s}, respectively.
We stress that if the orthodarkonium decay width, which is an effect at order $\alpha^3$, is included in the network of Boltzmann equations for the evolution of the dark matter pairs,
then one has to include at the same order the relevant matching coefficient of NRQED$_{\textrm{DM}}$ also in the scattering state annihilation cross section,
which has not always been the case in the literature.
Indeed the two contributions originate from the same four-fermion operator, the only difference consists in the two-particle states onto which they are projected. For the non-abelian model, the annihilation cross section for pairs in scattering states, which is a combination of singlet and adjoint contributions, can be found in eq.~\eqref{total_non_abelian_ann_colored_av}, whereas the paradarkonium and orthodarkonium widths are given in eqs.~\eqref{widthparaSUN} and \eqref{widthorthoSUN}, respectively.

For transitions between heavy pairs mediated by ultrasoft/thermal photons, the leading effect, as made manifest by pNRQED$_{\textrm{DM}}$, is due to an electric-dipole operator.
In the main body of the paper, we focus on the transition between a scattering and a bound state, and its reverse process, 
and compute the bound-state formation cross section, eq.~\eqref{bsfn}, for the general case and eq.~\eqref{our_gamma_1p} for the 1S case, 
and the bound-state dissociation width, eq.~\eqref{gamma_diss_n_state}, for the general case and eq.~\eqref{dis_width} for the 1S case.
We show how they can be derived from the self energy of the dark matter fermion pair in the thermal field theory version of pNRQED$_{\textrm{DM}}$.
The formation cross section and dissociation width turn out to depend on the temperature
through the (unsuppressed) statistical distributions of the involved particles, which appear naturally in the thermal field theory expression of the propagators.
For the non-abelian model of section~\ref{sec:non_abelian_model}, 
the transitions among pairs are also accounted for by an electric dipole operator. 
However, one finds more possibilities: transitions between (un-)bound singlet pairs and unbound adjoint pairs, 
as well as transitions between unbound adjoint pairs only. 
In this work, we focus only on transitions between singlet bound states and adjoint pairs, 
and we provide the bound-state formation cross section for the general case in eq.~\eqref{bsf_adjoint_to_bound} and for the 1S state in eq.~\eqref{bsf_adjoint_to_bound_1S}. 
In the appendices of the paper, we give the general expression for the electric dipole matrix element that can be used for the extraction of the bound-state formation cross section for any bound state, 
as well as for the dissociation width. 
For the abelian case only, we also provide scattering-state to scattering-state and bound-state to bound-state transitions for generic states.

Among the differences between abelian and non-abelian models, 
we remark here that, while in an abelian model a small value of the coupling at the hard scale is enough to guarantee a weak-coupling treatment for threshold observables, 
in a non-abelian dark matter model a weak-coupling treatment requires that the coupling remains small also at the ultrasoft scale. 
If we take the ultrasoft scale to be of the order of the temperature, then the smallest scale considered in this work is $T \approx 10^{-5}M$.
At one-loop running, $\alpha(2M)$ needs to be quite small to keep $\alpha(T)<1$.
For instance, in the SU(3) non-abelian model considered in section~\ref{sec:non_abelian_model}, the weak-coupling condition requires $\alpha(2M) \lesssim 0.04$.
This should caution about computing at weak coupling the bound-state formation cross section and dissociation width
entering the network of Boltzmann equations for the extraction of the DM energy density,
whenever dealing with QCD-like charged particles in coannihilation dark matter scenarios. 
Nevertheless, the EFT framework holds also if the ultrasoft scale is strongly coupled, 
which is a situation familiar to QCD~\cite{Brambilla:2004jw,Brambilla:2019tpt}.

Bound-state formation and dissociation rates are routinely used in the network of Boltzmann equations in order to extract the dark matter energy density.
Let us consider the abelian case as an example, but similar considerations apply to the non-abelian case as well.
The derivation of the rates relies on the evaluation of matrix elements of the electric-dipole operator following from the multipole expansion in pNRQED$_{\textrm{DM}}$.
The multipole expansion holds for thermal photons as long as the typical distance of the fermion-antifermion pair is smaller than $1/T$.
  At large temperatures, $T \sim M \alpha$, the multipole expansion breaks down.
  Estimating DM formation and dissociation in this situation requires computing thermal effects in NRQED$_{\textrm{DM}}$
  and matching to a version of pNRQED$_{\textrm{DM}}$ that does not contain thermal photons as degrees of freedom, but encodes them in a temperature dependent potential.\footnote{
      Similar scenarios have been examined in QED~\cite{Escobedo:2008sy,Escobedo:2010tu} and QCD~\cite{Brambilla:2008cx,Brambilla:2010vq}.}
  A quantitative assessment of such scenarios may be needed in order to solve the Boltzmann equations over a range of couplings
  that include values making $M\alpha$ smaller than the freeze-out temperature, i.e.~$\alpha \lesssim 0.04$.
  This range encompasses coupling values typical of the electroweak SM sector,
  which may be relevant for genuine WIMP dark matter particles, with or without coannihilating partners, e.g. supersymmetric model
  realizations~\cite{Hisano:2002fk,Hisano:2006nn,Cirelli:2007xd,Beneke:2012tg,Beneke:2014hja,Beneke:2016jpw} and the inert doublet model \cite{Deshpande:1977rw,Barbieri:2006dq,LopezHonorez:2006gr,LopezHonorez:2010tb}.

As a further outlook for future developments of this work, a richer dark sector, 
as in the case of the non-abelian model,
would modify, also significantly, the thermal dynamics, for instance by generating another thermal scale in the form of a Debye mass for the gauge boson.
The Debye mass modifies the thermal propagator of the gauge boson. 
Moreover, light degrees of freedom coupled to the gauge boson induce new processes for bound-state formation and dissociation.
A careful investigation of the Debye mass scale within the framework of non-relativistic effective field theories,
in particular its role in the bound-state formation via gauge boson emission, is being actively pursued~\cite{Binder:2021otw, Andrii}.

Finally, in the framework of the Boltzmann equations, thermal rates are just ingredients to be computed independently to fix the dynamics of the evolution equations.
It would be desirable, however, to derive out-of-equilibrium evolution equations for dark matter particles from non-relativistic effective field theories 
along the lines worked out for similar systems in the SM and QCD eventually leading
to Lindblad-like equations~\cite{Akamatsu:2014qsa,Braaten:2016sja,Brambilla:2016wgg,Brambilla:2017zei,Yao:2018nmy,Rothkopf:2019ipj,Akamatsu:2020ypb}.
It would then be the solution of these equations that provides the energy densities of the dark matter particles.

\section*{Acknowledgments}
G.Q. and A.V. would like to thank Miguel Angel Escobedo for discussions.
The work of S.B. is supported by the Swiss National Science Foundation (SNSF) under the Ambizione grant PZ00P2\_185783.
N.B., G.Q. and A.V. acknowledge support from the DFG (Deutsche Forschungsgemeinschaft, German Research Foundation) cluster of excellence ``ORIGINS''
under Germany's Excellence Strategy -  EXC-2094-390783311.

\appendix
\numberwithin{equation}{section}
\section{Bound-state formation and dissociation: general expressions}
\label{sec:app_A}
We consider the abelian model developed in sections~\ref{sec:NREFTs} to~\ref{sec:numerics}.
In this appendix, we give the bound-state formation (bsf) cross section and the bound-state dissociation (bsd) width for a bound state with generic quantum numbers $n$, $\ell$ and $m$:
$n$ is the principal quantum number, $\ell$ the orbital momentum quantum number and $m$ the magnetic quantum number.
The expression for the bsf cross section of a generic bound state follows from eq.~\eqref{bsfn} and reads
\begin{equation}
\begin{aligned}
(\sigma^{n\ell m}_{\textrm{bsf}}v_{\textrm{rel}})(\bm{p}) = \frac{g^2}{3\pi}\left[1+n_{\textrm{B}}(\Delta E_{n}^p)\right](\Delta E_{n}^p)^3 \left| \sum\limits_{\ell'=\ell\pm 1, \ell'\ge 0} \langle n\ell m|\bm{r}|\bm{p}\ell'\rangle\right|^2,
\label{general_bsf}
\end{aligned}
\end{equation}
where the dark photon energy is given in eq.~\eqref{energy_photon_bsf}.
The electric-dipole matrix element enters also in the reverse process, namely the dissociation of a bound pair into unbound DM particles by absorption of a dark photon from the plasma.
The thermal break-up width of a bound state with arbitrary quantum numbers $n$, $\ell$ and $m$  can be written as a convolution integral (see eq.~\eqref{photogammacrosssection})
\begin{eqnarray}
  \Gamma^{n\ell m}_{\textrm{bsd}} =  -2 \langle n\ell m  |  {\rm{Im}}\left[\Sigma^{11}(E_n) \right] | n\ell m \rangle
                             =  2 \int _{|\bm{k}|\geq |E^b_{n}|} \frac{d^3k}{(2\pi)^{3}}\,n_{\textrm{B}}(|\bm{k}|)\,\sigma^{n\ell m}_{\textrm{ion}}(|\bm{k}|)\,,
\label{gamma_diss_generic}
\end{eqnarray}
with the ionization cross section given by
\begin{equation}
  \sigma^{n\ell m}_{\textrm{ion}}(|\bm{k}|) =
  \frac{1}{2} \frac{g^{2}}{3\pi}\frac{M^{\frac{3}{2}}}{2}|\bm{k}|\sqrt{|\bm{k}|+E^b_{n}}\;\left.\left|\sum\limits_{\ell'=\ell\pm 1, \ell'\ge 0}  \langle n\ell m|\bm{r}|\bm{p}\ell'\rangle\right|^2\;\right \vert_{|\bm{p}| = \sqrt{M(|\bm{k}|+E^b_{n})}}.
\label{diss_generic}
\end{equation}
Both the bsf and ionization cross sections depend on the same electric-dipole matrix element.
We derive it in the following.

We build our derivation of the electric-dipole matrix element on refs.~\cite{gordon,stobbe}.\footnote{See a complementary derivation in parabolic coordinates in refs.~\cite{gordon,katkov}.}
The necessary ingredients are the wavefunctions of the Coulombic scattering and bound states.
The Coulomb wavefunction for a DM scattering state of positive energy $\bm{p}^2/M$ reads
\begin{eqnarray}
  \Psi_{\bm{p}}(\bm{r}) = \langle\bm{r}|\bm{p}\rangle &=& e^{\frac{\pi}{2}\frac{\alpha}{v_{\textrm{rel}}}}
                                                     \Gamma\left(1-i\frac{\alpha}{v_{\textrm{rel}}}\right)e^{i\bm{p}\cdot\bm{r}}~_{1}
                                                     F_{1}\left(i\frac{\alpha}{v_{\textrm{rel}}},1,i(pr-\bm{p}\cdot\bm{r})\right) \nonumber \\
           &= &\sqrt{\frac{2\pi\frac{\alpha}{v_{\textrm{rel}}}}{1-e^{-2\pi\frac{\alpha}{v_{\textrm{rel}}}}}}e^{i\bm{p}\cdot\bm{r}}~_{1}F_{1}\left(i\frac{\alpha}{v_{\textrm{rel}}},1,i(pr-\bm{p}\cdot\bm{r})\right),
\label{coulomb_wave0}                                                           
\end{eqnarray}
where $p \equiv |\bm{p}| \equiv M v_{\textrm{rel}}/2$ is the momentum in terms of the relative velocity of the pair and $_{1}F_{1}\left(  a,b,c\right)$ is the confluent hypergeometric function.
Choosing the coordinate system in such a way that $\bm{p}$ points to the $z$-direction,
the scattering wavefunction can be expanded into partial waves $\Psi_{\bm{p} \ell}(\bm{r})=\langle\bm{r}|\bm{p}\ell\rangle$ as  
\begin{eqnarray}
  \Psi_{\bm{p}}(\bm{r}) = \sum_{\ell=0}^{\infty} \Psi_{\bm{p} \ell}(\bm{r}) 
&=& \sqrt{\frac{2\pi \frac{\alpha}{v_{\textrm{rel}}}}{1-e^{-2\pi\frac{\alpha}{v_{\textrm{rel}}}}}}\sum_{\ell=0}^{\infty} e^{i\frac{\pi}{2}\ell}\frac{(2pr)^{\ell}}{(2\ell+1)!}(2\ell+1)P_{\ell}(\cos\theta)e^{ipr} \nonumber \\
&&\times ~_{1}F_{1}\left(\ell+1-i\frac{\alpha}{v_{\textrm{rel}}},2\ell+2,-2ipr\right)\prod \limits_{\kappa=1}^{\ell}\sqrt{\kappa^{2} + \left(\frac{\alpha}{v_{\textrm{rel}}}\right)^{2}}~,
\label{coulomb_wave}
\end{eqnarray}
where $P_\ell(x)$ are Legendre polynomials. 
The Coulomb wavefunction for a DM bound state of quantum numbers $n$, $\ell$ and $m$, Bohr radius $a_0=2/(M\alpha)$ and negative binding energy $E^b_n=-M\alpha^2/(4n^2)$ reads
\begin{equation}
\Psi_{n\ell m}(\bm{r}) = \langle\bm{r}|n\ell m \rangle = R_{n \ell}(r)Y_{\ell m}(\Omega) \,,
\label{boundwave}
\end{equation}
with $Y_{\ell m}(\Omega)$ being spherical harmonics and the radial functions given by
\begin{equation}
\begin{aligned}
  R_{n \ell}(r) = \frac{1}{(2\ell+1)!}\sqrt{\left(\frac{2}{na_0}\right)^3\frac{(n+\ell)!}{2n(n-\ell-1)!}}
  \left(\frac{2r}{na_0}\right)^\ell e^{-\frac{r}{na_0}}~_{1}F_{1}\left(\ell+1-n,2\ell+2,\frac{2r}{na_0}\right) \,.
\end{aligned}
\end{equation}

The matrix element $\langle n\ell m|\bm{r}|\bm{p}\rangle$ is then  
\begin{equation}
\begin{aligned}
&\langle n\ell m|\bm{r}|\bm{p}\rangle = \sum\limits_{\ell'=\ell\pm 1, \ell'\ge 0}\int d^3r\,\bm{r}\,\Psi_{n\ell m}^*(\bm{r})\Psi_{\bm{p} \ell'}(\bm{r}) \\
&~~~= N\left[\sqrt{\ell(\ell+1)}(\delta_{m,1}-\delta_{m,-1})\bm{e}_x - i\sqrt{\ell(\ell+1)}(\delta_{m,1}+\delta_{m,-1})\bm{e}_y + 2(\ell+1)\delta_{m,0}\bm{e}_z\right]X\, G_1 \\
&~~~~~+N\left[-\sqrt{\ell(\ell+1)}(\delta_{m,1}-\delta_{m,-1})\bm{e}_x + i\sqrt{\ell(\ell+1)}(\delta_{m,1}+\delta_{m,-1})\bm{e}_y + 2\ell\delta_{m,0}\bm{e}_z\right]Y\, G_2 \, ,
\label{matrixelement}
\end{aligned}
\end{equation}
where $\bm{e}_x$, $\bm{e}_y$ and $\bm{e}_z$ are the unit vectors in Cartesian coordinates.
Due to the selection rule for electric-dipole transitions, only the partial waves $\ell'=\ell \pm 1$ give a non-vanishing contribution.
The constants $N$, $X$ and $Y$ are
\begin{eqnarray}
&&N =  \frac{i^{\ell+3}(-1)^{n-\ell}}{(2\ell+1)!}\sqrt{\left(\frac{2}{na_0}\right)^3\frac{(n+\ell)!}{2n(n-\ell-1)!}}\left(\frac{2}{na_0}\right)^\ell \sqrt{\frac{2\pi \frac{\alpha}{v_{\textrm{rel}}}}{1-e^{-2\pi\frac{\alpha}{v_{\textrm{rel}}}}}} \nonumber
\\
  &&\phantom{xxxx} \times \sqrt{\frac{\pi}{2\ell+1}}\frac{1}
     {\left[M^2v_{\textrm{rel}}^2\left(1+\frac{\alpha^2}{n^2v_{\textrm{rel}}^2}\right)\right]^\ell} \, 
     e^{-2\left[i(\ell+1-n) + \frac{\alpha}{v_{\textrm{rel}}}\right]\arccot{\left(\frac{\alpha}{nv_{\textrm{rel}}}\right)}} \, ,\\
  &&X = \frac{i(M v_{\textrm{rel}})^{\ell+1}2^{2\ell+4}}{M^5v_{\textrm{rel}}^5\left(1+\frac{\alpha^2}{n^2v_{\textrm{rel}}^2}\right)^2}e^{-2i\arccot{\left(\frac{\alpha}{nv_{\textrm{rel}}}\right)}}
     \prod \limits_{\kappa=1}^{\ell+1}\sqrt{\kappa^{2} + \frac{\alpha^2}{v_{\textrm{rel}}^{2}}} \, , \\
  &&Y = \frac{n\ell(2\ell+1)(M v_{\textrm{rel}})^{\ell-1}2^{2\ell+3}}{\alpha M^3v_{\textrm{rel}}^2\left(1+\frac{\alpha^2}{n^2v_{\textrm{rel}}^2}\right)}
     \prod \limits_{\kappa=1}^{\ell-1}\sqrt{\kappa^{2} + \frac{\alpha^2}{v_{\textrm{rel}}^{2}}} \, ,
\end{eqnarray}
whereas $G_1$ and $G_2$ are combinations of hypergeometric functions, $_2F_1\left(a,b,c,d\right)$, 
\begin{eqnarray}
&&G_1 = ~_{2}F_{1}\left(\ell+2-i\frac{\alpha}{v_{\textrm{rel}}},\ell+1-n,2\ell+2,\frac{4i\alpha v_{\textrm{rel}}}{n\left(iv_{\textrm{rel}} + \frac{\alpha}{n}\right)^2}\right) \nonumber \\
&&\phantom{xxxx} -e^{4i\arccot{\left(\frac{\alpha}{nv_{\textrm{rel}}}\right)}}~_{2}F_{1}\left(\ell-i\frac{\alpha}{v_{\textrm{rel}}},\ell+1-n,2\ell+2,\frac{4i\alpha v_{\textrm{rel}}}{n\left(iv_{\textrm{rel}} + \frac{\alpha}{n}\right)^2}\right) \, ,
\\
&&G_2 = ~_{2}F_{1}\left(\ell+1-n,\ell-i\frac{\alpha}{v_{\textrm{rel}}},2\ell,\frac{4i\alpha v_{\textrm{rel}}}{n\left(iv_{\textrm{rel}} + \frac{\alpha}{n}\right)^2}\right) \nonumber
\\
&& \phantom{xxxx}-e^{4i\arccot{\left(\frac{\alpha}{nv_{\textrm{rel}}}\right)}}~_{2}F_{1}\left(\ell-1-n,\ell-i\frac{\alpha}{v_{\textrm{rel}}},2\ell,\frac{4i\alpha v_{\textrm{rel}}}{n\left(iv_{\textrm{rel}} + \frac{\alpha}{n}\right)^2}\right) \, .
\end{eqnarray}

The squared matrix element~\eqref{matrixelement} is
\begin{eqnarray}
  |\langle n\ell m|\bm{r}|\bm{p}\rangle|^2 &=& |N|^2  \Big\{\left[2\ell(\ell+1)(\delta_{m,1} + \delta_{m,-1}) + 4(\ell+1)^2\delta_{m,0}\right]|X|^2|G_1|^2 \nonumber
\\
&&~~ + \left[2\ell(\ell+1)(\delta_{m,1} + \delta_{m,-1}) + 4\ell^2\delta_{m,0}\right]|Y|^2|G_2|^2 \nonumber
\\
&&~~ + \left[-2\ell(\ell+1)(\delta_{m,1} + \delta_{m,-1}) + 4\ell(\ell+1)\delta_{m,0}\right](XY^{*}G_1G_2^{*}+X^{*}YG_1^{*}G_2)\Big\}.
\nonumber 
\\
\label{matrix_element_2}
\end{eqnarray}
In particular, the squared matrix element for the 1S state appearing in the bound-state formation cross section~\eqref{our_gamma_1p} and in the dissociation width~\eqref{dis_width} is 
\begin{equation}
  |\langle 1  \textrm{S}|\bm{r}|\bm{p}\rangle|^{2} =|\langle 1  \textrm{S}|\bm{r}|\bm{p}1\rangle|^{2} = \frac{2^9\pi^2a_0^4}{p(1+a_{0}^{2}p^{2})^{5}}\frac{e^{-\frac{4}{a_0p}\arctan(a_{0}p)}}{1-e^{-\frac{2\pi}{a_0p}}} \, .
\end{equation}
We notice that the relative momentum of the unbound pair, $p= M v_{\textrm{rel}}/2$ , and  the Bohr radius of the bound state, $a_{0} = 2/(M\alpha)$, appear in the matrix element in the combination  $a_{0}p=v_{\textrm{rel}}/\alpha$.

\subsection{Bound-state formation and dissociation for 2S, 2P and 3S states}
\label{sec:app_A_1}
In this section, we give the explicit expressions of the bound state formation cross section and dissociation width for 2S, 2P and 3S states
as an application of the general formulas \eqref{general_bsf}-\eqref{diss_generic} and \eqref{matrix_element_2}.

\subsubsection*{2S state}
A 2S bound state has quantum numbers $n=2$, $\ell=0$ and $m=0$, and binding energy $E^b_2=-M\alpha^2/16$.
The formation cross section reads
\begin{eqnarray}
  \sigma_{\hbox{\scriptsize bsf}}^{2 \textrm{S}} v_{\hbox{\scriptsize rel}}  &=& \frac{g^2}{3\pi} \left[ 1 + n_{\text{B}}(\Delta E_{2}^{p}) \right] |\langle 2  \textrm{S} | \bm{r} | \bm{p}1 \rangle|^2  (\Delta E_{2}^{p})^3 
    \nonumber \\
    &=& \frac{2^{7}\pi^{2} \alpha^{7}}{3 M^{2}v_{\textrm{rel}}^{5}}\frac{\left(1+\frac{\alpha^{2}}{v^{2}_{\textrm{rel}}}\right)}{\left(1+\frac{1}{4}\frac{\alpha^{2}}{v^{2}_{\textrm{rel}}}\right)^{3}}
    \frac{e^{-4\frac{\alpha}{v_{\textrm{rel}}}\arccot{\left(\frac{1}{2}\frac{\alpha}{v_{\textrm{rel}}}\right)}}}{1-e^{-2\pi\frac{\alpha}{v_{\textrm{rel}}}}}\left[ 1 + n_{\text{B}}(\Delta E_{2}^{p}) \right].
    \label{bsfresult2s}
\end{eqnarray}
In the second equality, we have inserted the expression of the electric-dipole matrix element squared  $|\left< 2 \textrm{S}|\bm{r}|\bm{p}1\right>|^2$ that we compute from \eqref{matrix_element_2} and find to be 
\begin{equation}
\begin{aligned}
    |\langle 2  \textrm{S}|\bm{r}|\bm{p}1\rangle|^{2} = \frac{2^{18}\pi^2a_0^4\left(1+a_0^2p^2\right)}{p\left(1+4a_0^2p^2\right)^{6}}\frac{e^{-\frac{4}{a_0p}\arctan{\left(2a_0p\right)}}}{1-e^{-\frac{2\pi}{a_0p}}}\,.
\end{aligned}
\end{equation} 

Thermal break-up of 2S darkonium can be triggered by the absorption of a photon from the background plasma.
The momentum threshold of the photon is smaller than for the 1S state because the 2S state is less tightly bound.
The bound-state dissociation width for the 2S state following from eqs.~\eqref{gamma_diss_generic} and \eqref{diss_generic} is
\begin{equation}
  \Gamma^{2 \textrm{S}}_{\textrm{bsd}}  =
2\int_{|\bm{k}|\geq |E^b_{2}|} \frac{d^{3}k}{(2\pi)^{3}}~n_{\textrm{B}}(|\bm{k}|)\frac{1}{2}\frac{g^{2}}{3\pi}\frac{M^{\frac{3}{2}}}{2}|\bm{k}|\sqrt{|\bm{k}|+E^b_{2}}~|\left< 2S|\bm{r}|\bm{p}1\right>|^{2}\bigg \vert_{|\bm{p}| = \sqrt{M(|\bm{k}|+E^b_{2})}}\,.
\label{eq:2Sbsd}
\end{equation}   
From eqs.~\eqref{eq:2Sbsd} and~\eqref{gamma_diss_generic} we can read the 2S state ionization cross section,
\begin{eqnarray}
     \sigma^{2 \textrm{S}}_{\hbox{\scriptsize ion}}(|\bm{k}|)= \alpha \frac{ 2^{12} \pi^2}{3} \left(4+w_2(|\bm{k}|)^2\right)\frac{|E^b_2|^4}{M  |\bm{k}|^5}  \frac{e^{-\frac{8}{w_2(|\bm{k}|)}\arctan(w_2(|\bm{k}|))} }{1-e^{-\frac{4\pi}{w_2(|\bm{k}|)}}} \, ,
\end{eqnarray}
where we have defined $w_2(|\bm{k}|) \equiv \sqrt{|\bm{k}|/|E^b_2|-1}$.

\subsubsection*{2P states}
We consider the formation of 2P bound states, whose quantum numbers are $n=2$, $\ell=1$ and  $m=0$, $\pm 1$.
Applying eq.~\eqref{matrix_element_2} for this special case, only the two partial waves with $\ell'=0$ and $\ell'=2$ contribute;
we obtain
\begin{equation}
\begin{aligned}
  \sigma_{\textrm{bsf}}^{2\textrm{P}_{m=0}}v_{\textrm{rel}} = \frac{2^{5}\pi^2\alpha^9}{3^3}\frac{\left(\sqrt{1+\frac{\alpha^2}{4v_{\textrm{rel}}^2}}
      -4\sqrt{1+\frac{\alpha^2}{v_{\textrm{rel}}^2}}\right)^2}{M^2v_{\textrm{rel}}^7
    \left(1+\frac{\alpha^2}{4v_{\textrm{rel}}^2}\right)^4}\frac{e^{-4\frac{\alpha}{v_{\textrm{rel}}}\arccot{\left(\frac{\alpha}{2v_{\textrm{rel}}}\right)}}}{1-e^{-2\pi\frac{\alpha}{v_{\textrm{rel}}}}}
  \left[1+n_{\textrm{B}}(\Delta E_{2}^p)\right] \, ,
\label{bsf_2p0}
\end{aligned}
\end{equation}
\begin{equation}
\begin{aligned}
  \sigma_{\textrm{bsf}}^{2\textrm{P}_{m=\pm 1}}v_{\textrm{rel}} = \frac{2^{5}\pi^2\alpha^9}{3^3}\frac{\left(\sqrt{1+\frac{\alpha^2}{4v_{\textrm{rel}}^2}}
      +2\sqrt{1+\frac{\alpha^2}{v_{\textrm{rel}}^2}}\right)^2}{M^2v_{\textrm{rel}}^7
    \left(1+\frac{\alpha^2}{4v_{\textrm{rel}}^2}\right)^4}\frac{e^{-4\frac{\alpha}{v_{\textrm{rel}}}\arccot{\left(\frac{\alpha}{2v_{\textrm{rel}}}\right)}}}{1-e^{-2\pi\frac{\alpha}{v_{\textrm{rel}}}}}
  \left[1+n_{\textrm{B}}(\Delta E_{2}^p)\right] \, .
\label{bsf_2p1}
\end{aligned}
\end{equation}
In order to account for the degeneracy in the magnetic quantum number $m$ and to compare explicitly the bsf of a $2P$ state with that one of a $2S$ state,
we sum up the results \eqref{bsf_2p0} and \eqref{bsf_2p1}
\begin{eqnarray}
  \sigma_{\textrm{bsf}}^{2\textrm{P}}v_{\textrm{rel}} &=& \sum \limits_{m=-1}^{m=1} \sigma_{\textrm{bsf}}^{2 \textrm{P}_m}v_{\textrm{rel}}
\nonumber 
\\
&=& \frac{2^{5}\pi^2\alpha^9}{3^3}\frac{ \left(\sqrt{1+\frac{\alpha^2}{4v_{\textrm{rel}}^2}} -4\sqrt{1+\frac{\alpha^2}{v_{\textrm{rel}}^2}}\right)^2 + 2\left(\sqrt{1+\frac{\alpha^2}{4v_{\textrm{rel}}^2}} +2\sqrt{1+\frac{\alpha^2}{v_{\textrm{rel}}^2}}\right)^2}{M^2v_{\textrm{rel}}^7\left(1+\frac{\alpha^2}{4v_{\textrm{rel}}^2}\right)^4} \nonumber
\\
&& \phantom{xxx}\times \frac{e^{-4\frac{\alpha}{v_{\textrm{rel}}}\arccot{\left(\frac{\alpha}{2v_{\textrm{rel}}}\right)}}}{1-e^{-2\pi\frac{\alpha}{v_{\textrm{rel}}}}}\left[1+n_{\textrm{B}}(\Delta E_{2}^p)\right] \, .
\end{eqnarray}
Since at $\mathcal{O}(\alpha^2)$ the binding energies depend only on the principal quantum number $n$, the states 2S and 2P$_{m=0,\pm 1}$ have the same energy. 

The thermal dissociation width, averaged over $m$, is 
\begin{equation}
\begin{aligned}
  \Gamma_{\textrm{bsd}}^{2\textrm{P}} = \frac{1}{3}\sum \limits_{m=0,\pm 1}&\int_{|\bm{k}|\geq |E^b_{2}|} \frac{d^{3}k}{(2\pi)^{3}}\,n_{\textrm{B}}(|\bm{k}|)\,\frac{g^{2}}{3\pi}\frac{M^{\frac{3}{2}}}{2}|\bm{k}|\sqrt{|\bm{k}|+E^b_{2}}
\\
  &\times \left.\left|\sum_{\ell'=0,2}\langle 2\textrm{P}_{m}|\bm{r}|\bm{p}\ell'\rangle \right|^2\;\right\vert_{|\bm{p}| = \sqrt{M(|\bm{k}|+E^b_{2})}}\,.
\label{diss_2p}
\end{aligned}
\end{equation}   
From eqs.~\eqref{diss_2p} and~\eqref{gamma_diss_generic} we can read the 2P$_{m=0,\pm 1}$ states ionization cross sections,
\begin{eqnarray}
  \sigma^{2\textrm{P}_{m=0}}_{\hbox{\scriptsize ion}}(|\bm{k}|)=
  \alpha \frac{ 2^{12} \pi^2}{3^3} \left(\sqrt{w_2(|\bm{k}|)^2 +1} - 4\sqrt{w_2(|\bm{k}|)^2+4}\right)^2\frac{|E^b_2|^5}{M  |\bm{k}|^6}  \frac{e^{-\frac{8}{w_2(|\bm{k}|)}\arctan(w_2(|\bm{k}|))} }{1-e^{-\frac{4\pi}{w_2(|\bm{k}|)}}} ,
\nonumber\\
\end{eqnarray}
\begin{eqnarray}
  \sigma^{2\textrm{P}_{m=\pm 1}}_{\hbox{\scriptsize ion}}(|\bm{k}|)=
  \alpha \frac{ 2^{12} \pi^2}{3^3} \left(\sqrt{w_2(|\bm{k}|)^2 +1} + 2\sqrt{w_2(|\bm{k}|)^2+4}\right)^2\frac{|E^b_2|^5}{M  |\bm{k}|^6}  \frac{e^{-\frac{8}{w_2(|\bm{k}|)}\arctan(w_2(|\bm{k}|))} }{1-e^{-\frac{4\pi}{w_2(|\bm{k}|)}}} .
  \nonumber\\
\end{eqnarray}

\subsubsection*{3S state}
A 3S bound state has quantum numbers $n=3$, $\ell=0$ and  $m=0$, and binding energy $E^b_{3}=-M \alpha^2/(4 \cdot 3^2)$.
The formation cross section reads
\begin{equation}
  \sigma_{\textrm{bsf}}^{3\textrm{S}}v_{\textrm{rel}} = \frac{2^{10}\pi^2}{3^4}\frac{\alpha^7\left(1+\frac{\alpha^2}{v_{\textrm{rel}}^2}\right)}{M^2v_{\textrm{rel}}^5}
  \frac{\left(1+\frac{7\alpha^2}{3^3v_{\textrm{rel}}^2}\right)^2}{\left(1+\frac{\alpha^2}{3^2v_{\textrm{rel}}^2}\right)^5}
  \frac{e^{-4\frac{\alpha}{v_{\textrm{rel}}}\arccot{\left(\frac{\alpha}{3v_{\textrm{rel}}}\right)}}}{1-e^{-2\pi\frac{\alpha}{v_{\textrm{rel}}}}}
  \left[1+n_{\textrm{B}}(\Delta E_{3}^{p})\right]
 \,.
\end{equation}
The bound-state dissociation width for the 3S state is 
\begin{equation}
  \Gamma^{3 \textrm{S}}_{\textrm{bsd}} =
  2\int_{|\bm{k}|\geq |E^b_{3}|} \frac{d^{3}k}{(2\pi)^{3}}\;n_{\textrm{B}}(|\bm{k}|)\,\frac{1}{2}\frac{g^{2}}{3\pi}\frac{M^{\frac{3}{2}}}{2}|\bm{k}|\sqrt{|\bm{k}|+E^b_{3}}\,
  \left.|\langle 3S|\bm{r}|\bm{p}1\rangle|^{2}\right\vert_{|\bm{p}| = \sqrt{M(|\bm{k}|+E^b_{3})}}\,,
\label{eq:3Sbsd}
\end{equation}   
and from eqs.~\eqref{eq:3Sbsd} and~\eqref{gamma_diss_generic} we can read the 3S state ionization cross section,
\begin{eqnarray}
  \sigma^{3 \textrm{S}}_{\hbox{\scriptsize ion}}(|\bm{k}|)= \alpha 2^9 \pi^2 3^2 (3^2+w_3(|\bm{k}|)^2)\left(\frac{7}{3}+w_3(|\bm{k}|)^2\right)^2
  \frac{|E^b_3|^6}{M  |\bm{k}|^7}  \frac{e^{-\frac{12}{w_3(|\bm{k}|)}\arctan(w_3(|\bm{k}|))} }{1-e^{-\frac{6\pi}{w_3(|\bm{k}|)}}} \, ,
  \nonumber
  \\
\end{eqnarray}
where we have defined $w_3(|\bm{k}|) \equiv \sqrt{\bm{k}|/|E^b_3|-1}$.

\section{Bound-state to bound-state transitions}
\label{sec:app_B}
In this section, we derive the transition rates between two DM bound states in the abelian model.
In pNR\-QED$_{\textrm{DM}}$ the relevant diagram is the middle diagram of figure~\ref{fig:pNREFT_DM_fig0}, 
which shows the transition from a bound state with quantum numbers $n$ to a bound state with quantum numbers $n'$ mediated by an electric-dipole vertex.
Excitation refers to the process where a DM bound state absorbs an ultrasoft photon from the thermal bath.\footnote{
  The energy of the incoming photon is not large enough to break the bound state into an unbound DM pair,
  thus the excitation process can be distinguished from the thermal break-up process by requiring the energy $E_\gamma + E_{n}$ to be negative.}
Due to the energy gain, the state jumps to a higher energy configuration, like in the transition $\gamma + (X\bar{X})_\textrm{1S} \rightarrow  (X\bar{X})_\textrm{2P}$. 
De-excitation by emission of a photon is the reverse process, it involves an excited state that looses energy to a more tightly bound state. 
Since these processes are mediated by the electric-dipole operator,
the  angular momentum of the bound state must change by one unit, $|\Delta \ell |=1 $, whereas the spin is left unchanged. 

\begin{figure}[ht]
    \centering
    \includegraphics[scale=0.65]{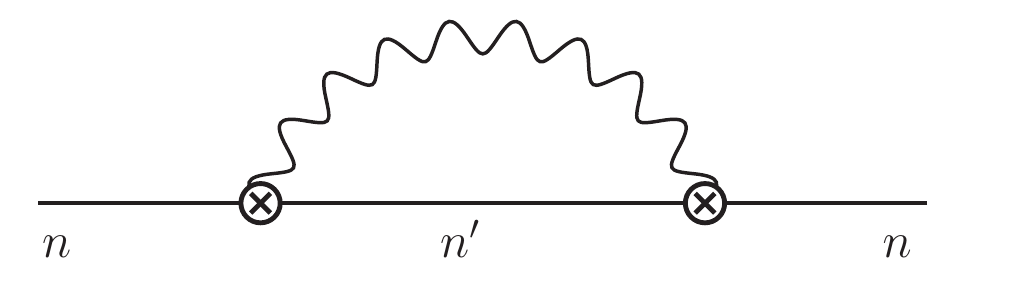}
    \caption{One-loop self-energy diagram for a bound state in pNRQED$_\textrm{DM}$.
      The incoming bound state has quantum numbers $n$, whereas the bound state in the loop carries quantum numbers $n'$.}
    \label{fig:bound_to_bound}
\end{figure}

In pNRQED$_{\textrm{DM}}$, both excitation and de-excitation processes can be described at leading order
by cutting the self-energy diagram shown in figure~\ref{fig:bound_to_bound}.
Here, the matter states inside and outside the loop are bound states, at variance with the diagrams in figure~\ref{fig:pnEFT_DM_self} 
that show transitions between bound states and scattering states.
The computation of the transition width closely follows the one carried out in section~\ref{sec:Electric_transitions}.

De-excitation transitions require the energy conservation condition $E_n = E_{n'}+E_\gamma$,
with $E_n$ the energy of the incoming bound state, $E_{n'}$ the energy of the outcoming bound state and $E_\gamma>0$ the energy of the on-shell dark photon.
For Coulombic bound states, energy conservation can be fulfilled if $n>n'$.
De-excitation transitions may happen both in vacuum and in the thermal medium, although in the thermal medium they are enhanced by the {\it stimulated emission}.
The de-excitation width $\displaystyle \Gamma^{n\ell m}_{\textrm{de-ex.}} \equiv \sum_{n'< n, \ell', m'} \Gamma^{n\ell m\rightarrow n'\ell' m'}_{\textrm{de-ex.}}$ reads 
\begin{eqnarray}
  \Gamma^{n\ell m}_{\textrm{de-ex.}} =
  \sum_{n'< n, \ell',m'}\frac{g^2}{3 \pi}\left(\Delta E_{n'}^{n}\right)^3
  \left[1+n_{\text{B}}\left(\Delta E_{n'}^{n}\right)\right] \left| \langle \, n'\ell'm' \, |  \bm{r}  | \, n \ell m\, \rangle \right|^2 \, ,
    \label{gamma_deexcitation1}
\end{eqnarray}
where we have specified the quantum numbers, $n$, $\ell$ and $m$, of the decaying bound state and the quantum numbers, $n'$, $\ell'$ and $m'$, of the final bound state.
The dark photon carries at leading order an energy $\Delta E^n_{n'} = E_n-E_{n'} = (M\alpha^2/4)\left(1/{n'}^2-1/n^2\right)$.

Excitation transitions require the energy conservation condition $E_n +E_\gamma=E_{n'}$, which can be fulfilled if $n<n'$.
Excitation transitions may happen only in the thermal medium.
The excitation width $\displaystyle \Gamma^{n\ell m}_{\textrm{ex.}} \equiv \sum_{n'> n, \ell', m'} \Gamma^{n\ell m\rightarrow n'\ell' m'}_{\textrm{ex.}}$ reads
\begin{eqnarray}
      \Gamma^{n\ell m}_{\textrm{ex.}} =
      \sum_{n'> n, \ell',m'}\frac{g^2}{3 \pi}\left|\Delta E_{n'}^{n}\right|^3  n_{\text{B}}
      \left(\left|\Delta E_{n'}^{n}\right|\right) \left| \langle \, n'\ell'm' \, |  \bm{r}  | \, n \ell m \, \rangle \right|^2 \, .
    \label{gamma_excitation1}
\end{eqnarray}
Note that in these expressions we have summed over all polarizations of the dark photon either if the photon is emitted into or absorbed from the medium.

Projected on a specific final state,
the transition widths \eqref{gamma_deexcitation1} and \eqref{gamma_excitation1} satisfy the detailed balance equation (for $n>n'$)
\begin{equation}
n_\textrm{B}(E_{n})\,\Gamma^{n\ell m \to n'\ell'm'}_{\textrm{de-ex.}}
= n_\textrm{B}(E_{n'})\,\Gamma^{n'\ell' m' \to n\ell m}_{\textrm{ex.}}\,,
\end{equation}
where the distribution of the bound states may be approximated by the Maxwell--Boltz\-mann distribution, as $M \gg T$, so that 
$\displaystyle n_\textrm{B}(E_{n'})/n_\textrm{B}(E_{n}) \approx e^{\Delta E^n_{n'}/T}$. 
Radiative transitions have been considered also in ref.~\cite{Garny:2021qsr}, however, there the photon distribution has been set to 0 in the de-excitation width, 
which amounts at ignoring stimulated emission, and to $e^{-|\Delta E^n_{n'}|/T}$ in the excitation width. 
Both approximations are valid only at very small temperatures, $T \ll |\Delta E^n_{n'}|$. 
In our numerical computations, we employ the transition widths in eqs.~\eqref{gamma_deexcitation1} and \eqref{gamma_excitation1}.

In order to evaluate the transition widths \eqref{gamma_deexcitation1} and \eqref{gamma_excitation1},
we have to compute the same quantum-mechanical matrix element  $\langle \, n\ell m \, |  \bm{r}  | \, n'\ell'm' \, \rangle$.
We obtain
\begin{equation}
\begin{aligned}
&\langle n\ell m |\bm{r}|n'\ell'm' \rangle = \int d^3r\,\bm{r}\,\Psi_{n\ell m}^*(\bm{r})\Psi_{n'\ell'm'}(\bm{r}) \\
&~~~= \mathcal{N} \Bigg\{-\delta_{\ell,\ell'-1}\sqrt{\frac{(n'+\ell+1)!}{(n'-\ell-2)!}}\sqrt{\frac{1}{4\pi(2\ell + 3)}} \\
&~~~~~~~~~~~\times \Bigg[\sqrt{\frac{(\ell + m)!}{(\ell-m+2)!}}(\ell - m+1)(\ell - m+2)\delta_{m,m'+1}(-\bm{e}_x+i\bm{e}_y) \\
&~~~~~~~~~~~~~~+ \sqrt{\frac{(\ell + m+2)!}{(\ell-m)!}}\delta_{m,m'-1}(\bm{e}_x+i\bm{e}_y) + 2\sqrt{\frac{(\ell + m+1)!}{(\ell-m+1)!}}(\ell - m+1)\delta_{m,m'}\bm{e}_z\Bigg]\mathcal{G}_1 \\
&~~~~~~~~~+ \delta_{\ell,\ell'+1}nn'\sqrt{\frac{(n'+\ell-1)!}{(n'-\ell)!}}\sqrt{\frac{2\ell - 1}{4\pi}}\frac{(2\ell + 1)\ell}{2(2\ell - 1)} \\
&~~~~~~~~~~~\times \Bigg[\sqrt{\frac{(\ell + m-2)!}{(\ell-m)!}}(\ell + m-1)(\ell + m)\delta_{m,m'+1}(\bm{e}_x-i\bm{e}_y) \\
&~~~~~~~~~~~~~~+ \sqrt{\frac{(\ell + m)!}{(\ell-m-2)!}}\delta_{m,m'-1}(-\bm{e}_x-i\bm{e}_y) + 2\sqrt{\frac{(\ell + m-1)!}{(\ell-m-1)!}}(\ell + m)\delta_{m,m'}\bm{e}_z\Bigg]\mathcal{G}_2 \Bigg\}~,
\label{b_to_b_01}
\end{aligned}
\end{equation}
where
\begin{equation}
\begin{aligned}
  \mathcal{N} \equiv \frac{(-1)^{n'-\ell}}{M\alpha}\frac{2^{2\ell+4}\pi}{2\ell + 1}
  \sqrt{\frac{2\ell+1}{4\pi}\frac{(\ell-m)!}{(\ell+m)!}}\left(\frac{n-n'}{n+n'}\right)^{n+n'}\sqrt{\frac{(n+\ell)!}{(n-\ell-1)!}}\frac{1}{(2\ell+1)!}\frac{n^\ell n'^\ell}{(n-n')^{2\ell+2}},
\end{aligned}
\label{b_to_b_02}
\end{equation}
\begin{equation}
\begin{aligned}
\mathcal{G}_1 &\equiv n^2n'^2\Bigg[\frac{_2F_1\left(\ell+2-n',\ell+1-n,2\ell+2,-\frac{4n'n}{(n'-n)^2}\right)}{(n-n')^2} \\
&~~~~~~~~~~~- \frac{_2F_1\left(\ell-n',\ell+1-n,2\ell+2,-\frac{4n'n}{(n'-n)^2}\right)}{(n+n')^2} \Bigg]~,
\end{aligned}
\label{b_to_b_03}
\end{equation}
\begin{equation}
\begin{aligned}
\mathcal{G}_2 &\equiv \Bigg[~_2F_1\left(\ell+1-n,\ell-n',2\ell,-\frac{4n'n}{(n'-n)^2}\right) \\
&~~~~~- \left(\frac{n-n'}{n+n'}\right)^2~_2F_1\left(\ell-1-n,\ell-n',2\ell,-\frac{4n'n}{(n'-n)^2}\right) \Bigg].
\end{aligned}
\label{b_to_b_04}
\end{equation}

As a special example, we consider transitions between 1S or 2S states and 2P or 3P states.
The transition matrix elements squared are
\begin{eqnarray}
&&|\langle \textrm{1S}|\bm{r}|\textrm{2P}_{m=0}\rangle|^2 = |\langle \textrm{1S}|\bm{r}|\textrm{2P}_{m=\pm 1}\rangle|^2  =  \frac{2^{17}}{3^{10}}\frac{1}{M^2\alpha^2}~, \\
&& |\langle \textrm{1S}|\bm{r}|\textrm{3P}_{m=0}\rangle|^2 = |\langle \textrm{1S}|\bm{r}|\textrm{3P}_{m=\pm 1}\rangle|^2 = \frac{3^6}{2^{11}}\frac{1}{M^2\alpha^2}~, \\
&&|\langle \textrm{2S}|\bm{r}|\textrm{3P}_{m=0}\rangle|^2 = |\langle \textrm{2S}|\bm{r}|\textrm{3P}_{m=\pm 1}\rangle|^2=  \frac{2^{22}3^6}{5^{12}}\frac{1}{M^2\alpha^2}.
\end{eqnarray}
The excitation widths read
\begin{align}
  \Gamma^{1\textrm{S}\rightarrow 2\textrm{P}}_{\textrm{ex.}} &=
  \frac{2^7}{3^7}\frac{M\alpha^5}{e^{\frac{3M\alpha^2}{16T}}-1}\,, \\
  \Gamma^{1\textrm{S}\rightarrow 3\textrm{P}}_{\textrm{ex.}} &= \frac{1}{2^6}\frac{M\alpha^5}{e^{\frac{2M\alpha^2}{9T}}-1}\,,\\
  \Gamma^{2\textrm{S}\rightarrow 3\textrm{P}}_{\textrm{ex.}} &= \frac{2^{12}}{5^9}\frac{M\alpha^5}{e^{\frac{5M\alpha^2}{144T}}-1}\,,
\end{align}
where we have summed over the polarizations of the final states.
The de-excitation widths read
\begin{align}
  \Gamma^{2\textrm{P}\rightarrow 1\textrm{S}}_{\textrm{de-ex.}}
  &=\frac{2^7}{3^8}\frac{M\alpha^5}{1-e^{-\frac{3M\alpha^2}{16T}}} \, , \\
  \Gamma^{3\textrm{P}\rightarrow 1\textrm{S}}_{\textrm{de-ex.}} &=\frac{1}{3}\frac{1}{2^6}\frac{M\alpha^5}{1-e^{-\frac{2M\alpha^2}{9T}}}\,,\\
  \Gamma^{3\textrm{P}\rightarrow 2\textrm{S}}_{\textrm{de-ex.}} &=\frac{1}{3}\frac{2^{12}}{5^9}\frac{M\alpha^5}{1-e^{-\frac{5M\alpha^2}{144T}}}\,,
\end{align}
where we have now averaged over the magnetic quantum numbers of the decaying states.

\begin{figure}[ht]
		\centering
		\includegraphics[width=.47\textwidth]{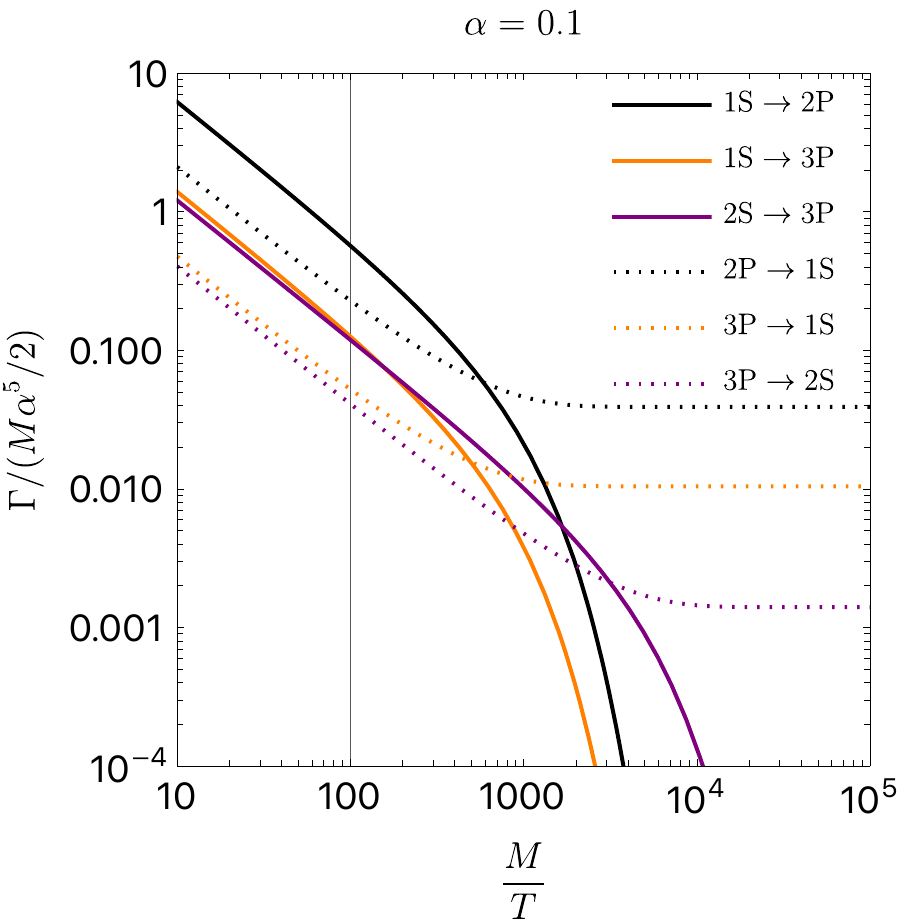}
		\includegraphics[width=.47\textwidth]{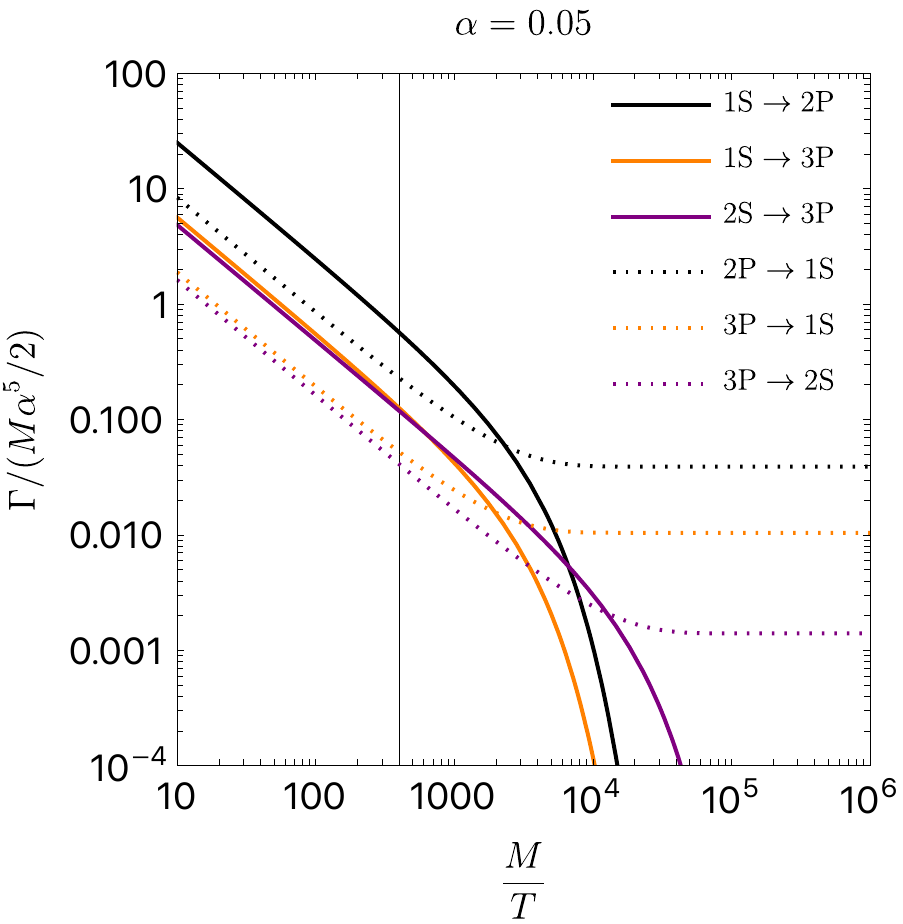}
		\caption{Ratios of excitation (continuous lines) and de-excitation (dotted lines) widths 
                  over the paradarkonium annihilation width at leading order for the first few excited states.
                  The vertical lines mark the position where $T=M\alpha^2$.}
		\label{bound-to-bound_pic}
\end{figure}

We plot the ratio of the excitation and de-excitation widths over the paradarkonium annihilation width in figure~\ref{bound-to-bound_pic}.
One recognizes that in the region $M\alpha^2 \gtrsim T$ bound-state to bound-state widths are suppressed
with respect to the leading-order annihilation width $\Gamma^{1\textrm{S},\hbox{\scriptsize para}}_{\textrm{ann}}=M\alpha^5/2$.
Moreover, one can notice the qualitative different dependence on the temperature of the excitation and de-excitation processes.
The former process becomes exponentially suppressed for smaller temperatures (solid lines),
since the photons that allow for the excitation are progressively less abundant in the plasma,
whereas the de-excitation has an in-vacuum contribution that dominates at small temperatures (dotted lines).

\section{Scattering-state to scattering-state transitions}
\label{sec:app_C}
Similarly to what happens for bound-state to bound-state transitions, scattering states can undergo processes of emission (or bremsstrahlung)
and absorption of an ultrasoft/thermal photon.
In  pNRQED$_{\textrm{DM}}$ the relevant diagram is the most right diagram of figure~\ref{fig:pNREFT_DM_fig0},
which shows the transition from a scattering state of relative momentum $p$ to a scattering state of relative momentum $p'$ mediated by an electric-dipole vertex.

\begin{figure}[ht]
    \centering
    \includegraphics[scale=0.65]{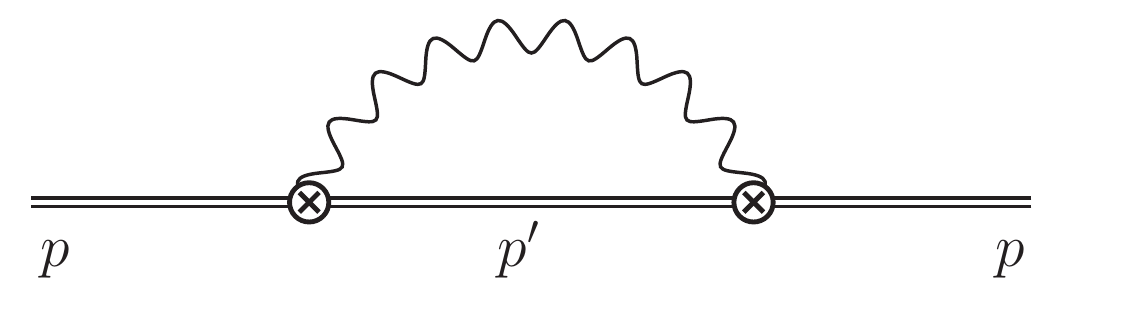}
    \caption{One-loop self-energy diagram for a scattering state in pNRQED$_\textrm{DM}$.
      The incoming scattering state has a relative momentum $p$, whereas the scattering state in the loop carries a momentum $p'$.}
    \label{fig:scat_to_scat}
\end{figure}

The scattering state transition cross section can be evaluated from the imaginary part of the self-energy diagram for an unbound DM pair, which is shown in figure~\ref{fig:scat_to_scat}.
For the differential cross section of the emission process, we obtain for $|\bm{p}'|<|\bm{p}|$
\begin{equation}
\begin{aligned}
  \frac{d(\sigma_{\textrm{emi}} v_{\textrm{rel}})(\bm{p},\bm{p}')}{d^3p'} &= \frac{g^2}{24\pi^4}
  (E_p-E_{p'})^3\left[1+n_{\text{B}}(E_p-E_{p'})\right]|\langle \bm{p}'|\bm{r}|\bm{p}\rangle|^2,
    \label{emission}
\end{aligned}
\end{equation}
where $E^s_p \equiv E_p - 2M$.
%The emission width follows from integrating $(\sigma_{\textrm{emi}} v_{\textrm{rel}})(\bm{p},\bm{p}')$ over the momenta $\bm{p}$ and  $\bm{p}'$.
For the differential cross section of the absorption process, we obtain for $|\bm{p}'|>|\bm{p}|$
\begin{equation}
\begin{aligned}
 \frac{d(\sigma_{\textrm{abs}} v_{\textrm{rel}})(\bm{p},\bm{p}')}{d^3p'} &= \frac{1}{2}\frac{g^2}{24\pi^4}
  |E_p-E_{p'}|^3n_{\text{B}}(|E_p-E_{p'}|)|\langle \bm{p}'|\bm{r}|\bm{p}\rangle|^2 ,
    \label{absorption}
\end{aligned}
\end{equation}
where we have averaged over the polarizations of the incoming dark photon.
%The absorption width follows from integrating $(\sigma_{\textrm{abs}} v_{\textrm{rel}})(\bm{p},\bm{p}')$ over the momenta $\bm{p}$ and $\bm{p}'$, and summing over the two photon polarizations.
Notice that the absorption process cannot happen in vacuum.
%, and the integration in the photon momentum $k$ does not have a threshold, at variance with the emission process of eq.~\eqref{emission}.

Choosing the reference frame such that the relative momentum $\bm{p}=M\bm{v}_{\textrm{rel}}/2$ of the incoming scattering state is oriented along the $z$-direction,
the outcoming relative momentum $\bm{p}'$ can be written as $\bm{p}'= (\cos{(\phi_{p'})}\sin{(\theta_{p'})},\sin{(\phi_{p'})}\sin{(\theta_{p'})},\cos{(\theta_{p'})})Mv_{\textrm{rel}}'/2$.
The matrix element $\langle \bm{p}|\bm{r}|\bm{p}' \rangle$ reads
\begin{equation} 
    \begin{aligned}
    &\langle \bm{p}|\bm{r}|\bm{p}' \rangle = \int d^3r\,\bm{r}\,\Psi_{\bm{p}}^*(\bm{r})\Psi_{\bm{p}'}(\bm{r}) \\
&~~~= \mathcal{M}\sum \limits_{\ell=0}^\infty \mathcal{A}_\ell \Bigg\{ \frac{\ell +1}{2\ell +1}\sqrt{(\ell+1)^2+(\alpha/v_{\textrm{rel}}')^2} \\
&~~~\times \left[P_{\ell+1}(\cos{(\theta_{p'})})\bm{e}_z+\frac{P^1_{\ell+1}(\cos{(\theta_{p'})})}{\ell+1}\left[\cos{(\phi_{p'})}\bm{e}_x + \sin{(\phi_{p'})}\bm{e}_y\right]\right]\left(\frac{2}{v_{\textrm{rel}}'+v_{\textrm{rel}}}\right)^2\mathcal{X}_1 \\
&~~~-\frac{2\ell^2}{v_{\textrm{rel}}' v_{\textrm{rel}}\sqrt{\ell^2+(\alpha/v_{\textrm{rel}}')^2}}\\
&~~~\times \left[P_{\ell-1}(\cos{(\theta_{p'})})\bm{e}_z+\frac{1}{\ell}P^1_{\ell-1}(\cos{(\theta_{p'})})\left[\cos{(\phi_{p'})}\bm{e}_x + \sin{(\phi_{p'})}\bm{e}_y\right]\right]\mathcal{X}_2\Bigg\} \, ,
    \end{aligned}
\end{equation}
with 
\begin{equation}
    \mathcal{M} \equiv \sqrt{\frac{2\pi \alpha/v_{\textrm{rel}}}{1-e^{-2\pi \alpha/v_{\textrm{rel}}}}}\sqrt{\frac{2\pi \alpha/v_{\textrm{rel}}'}{1-e^{-2\pi \alpha/v_{\textrm{rel}}'}}}e^{-\frac{\pi}{2}\alpha\left(\frac{1}{v_{\textrm{rel}}'}+\frac{1}{v_{\textrm{rel}}}\right)}e^{\frac{\pi}{2}\alpha\left|\frac{1}{v_{\textrm{rel}}'}-\frac{1}{v_{\textrm{rel}}}\right|}\left|\frac{v_{\textrm{rel}}'+v_{\textrm{rel}}}{v_{\textrm{rel}}'-v_{\textrm{rel}}}\right|^{i\alpha\left(\frac{1}{v_{\textrm{rel}}'}-\frac{1}{v_{\textrm{rel}}}\right)} ,
\end{equation}
\begin{equation}
    \mathcal{A}_\ell \equiv \frac{2^{2\ell+4}\pi}{M^4(2\ell)!}\frac{v_{\textrm{rel}}'^{\ell}v_{\textrm{rel}}^{\ell}}{(v_{\textrm{rel}}'+v_{\textrm{rel}})^{2\ell +2}}\prod \limits_{s=1}^{\ell}\sqrt{s^2+(\alpha/v_{\textrm{rel}})^2}\prod \limits_{s'=1}^{\ell}\sqrt{s'^2+(\alpha/v_{\textrm{rel}}')^2} \, ,
\end{equation}
\begin{equation}
\begin{aligned}
    \mathcal{X}_1 &\equiv \Bigg[~_2F_1\left(\ell+2-i\frac{\alpha}{v_{\textrm{rel}}'},\ell+1+i\frac{\alpha}{v_{\textrm{rel}}},2\ell+2,\frac{4v_{\textrm{rel}}'v_{\textrm{rel}}}{(v_{\textrm{rel}}'+v_{\textrm{rel}})^2}\right) \\
&~~~~~- \left(\frac{v_{\textrm{rel}}'+v_{\textrm{rel}}}{v_{\textrm{rel}}'-v_{\textrm{rel}}}\right)^2~_2F_1\left(\ell-i\frac{\alpha}{v_{\textrm{rel}}'},\ell+1+i\frac{\alpha}{v_{\textrm{rel}}},2\ell+2,\frac{4v_{\textrm{rel}}'v_{\textrm{rel}}}{(v_{\textrm{rel}}'+v_{\textrm{rel}})^2}\right) \Bigg] \, ,
\end{aligned}
\end{equation}
\begin{equation}
\begin{aligned}
    \mathcal{X}_2 &\equiv \Bigg[~_2F_1\left(\ell+1+i\frac{\alpha}{v_{\textrm{rel}}},\ell-i\frac{\alpha}{v_{\textrm{rel}}'},2\ell,\frac{4v_{\textrm{rel}}'v_{\textrm{rel}}}{(v_{\textrm{rel}}'+v_{\textrm{rel}})^2}\right) \\
&~~~~~- \left(\frac{v_{\textrm{rel}}'+v_{\textrm{rel}}}{v_{\textrm{rel}}'-v_{\textrm{rel}}}\right)^2~_2F_1\left(\ell-1+i\frac{\alpha}{v_{\textrm{rel}}},\ell-i\frac{\alpha}{v_{\textrm{rel}}'},2\ell,\frac{4v_{\textrm{rel}}'v_{\textrm{rel}}}{(v_{\textrm{rel}}'+v_{\textrm{rel}})^2}\right) \Bigg] \, .
\end{aligned}
\end{equation}

\section{Electric dipole matrix elements for the non-abelian SU($N$) model}
\label{sec:app_D}
In this section, we compute the electric dipole matrix element for the non-abelian dark matter model that is discussed in section~\ref{sec:non_abelian_model}. 
We give the result in full generality, as in the abelian case \eqref{matrixelement}, with $\bm{p}$ chosen along the $z$-direction,
\begin{equation}
\begin{aligned}
&\langle n\ell m|\bm{r}|\bm{p}\rangle^{[\textbf{adj}]} = \int d^3r\,\bm{r}\,\Psi_{n\ell m}^*(\bm{r})\Psi^{[\textbf{adj}]}_{\bm{p}}(\bm{r}) \\
&~~~= N\left[\sqrt{\ell(\ell+1)}(\delta_{m,1}-\delta_{m,-1})\bm{e}_x - i\sqrt{\ell(\ell+1)}(\delta_{m,1}+\delta_{m,-1})\bm{e}_y + 2(\ell+1)\delta_{m,0}\bm{e}_z\right]XG_1 \\
&~~~~~+N\left[-\sqrt{\ell(\ell+1)}(\delta_{m,1}-\delta_{m,-1})\bm{e}_x + i\sqrt{\ell(\ell+1)}(\delta_{m,1}+\delta_{m,-1})\bm{e}_y + 2\ell\delta_{m,0}\bm{e}_z\right]YG_2 \, ,
\label{matrixelement_su(N)}
\end{aligned}
\end{equation}
where
\begin{align}
N \equiv&  \frac{i^{\ell+3}(-1)^{n-\ell}}{(2\ell+1)!}\sqrt{\left(\frac{2}{na_0}\right)^3\frac{(n+\ell)!}{2n(n-\ell-1)!}}\left(\frac{2}{na_0}\right)^\ell
     \sqrt{\frac{\frac{\pi}{N} \frac{\alpha}{v_{\textrm{rel}}}}{e^{\frac{\pi}{N}\frac{\alpha}{v_{\textrm{rel}}}}-1}} \nonumber\\
  &\times \sqrt{\frac{\pi}{2\ell+1}}\frac{1}
    {\left[M^2v_{\textrm{rel}}^2\left(1+\frac{C_F^2\alpha^2}{n^2v_{\textrm{rel}}^2}\right)\right]^\ell}\,
    e^{-2\left[i(\ell+1-n) - \frac{\alpha}{2 N v_{\textrm{rel}}} \right] \arccot{\left(\frac{C_F\alpha}{nv_{\textrm{rel}}}\right)}} ,
\end{align}
\begin{align}
X &\equiv \frac{i(M v_{\textrm{rel}})^{\ell+1}2^{2\ell+4}}{M^5v_{\textrm{rel}}^5\left(1+\frac{C_F^2\alpha^2}{n^2v_{\textrm{rel}}^2}\right)^2}
     e^{-2i\arccot{\left(\frac{C_F\alpha}{nv_{\textrm{rel}}}\right)}}\prod \limits_{\kappa=1}^{\ell+1}\sqrt{\kappa^{2} + \left(\frac{\alpha}{2N v_{\textrm{rel}}}\right)^2} ,
\end{align}
\begin{align}
Y &\equiv \frac{n\ell(2\ell+1)(M v_{\textrm{rel}})^{\ell-1}2^{2\ell+3}}{C_F\alpha M^3v_{\textrm{rel}}^2\left(1+\frac{C_F^2\alpha^2}{n^2v_{\textrm{rel}}^2}\right)}
     \prod \limits_{\kappa=1}^{\ell-1}\sqrt{\kappa^{2} + \left(\frac{\alpha}{2N v_{\textrm{rel}}}\right)^2} ,
\end{align}
\begin{align}
G_1 &\equiv \left(1+i\frac{N \alpha}{2v_{\textrm{rel}}}\right)\,_{2}F_{1}\left(\ell+2+i\frac{\alpha}{2N v_{\textrm{rel}}},\ell+1-n,2\ell+2,\frac{-4iC_F\alpha}{nv_{\textrm{rel}}
      \left(1-i\frac{C_F\alpha}{nv_{\textrm{rel}}}\right)^2}\right) \nonumber\\
& -iN\frac{\alpha}{v_{\textrm{rel}}}e^{2i\arccot{\left(\frac{C_F\alpha}{nv_{\textrm{rel}}}\right)}}\,_{2}F_{1}\left(\ell+1+i\frac{\alpha}{2N v_{\textrm{rel}}},\ell+1-n,2\ell+2,
  \frac{-4iC_F\alpha}{nv_{\textrm{rel}}\left(1-i\frac{C_F\alpha}{nv_{\textrm{rel}}}\right)^2}\right) \nonumber\\
& -\left(1-i\frac{N \alpha}{2v_{\textrm{rel}}}\right)e^{4i\arccot{\left(\frac{C_F\alpha}{nv_{\textrm{rel}}}\right)}}\,_{2}F_{1}\left(\ell+i\frac{\alpha}{2N v_{\textrm{rel}}},
  \ell+1-n,2\ell+2,\frac{-4iC_F\alpha }{nv_{\textrm{rel}}\left(1-i\frac{C_F\alpha}{nv_{\textrm{rel}}}\right)^2}\right), 
\end{align} 
\begin{align}
G_2 &\equiv \left(1-\frac{N n}{2C_F}\right)\,_{2}F_{1}\left(\ell+1-n,\ell+i\frac{\alpha}{2N v_{\textrm{rel}}},2\ell,
    \frac{-4iC_F\alpha }{nv_{\textrm{rel}}\left(1-i\frac{C_F\alpha}{nv_{\textrm{rel}}}\right)^2}\right) \nonumber\\
&+N\frac{n}{C_F}e^{2i\arccot{\left(\frac{C_F\alpha}{nv_{\textrm{rel}}}\right)}}\,_{2}F_{1}\left(\ell-n,\ell+i\frac{\alpha}{2N v_{\textrm{rel}}},2\ell,
  \frac{-4iC_F\alpha }{nv_{\textrm{rel}}\left(1-i\frac{C_F\alpha}{nv_{\textrm{rel}}}\right)^2}\right) \nonumber\\
&-\left(1+\frac{N n}{2C_F}\right)e^{4i\arccot{\left(\frac{C_F\alpha}{nv_{\textrm{rel}}}\right)}}\,_{2}F_{1}\left(\ell-1-n,\ell+i\frac{\alpha}{2N v_{\textrm{rel}}},2\ell,
  \frac{-4iC_F\alpha }{nv_{\textrm{rel}}\left(1-i\frac{C_F\alpha}{nv_{\textrm{rel}}}\right)^2}\right) \, .
\end{align}
In the dipole matrix element the natural renormalization scale of the coupling $\alpha$ is $\mu_{\textrm{s}}$, which is of the order of the soft scale.

\bibliographystyle{JHEP.bst}
\bibliography{DMcsw.bib}

\end{document}